\documentclass[12pt]{iopart}
\pdfoutput=1
\usepackage[T1]{fontenc}\usepackage[latin1]{inputenc}
\usepackage{graphicx,color,bm,booktabs,wrapfig,microtype,afterpage,cite,booktabs} \graphicspath{{Figures/}}
\usepackage[charter,greekuppercase=italicized]{mathdesign}

\graphicspath{{Figures/}}

\usepackage{fancyhdr}
\renewcommand{\sectionmark}[1]{\markboth{}{}}
\renewcommand{\subsectionmark}[1]{\markright{}}
\usepackage{hyperref}
\hypersetup{
    colorlinks,
    citecolor=black,
    filecolor=black,
    linkcolor=black,
    urlcolor=black
}

\begin{document}
\title{Multipolar phases and magnetically hidden order: \\ Review of the heavy-fermion compound Ce$_{1-x}$La$_{x}$B$_{6}$}

\author{Alistair~S.~Cameron$^{1}$, Gerd~Friemel$^{2}$, Dmytro~S.~Inosov$^{1}$}

\address{$^{1}$Institut f\"ur Festk\"orperphysik, TU Dresden, D-01069 Dresden, Germany}
\address{$^{2}$Max-Planck-Institut f\"ur Festk\"orperforschung, Heisenbergstr. 1, 70569 Stuttgart, Germany}

\begin{abstract}
Cerium hexaboride is a cubic \textit{f}-electron heavy-fermion compound that displays a rich array of low-temperature magnetic ordering phenomena which have been the subject of investigation for more than 50 years. Its complex behaviour is the result of competing interactions, with both itinerant and local electrons playing important roles. Investigating this material has proven to be a substantial challenge, in particular because of the appearance of a ``magnetically hidden order'' phase, which remained elusive to neutron-scattering investigations for many years. It was not until the development of modern x-ray scattering techniques that the long suspected multipolar origin of this phase was confirmed. Doping with non-magnetic lanthanum dilutes the magnetic cerium sublattice and reduces the \textit{f}-electron count, bringing about substantial changes to the ground state with the emergence of new phases and quantum critical phenomena. To this day, Ce$_{1-x}$La$_{x}$B$_{6}$ and its related compounds remain a subject of intense interest. Despite the substantial progress in understanding their behaviour, they continue to reveal new and unexplained physical phenomena. Here we present a review of the accumulated body of knowledge on this family of materials in order to provide a firm standpoint for future investigations.\bigskip\bigskip\\
(Some figures may appear in colour only in the online journal)
\end{abstract}


\date{\today}
\maketitle


\tableofcontents
\pagebreak

\section{Introduction}


CeB$_{6}$ has been the subject of continued examination for many years, its simple cubic structure containing a rich plethora of low-temperature ordering phenomena. Aside from a more conventional low-temperature antiferromagnetic (AFM) phase it is well known for displaying a ``magnetically hidden order'' phase, so called because it is invisible to neutron scattering in no applied field, and systematically investigating this system has proven to be a challenge for both experiment and theory. Interest originates not only from the complexity of its low-temperature phase diagram but also in its origin: a variety of intricate spin- and orbital-ordered phases in a system with a simple, high-symmetry crystal structure indicates complex electronic correlations arising from an interplay of several competing energy scales. Indeed, as an \textit{f}-electron heavy-fermion compound it is well placed to display some of the more exotic phenomena which are of interest to modern condensed matter physics. The goal of this review is to draw a coherent picture from the myriad of investigations which have been done into CeB$_{6}$ and its doped derivatives such as Ce$_{1-x}$La$_{x}$B$_{6}$ since its original synthesis in 1932 \cite{Sta32}. Following its discovery, much attention was paid to its potential as a cathode material \cite{Laf51}, with some early interest in the following decades regarding its physical behaviour \cite{Pad61, Geb68, Nic69, Win75} before modern condensed matter physics became interested in its magnetic ordering phenomena.

\begin{figure}[t]
	\includegraphics[width=\textwidth]{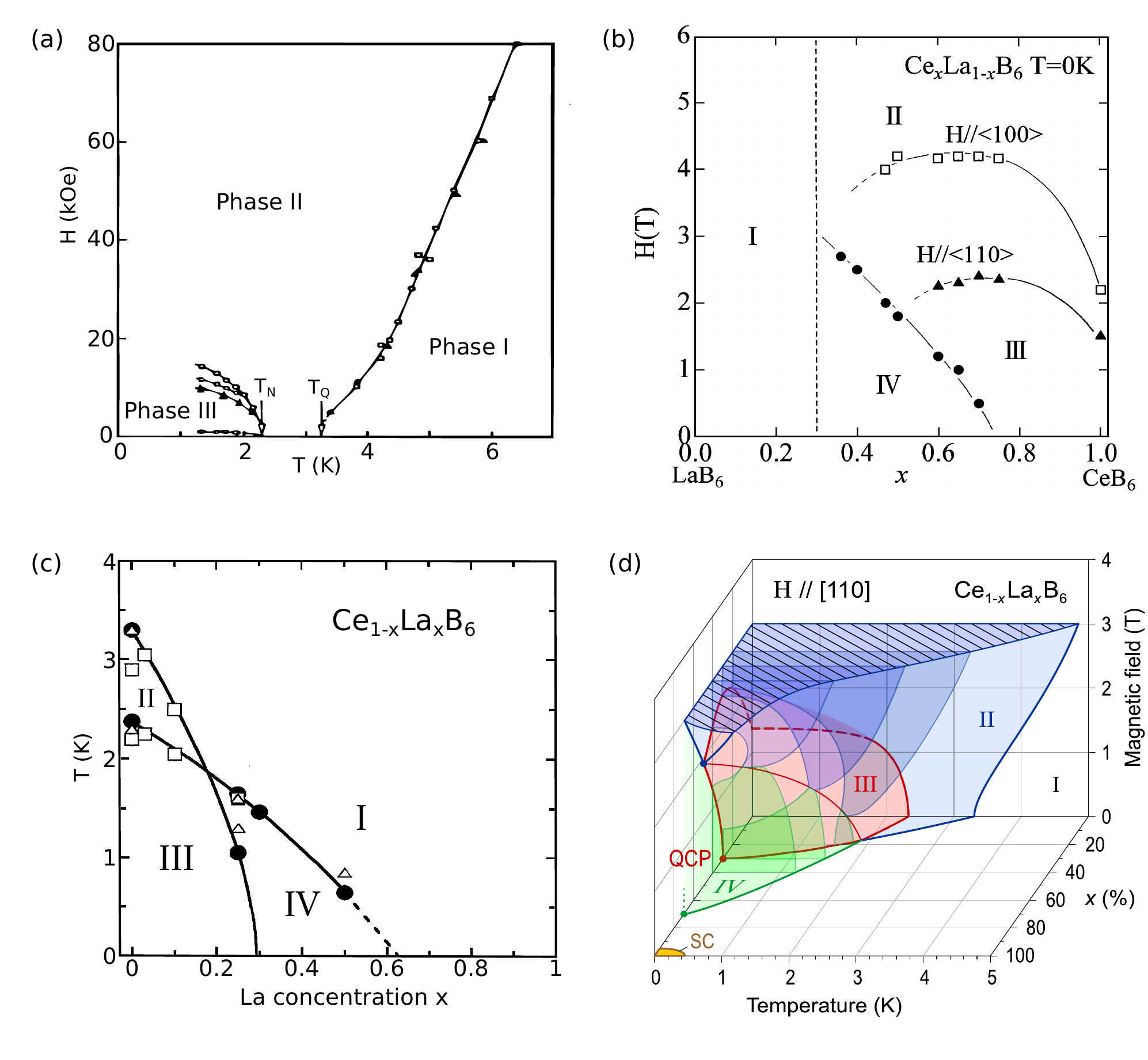}
	\caption{(a) The temperature\,--\,magnetic field phase diagram for CeB$_{6}$, reproduced from Ref.~\cite{Tak80}. Squares are for $\mathbf{B} \parallel \langle100\rangle$, circles for $\mathbf{B} \parallel \langle110\rangle$ and triangles are for $\mathbf{B} \parallel \langle111\rangle$. (b) Doping and applied magnetic field diagram for measurements extrapolated to $T=0$, taken from Ref.~\cite{Kob00a}. (c) Temperature and doping phase diagram, reproduced from Ref.~\cite{Tay97}, for the system in no applied field. (d) Summary of the temperature, magnetic field and doping phase diagrams for the field direction along the $\langle110\rangle$ direction, reproduced from Ref.~\cite{Fri15}. Approximate location of phases I (clear), II (blue), III (red) and IV (green) are shown, with the small superconducting dome of LaB$_{6}$ schematically shown in yellow \cite{Mat68}. Quantum critical points are indicated at $x \approx 0.3$ and $x \approx 0.7$ for zero field, with a third shown at $x \approx 0.5$ in an applied field of $\approx 2$~T. The cuts in the phase diagram are shown for reported data as a function of lanthanum doping \cite{Kun10, Erk87, Suz05, Hir98}.}\label{Phase_diagram}
\end{figure}

Setting the scene for this review, we present the phase diagrams for Ce$_{1-x}$La$_{x}$B$_{6}$ in Fig.~\ref{Phase_diagram}. Panel (a) shows the applied magnetic field vs. temperature phase diagram for pure CeB$_{6}$ \cite{Tak80}. We see that in zero field the ground state of CeB$_{6}$ is antiferromagnetic (phase III). It gives way to an antiferroquadrupolar (AFQ) phase (phase II) at $T_{\mathrm{N}} = 2.4$~K, which survives up until $T_{\mathrm{Q}} = 3.2$~K \cite{Zir84}, above which the system is in the paramagnetic state (phase I). The transition temperatures in early measurements were slightly sample dependent, with noticeable changes arising from crystal growth technique and annealing \cite{Lee72, Fuj80, Bur82, Zir84, Eff85a}. As a function of increasing field, $\mathbf{B}$, the AFM state evolves from phase III to phase III$^{\prime}$, which shall be discussed later in section \ref{section_magnetic_structure}, and at low fields $T_{\mathrm{Q}}(\textbf{B})$ possesses a \textit{positive} slope indicating that this phase is stabilised in field. Panel (b) shows the lanthanum doping\,--\,applied field phase diagram extrapolated to zero temperature~\cite{Kob00}. We see that for concentrations of lanthanum higher than $\sim$\,0.3, we enter phase IV at low field, the nature of which is discussed in section \ref{Sec:La_Doped}. This phase is stabilised against the field with increasing lanthanum doping, competing with phase III. Panel (c) shows the lanthanum doping vs. temperature phase diagram in zero field~\cite{Tay97}. Panel (d) summarizes the various phases of Ce$_{1-x}$La$_{x}$B$_{6}$ as a function of temperature, applied magnetic field and doping parameter $x$ in a 3D sketch \cite{Fri15}.

Pure CeB$_{6}$ expresses three distinct phases, labelled as phases I, II and III, and with lanthanum doping a further phase IV appears. There are zero-field quantum critical points (QCP) at $x \approx 0.3$, where phase III gives way to phase IV, and at $x \approx 0.7$, where phase IV disappears. In an applied field a tri-critical point between phases II, III and IV appears, whose position in doping depends on the crystal direction along which the magnetic field is applied. For the field applied in the $\langle110\rangle$ direction this point sits at around 2~T and $x \approx 0.5$, however for fields applied along the $\langle100\rangle$ direction, shown in panel (b), this point is shifted to higher fields and lower cerium content.
The phase diagrams reproduced here are only a small selection of the data available in the published literature. In table \ref{table}, we present a list of all the published phase diagrams known to us, covering a wide variety of techniques and experimental configurations. With the available parameters of magnetic field (\textbf{B}), temperature (\textit{T}), pressure (\textit{P}) and lanthanum concentration (\textit{x}), as well as the changing behaviour of the phase diagram with magnetic field applied along the three principal crystal directions, this represents a multidimensional parameter space which no single investigation can cover.

\begin{table}
\scriptsize\begin{center}
    \begin{tabular}{lllll}\toprule
\textbf{Publication} & \textbf{Type} & \textbf{Field direction} & \textbf{Technique} & \textbf{Ref.}\\ \midrule
	Magnetic ordering in cerium hexaboride & \textbf{B}\,--\,\textit{T} & $\langle110\rangle$ & Neutron scattering & \cite{Bur82} \\ \midrule
    Thesis: Jean-Michelle Effantin & \textbf{B}\,--\,\textit{T} & $\langle100\rangle$, $\langle111\rangle$ & Neutron scattering & \cite{Eff85} \\ \midrule
	Magnetic form factor measurements in cerium hexaboride & \textbf{B}\,--\,\textit{T} & $\langle111\rangle$ & Neutron scattering & \cite{Bur82a} \\ \midrule
    Magnetic phase diagram of CeB$_{6}$ & \textbf{B}\,--\,\textit{T} & $\langle100\rangle$, $\langle111\rangle$ & Neutron scattering & \cite{Eff85a} \\ \midrule
    \begin{tabular}{@{}l@{}} Extension of the temperature\,--\,magnetic field phase \\ diagram of CeB$_{6}$ \end{tabular}& \textbf{B}\,--\,\textit{T} & $\langle100\rangle$, $\langle110\rangle$ & Magnetometry & \cite{Goo04} \\ \midrule
   	\begin{tabular}{@{}l@{}} The magnetic behaviour of CeB$_{6}$: Comparison between \\ elastic and inelastic neutron scattering, initial susceptibility \\ and high-field magnetization \end{tabular} & \textbf{B}\,--\,\textit{T} & $\langle100\rangle$, $\langle110\rangle$, $\langle111\rangle$ & \begin{tabular}{@{}l@{}}  Neutron scattering \\ and magnetisation\end{tabular} & \cite{Hor81} \\ \midrule
	\begin{tabular}{@{}l@{}}Magnetoresistance and magnetisation anomalies in CeB$_{6}$  \end{tabular}& \textbf{B}\,--\,\textit{T} & $\langle110\rangle$, $\langle111\rangle$ & Magnetoresistance & \cite{Bog06} \\ \midrule
	Magnetic properties of a CeB$_{6}$ single crystal & \textbf{B}\,--\,\textit{T} & $\langle100\rangle$, $\langle110\rangle$, $\langle111\rangle$ & Magnetisation & \cite{Kaw80} \\ \midrule
	Electrical resistivity and magnetoresistance & \textbf{B}\,--\,\textit{T} & $\langle100\rangle$, $\langle110\rangle$, $\langle111\rangle$ & Magnetoresistance  & \cite{Tak80} \\ \midrule	 
	Anomalous specific heat in CeB$_{6}$ & \textbf{B}\,--\,\textit{T} & $\langle100\rangle$, $\langle110\rangle$, $\langle111\rangle$ & Heat capacity & \cite{Fuj80} \\ \midrule
	\begin{tabular}{@{}l@{}}Enhancement of band magnetism and features of the \\ magnetically ordered state in the CeB$_{6}$ compound with \\ strong electron correlations \end{tabular}& \textbf{B}\,--\,\textit{T} & $\langle111\rangle$, $\langle110\rangle$ & Magnetisation & \cite{Slu07} \\ \midrule
	\begin{tabular}{@{}l@{}}Pressure dependence of quadrupole ordering temperature \\ $T_{\mathrm{Q}}$ in CeB$_{6}$ \end{tabular}& $T\!$\,--\,$P$ & --- & Resistivity & \cite{Kob00} \\ \midrule
		High pressure studies of cerium hexaboride & \textbf{B}\,--\,\textit{T}, $T\!$\,--\,$P$ & $\langle100\rangle$, $\langle110\rangle$ & \begin{tabular}{@{}l@{}} Magnetic \\ susceptibility and \\ magnetoresistance \end{tabular} & \cite{Bra85} \\ \midrule
    Specific heat of CeB$_{6}$ under high pressure & $T\!$\,--\,$P$ & --- & Heat capacity & \cite{Sul94} \\ \midrule
	Dense Kondo behavior in CeB$_{6}$ and its alloys & \textbf{B}\,--\,\textit{T} & $\langle111\rangle$ & Ultrasound & \cite{Kom83} \\ \midrule	
	\begin{tabular}{@{}l@{}}Magnetic phase diagram of Ce$_{0.5}$La$_{0.5}$B$_{6}$ under \\  high pressure \end{tabular}& \textbf{B}\,--\,\textit{T} & $\langle100\rangle$ & Magnetization & \cite{Han01} \\ \midrule		
	\begin{tabular}{@{}l@{}}Stable Existence of phase IV inside phase II under pressure\\  in Ce$_{0.8}$La$_{0.2}$B$_{6}$ \end{tabular}& \textbf{B}\,--\,\textit{T} & $\langle100\rangle$, $\langle110\rangle$ & Magnetization & \cite{Kun10} \\ \midrule		
	\begin{tabular}{@{}l@{}}Neutron scattering study of the antiferroquadrupolar \\ ordering in CeB$_{6}$ and Ce$_{0.75}$La$_{0.25}$B$_{6}$ \end{tabular}& \textbf{B}\,--\,\textit{T} & $\langle110\rangle$ & Neutron scattering & \cite{Erk87} \\ \midrule
	\begin{tabular}{@{}l@{}} Magnetic phase diagram of Ce$_{x}$La$_{1-x}$B$_{6}$ studied by static \\ magnetization measurement at very low temperatures \end{tabular}& \textbf{B}\,--\,\textit{T}\,--\,$x$ & $\langle100\rangle$ & Magnetisation & \cite{Tay97} \\ \midrule
	 \begin{tabular}{@{}l@{}}Elastic properties and magnetic phase diagrams of dense \\ Kondo compound Ce$_{0.75}$La$_{0.25}$B$_{6}$ \end{tabular}& \textbf{B}\,--\,\textit{T}  & $\langle100\rangle$, $\langle110\rangle$, $\langle100\rangle$ & Ultrasound & \cite{Suz05} \\ \midrule
		\begin{tabular}{@{}l@{}}Evidence for hidden quadrupolar fluctuations behind the\\ octupole order in Ce$_{0.7}$La$_{0.3}$B$_{6}$ from resonant x-ray \\ diffraction in magnetic fields \end{tabular}& \textbf{B}\,--\,\textit{T} & $\langle100\rangle$ & RXS & \cite{Mat14} \\ \midrule
		\begin{tabular}{@{}l@{}} Magnetic phase diagrams of kondo compounds \\ {Ce}$_{0.75}${L}a$_{0.25}${B}$_{6}$  and {Ce}$_{0.6}${L}a$_{0.4}${B}$_{6}$ \end{tabular}& \textbf{B}\,--\,\textit{T} & $\langle100\rangle$ & \begin{tabular}{@{}l@{}} Ultrasound and \\ heat capacity \hfill \end{tabular} & \cite{Suz98} \\ \midrule	
		\begin{tabular}{@{}l@{}} Appearance of the phase {IV} in {C}e$_{x}${L}a$_{1-x}${B}$_{6}$ at $x \approx 0.8$ \end{tabular}& \textbf{B}\,--\,\textit{T} & $\langle100\rangle$, $\langle110\rangle$ & \begin{tabular}{@{}l@{}}	Magnetisation, \\ resistivity \\ and ultrasound  \end{tabular} & \cite{Kob03} \\ \midrule
		\begin{tabular}{@{}l@{}} Drastic change of the magnetic phase diagram of \\ {C}e$_{x}${L}a$_{1-x}${B}$_{6}$ between $x = 0.75$ and 0.5 \end{tabular} & \textbf{B}\,--\,\textit{T} & $\langle100\rangle$, $\langle110\rangle$ & Resistivity & \cite{Hir98} \\ \bottomrule	
    \end{tabular}\vspace{3pt}
        \caption{This table lists all the known phase diagrams of CeB$_{6}$ and Ce$_{1-x}$La$_{x}$B$_{6}$, indicating the type of phase diagram, the crystal direction along which the magnetic field was applied and the technique of measurement.}
    \label{table}\end{center}

\end{table}

Despite the early classification of the ground state of CeB$_{6}$ as an AFM state, this material continues to reveal unexpected behaviour, such as the ferromagnetic low-energy modes revealed by neutron scattering which dominate the low energy excitation spectrum \cite{Jan14}. Such measurements illustrate that there is still much to be discovered regarding CeB$_{6}$ and related heavy-fermion systems, both in terms of a complete set of experimentally observed behaviour and the theoretical background with which to describe it. Here we present a review of published information, alongside new measurements, in order to build a comprehensive picture of our current understanding of Ce$_{1-x}$La$_{x}$B$_{6}$.

\section{Parent Compound -- CeB$_{6}$}

\subsection{Crystal and electronic structure of CeB$_{6}$}
\label{section:crystal_and_gs}

\begin{figure}[b]
\begin{center}
	\includegraphics[width=0.6\linewidth]{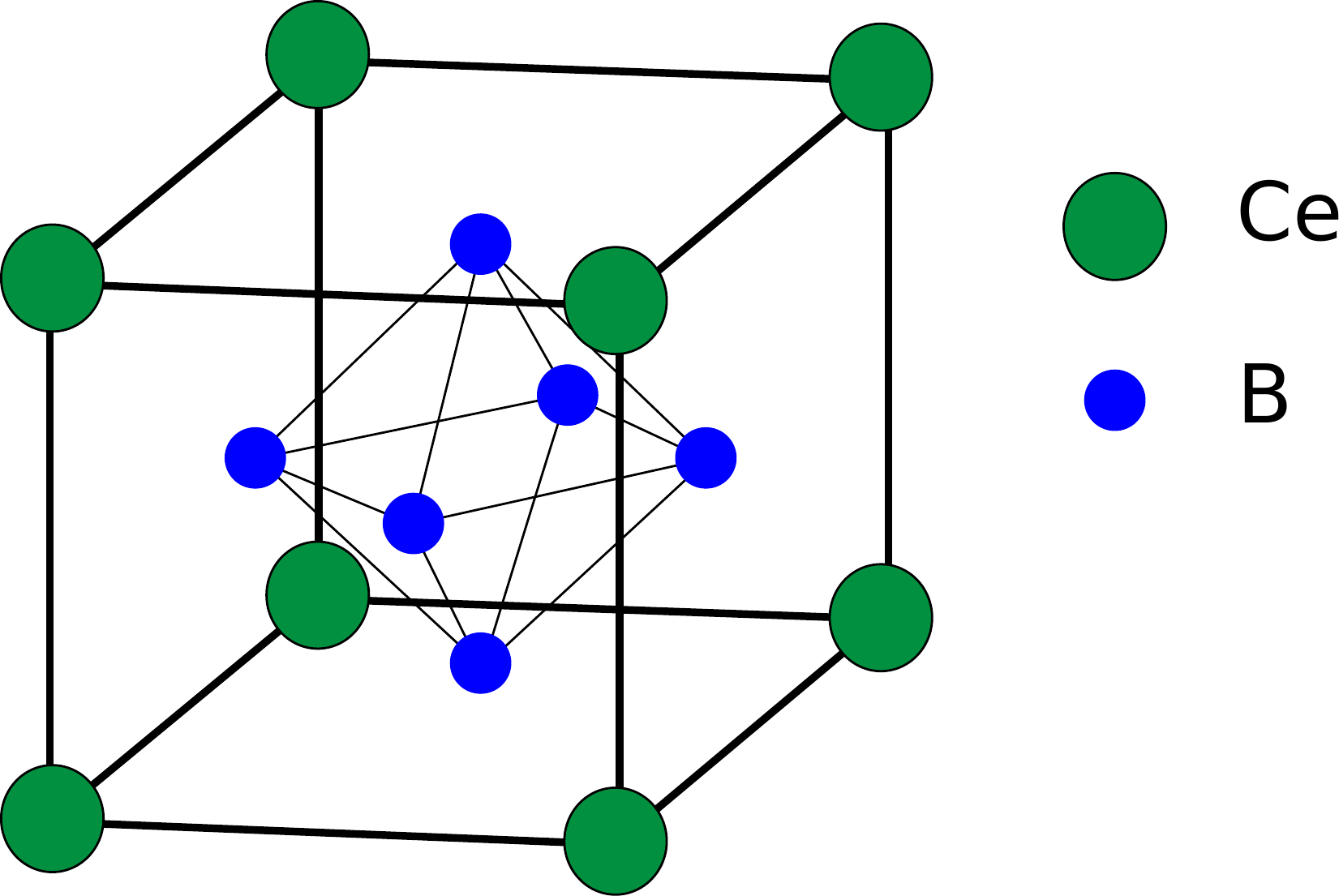}
\end{center}
	\caption{Sketch of the crystal structure of CeB$_{6}$, showing the cubic unit cell. }
		\label{Crystal_structure}
\end{figure}

CeB$_{6}$ possesses a simple-cubic structure comprised of Ce$^{3+}$ ions in a cubic lattice separated by B$_{6}$ octahedra, shown in Fig.~\ref{Crystal_structure}, characterized by the $Pm\bar{3}m$ space group and a unit cell parameter $a = 4.14$~\AA. The structure can be considered as a combination of two cubic lattices, each arranged so that it is body-centred to the other, with one lattice consisting of the B$_{6}$ octahedra and the other of cerium ions. It was noted in early research that the B$_{6}$ octahedra, which form a covalently bonded structure, require two additional electrons from the cerium ions in order to stabilise \cite{Gru85, Lan54}. The outer electronic structure of cerium is [Xe]$4f^{1}5d^{1}6s^{2}$, with the two \textit{s}-electrons being donated to the boron octahedra, and it is generally considered that the \textit{f}-electron states remain localised with the \textit{d}-electrons forming the conduction band, resulting in the Ce$^{3+}$ ion. However, the nature of the \textit{f}-electron states, whether they are localised or itinerant, is a topic still under debate. Whilst much of the low-temperature ordering phenomena can be understood in the more simplistic picture of localised \textit{f}-electrons and itinerant \textit{d}-electrons, it is important to keep in mind that under a more rigorous scrutiny the behaviour of the \textit{f}-electrons seems far less trivial, and discussion of this is reserved for section~\ref{Itinerant}. It is, however, the capacity for the \textit{f}-electron states to support various types of multipolar ordering that lead to the rich low-temperature phase diagram of CeB$_{6}$.

Initially, it was thought that the Ce$^{3+}$ multiplet 4$f^{1}$ is split by the crystalline electric field into a $\mathrm{\Gamma}_{7}$ ground state with a $\mathrm{\Gamma_{8}}$ excited state \cite{Nic69}. However, this was later reversed to the $\mathrm{\Gamma}_{8}$ quartet ground state, which is fourfold degenerate and possesses 2 orbital and 2 spin degrees of freedom, located 46~meV below the $\mathrm{\Gamma}_{7}$ doublet state \cite{Zir84, Loe85, Sat84}. This was not able to fully explain experimental observations, such as an anomalous change in entropy at around 30~K, seen in specific heat data \cite{Hor81}, or the magnetic moment of the ground state \cite{Kas81, Got83, Kaw80}. The Raman scattering measurements provide an explanation for these observations, indicating that the $\mathrm{\Gamma}_{8}$ quartet is further split into two doublets, $\mathrm{\Gamma}_{8,1}$ and $\mathrm{\Gamma}_{8,2}$, separated by around 30~K \cite{Zir84}. The ground state of CeB$_{6}$ and its interpretation in various theoretical models is discussed in more detail in sections \ref{Mean_field} and \ref{section:Alternative_II}.

\subsection{Magnetic structure in the antiferromagnetic phase of CeB$_{6}$}
\label{section_magnetic_structure}

\begin{figure}[t]
\begin{center}
	\includegraphics[width=0.9\linewidth]{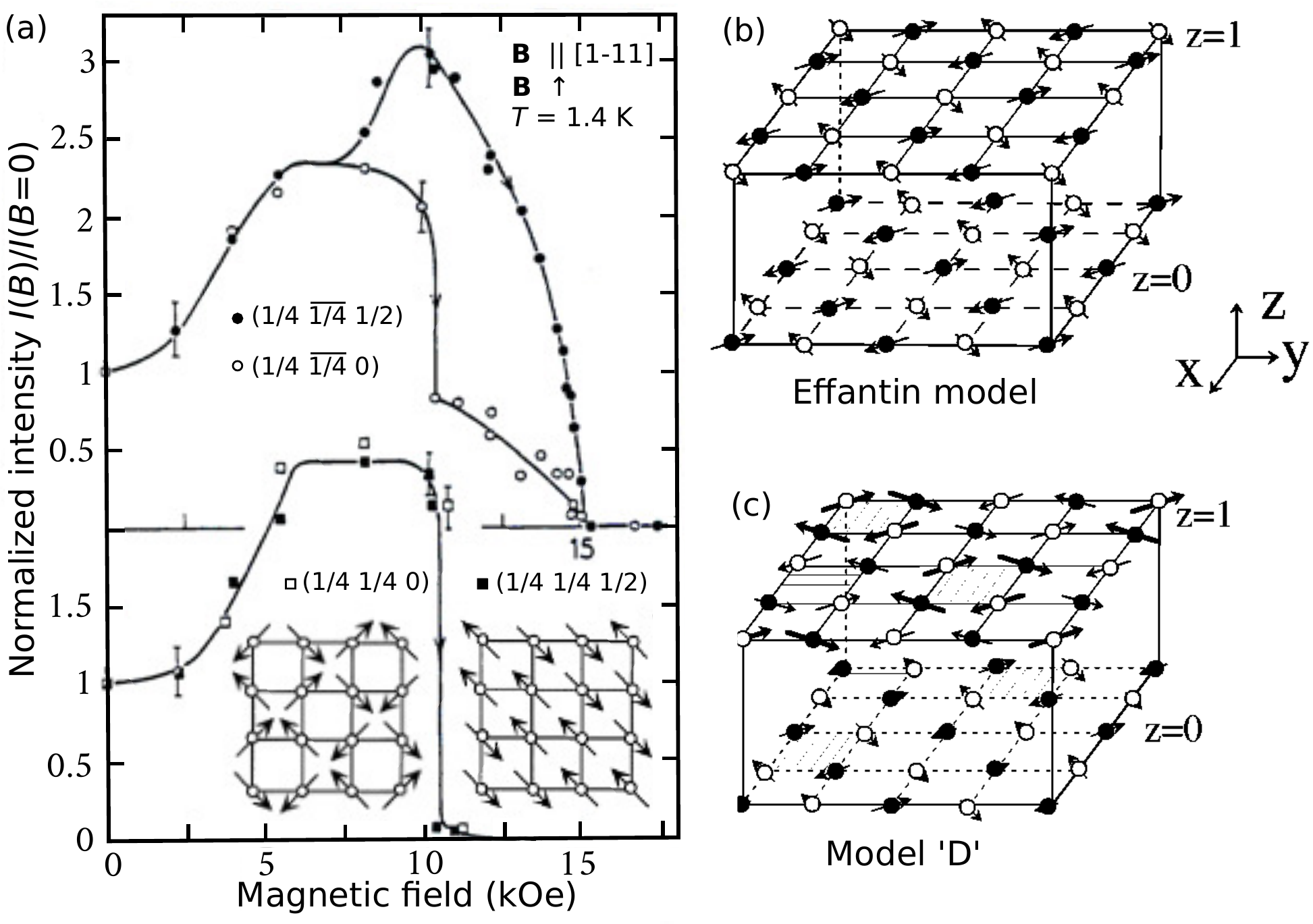}
\end{center}
	\caption{(a) Magnetic field dependence of the AFM Bragg peaks originating from the $K_{xy}$ domain. Insets depict the ordering of the spins in phases III and III$^{\prime }$ in the (0 0 1) plane, reproduced from Ref.~\cite{Eff82}. (b) Effantin model of spin alignment in the AFM phase. (c) Model `D' proposed by Zharko \textit{et~al.}, reproduced from Ref.~\cite{Zah03}, copyright by the American Physical Society.}
		\label{AFM_structure}
\end{figure}

In zero field, CeB$_{6}$ orders antiferromagnetically at a temperature $T_{\mathrm{N}} = $ 2.4 K, with this phase labelled `phase III'. From neutron diffraction measurements in zero field \cite{Eff82, Eff85, Bur82, Zah03}, the AFM ordering vectors $\textbf{q}_{1} = \big ( \frac{1}{4} \, \frac{1}{4} \, 0 \big )$, $\textbf{q}_{2} =  \big ( \frac{1}{4} \, \bar{\frac{1}{4}}  \, 0 \big )$ and $\textbf{q}_{1}^{\prime} = \big ( \frac{1}{4} \, \frac{1}{4} \, \frac{1}{2} \big )$, $\textbf{q}_{2}^{\prime} = \big (\frac{1}{4} \, \bar{\frac{1}{4}} \, \frac{1}{2} \big )$ have been established. These vectors describe a double-\textbf{q}, or  2\textbf{q}$_{1}$ -- \textbf{q}$_{1}^{\prime}$ structure. A variety of models have been proposed to describe the spin structure which results in these ordering vectors, with some of the earliest work being undertaken by Effantin \textit{et~al.}, where they proposed a transverse sine modulated non-collinear order to describe their results \cite{Eff82, Bur82}. This is shown as an inset in Fig~\ref{AFM_structure}\,(a) and reproduced in greater detail in Fig.~\ref{AFM_structure}\,(b), with magnetic moments found to be orthogonal and aligned along the (110) and (1$\bar{1}$0) directions. This original model proposed wave amplitudes associated with the $\textbf{q}$ and $\textbf{q}^{\prime}$ vectors to be almost equal, giving an ordered magnetic moment of 0.28\,$\mu_{\mathrm{B}}$ at 1.3\,K. This, however, was unable to explain later $\mu$SR measurements in zero field which found eight precession frequencies for the muons, whereas the model would provide only three \cite{Fey94, Fey95}. Later work by Zaharko \textit{et~al.} proposed a selection of new models to describe the spin pattern in the AFM phase, with their favoured model  `D' shown in Fig.~\ref{AFM_structure}\,(c) \cite{Zah03}. This model was able to describe the available neutron scattering data, could also account for the $\mu$SR results and was a feasible conclusion from theoretical predictions \cite{Zah03, Eff82, Kus01, Fey95}. The moments in this model are not equal, and they are modulated along the (001) direction as well as in-plane. Using the notation of Fig.~\ref{AFM_structure}\,(b) the moments were found to be $\mu = 0.01\,\mu_{\mathrm{B}}$ and $0.136(7)\,\mu_{\mathrm{B}}$ for the $z = 0$ plane and $\mu = 0.744(16)\,\mu_{\mathrm{B}}$ and $0.543(16)\,\mu_{\mathrm{B}}$ for the $z = 1$ plane. This complicated structure is attributed to competition between the dipolar, quadrupolar and octupolar interactions \cite{Zah03}.

In zero field there exist three magnetic domains within the AFM state, $K_{yz}$, $K_{zx}$ and $K_{xy}$, with the indices denoting the planes in which spins are ordered, and therefore the illustrations in Fig.~\ref{AFM_structure} are of the $K_{xy}$ domain. Neutron scattering reveals that in an applied field the degeneracy of these domains can be broken and a selection process takes place \cite{Eff82}. Fig.~\ref{AFM_structure}\,(a) shows the field dependence of the normalised intensity of the $\big ( \frac{1}{4} \, \frac{1}{4} \, 0 \big )$, $\big ( \frac{1}{4} \, \bar{\frac{1}{4}}  \, 0 \big )$, $ \big ( \frac{1}{4} \, \frac{1}{4} \, \frac{1}{2} \big )$ and $ \big (\frac{1}{4} \, \bar{\frac{1}{4}} \, \frac{1}{2} \big )$ peaks as a function of magnetic field applied along the (1$\bar{1}$1) direction. Whilst one would not normally expect domain selection for a field applied along a $\langle111\rangle$ direction when the domains exist in (001) planes, in this measurement a slight misalignment of the magnetic field direction, such that $B_{z} > B_{x}$ and $B_{y}$, was sufficient to break the degeneracy. At low fields, the intensity of all of the Bragg peaks in this figure increases as this domain is preferentially selected, until around 6~kOe we are left in a single domain state and the intensity as a function of field flattens out. Further measurements in Ref.~\cite{Eff82} show that a single domain is selected for fields applied along a $\langle001\rangle$ direction, the equivalent situation to Fig.~\ref{AFM_structure}\,(a), whilst two domains are selected for fields applied along the $\langle110\rangle$ directions. However, this figure also indicates that at higher fields a magnetic phase transition takes place. Here we see that the Bragg reflections at $\big ( \frac{1}{4} \, \frac{1}{4} \, 0 \big )$ and $ \big ( \frac{1}{4} \, \frac{1}{4} \, \frac{1}{2} \big )$ are sharply suppressed in an applied field of around 1~T, whilst the two other Bragg peaks remain in higher fields and the ratio of their moments is strongly modified. This indicates a change in magnetic structure from a double-\textbf{q} to a single-\textbf{q} type, with the new spin arrangement illustrated as an inset to Fig.~\ref{AFM_structure}\,(a), and this new collinear AFM phase is referred to as phase III$^{\prime}$. The domain selection and emergence of phase III$^{\prime}$ are observed for all applied field directions, and have also been seen in lanthanum doped samples \cite{Kun11}.

\subsection{Phase II -- what orders?}
\label{Low_temp_phases}

The order parameter responsible for phase II proved difficult to determine, since it is `hidden' to neutron diffraction measurements in zero field \cite{Eff85}. It is stabilised by an applied magnetic field, as evidenced by the enhanced transition temperature at low field, and in field dipolar moments at the Ce$^{3+}$ sites are induced, which were seen to order with a wavevector of $( \frac{1}{2} \, \frac{1}{2} \, \frac{1}{2} )$  by neutron scattering \cite{Mig87, Eff85a}. This discovery provided a helpful clue to the origin of this phase, and was concurrent with an antiferro-orbital ordering mechanism that was proposed for phase II \cite{Han84}. Since it is related to an ordering of the quadrupolar moments, it is referred to as an antiferroquadrupolar (AFQ) state \cite{Ohk83}. This modulation of the spins is a result of the Zeeman interaction of the $\mathrm{\Gamma}_{8}$ quartet in a magnetic field \textbf{B}, $\mathcal{H}_{\mathrm{Z}} = g \mu_{\mathrm{B}}\,\mathbf{J}\cdot\mathbf{B}$, where \textit{g} is the Land\'{e} factor ($g = \frac{6}{7}$) and $\mathbf{J}$ is the total angular momentum. With the total angular momentum written in terms of the orbital operator $\tau$ and the spin operator $\sigma$, the Zeeman term is written as \cite{Ohk85}
\begin{equation}
\mathcal{H}_{\mathrm{Z}} = 2 \mu_{\mathrm{B}} \bigg[ \sigma_{\!x} (1 + \frac{8}{7} T^{x})B_{x} + \sigma_{\!y}(1 + \frac{8}{7}T^{y})B_{y} + \sigma_{\!z}(1 + \frac{8}{7}T^{y})B_{z} \bigg]
\end{equation}
where
\begin{equation}
T^{x} = \frac{\sqrt{3}}{2} \tau^{x} - \frac{1}{2}\tau^{z}, \; \; \; T^{y} = -\frac{\sqrt{3}}{2} \tau^{x} - \frac{1}{2}\tau^{z}, \; \; \; T^{z} = \tau^{z}.
\end{equation}
For an $O_{xz}$ type quadrupolar order (orbital operator $\tau \neq 0$), with the magnetic field applied along (110), an AFM dipolar moment $J_{z}$ is induced \cite{Shi97}, where for low enough fields this moment increases linearly with field and disappears in zero applied field. This has been observed in experiment \cite{Mig87, Eff85a, Erk87}. Alongside this a ferromagnetic moment is induced, seen in magnetisation measurements \cite{Slu07}. To date, the appearance of this ferromagnetic moment has not been well studied, but inelastic neutron scattering in zero field has indicated that the system is close to a ferromagnetic instability \cite{Jan14}. Further evidence supporting AFQ order came from NMR experiments in an applied field, where a splitting of the $^{11}$B resonance line indicated the presence of AFQ order \cite{Kaw81, Tak83}. The underlying quadrupolar order, $\tau_{z} = \pm\tau$, results in an anisotropic Zeeman splitting for different directions of applied magnetic field. This causes an anisotropy in the phase diagram for the phase II -- phase III boundary such that $B_{\mathrm{c}}^{\langle 001 \rangle} \geq B_{\mathrm{c}}^{\langle 110 \rangle} \geq B_{\mathrm{c}}^{\langle 111 \rangle}$ and the AFQ transition $T_{\mathrm{Q}}^{\langle 001 \rangle} \leq T_{\mathrm{Q}}^{\langle 110 \rangle} \leq T_{\mathrm{Q}}^{\langle 111 \rangle}$ \cite{Hir98, Suz05}. Furthermore, the temperature dependence of the field-induced neutron scattering intensity in the AFQ phase follows an order-parameter like increase below $T_{\mathrm{Q}}$, preceding a suppression towards zero at lower temperatures, which suggests that the AFM and AFQ phases are competing \cite{Ser01}.

The first direct observation of orbital ordering came from resonant x-ray scattering (RXS), where probing of the L$_{3}$ edge of the Ce ion indicated a splitting of the $5d$ orbital, induced by the Coulomb interaction between the $5d$ and $4f$ orbitals upon orbital ordering \cite{Nak01, Yak01}. This ordering was found to have the same wavevector of $( \frac{1}{2} \, \frac{1}{2} \, \frac{1}{2} )$ as seen from the in-field neutron scattering measurements, which indicated that the dipole order was indeed modulated to the underlying AFQ order. High-field measurements reveal a maximum of $T_{\mathrm{Q}}$ of 10 K at around 35 T, subsequently decreasing to about 8 K in an applied field of 60 T, with full suppression of $T_{\mathrm{Q}}$ not yet achieved \cite{Mur98}. Theoretical models \cite{Shi98, Kas97, Lov02} have predicted that AFQ ordering should introduce distortions to the crystal lattice. These are expected to be small, as the quadrupole coupling to the lattice strain is weak due to differing symmetries. In both pure and lanthanum-doped samples a small tetragonal distortion upon the transition to the antiferromagnetic phase was noted in magnetoelastic measurements \cite{Ser88}, however no lattice distortion on the transition to the AFQ phase has so far been seen. For a more comprehensive discussion of multipolar ordering than will be presented here, see for instance Refs.~\cite{San09, Kur09}, as the topic is too large to cover in a dedicated review on Ce$_{1-x}$La$_{x}$B$_{6}$ and has already been well described elsewhere.

\subsection{Mean-field description of the ordering phenomena in CeB$_{6}$}
\label{Mean_field}

Compounds with 3\textit{d} or 4\textit{f} ions show not only common types of dipolar magnetic order but are also able to express multipolar order from the available orbital degrees of freedom. The AFQ phase in CeB$_{6}$ is of similar character, however it displays some unusual behaviour and describing this theoretically has been far from straightforward. In an applied magnetic field the AFQ phase stabilises with increasing magnetic field, with $T_{\mathrm{Q}}$ showing a \textit{positive} slope at low field (see Fig.~\ref{Phase_diagram}). Several mechanisms were proposed to explain this unusual behaviour of $T_{\mathrm{Q}}(\textbf{B})$ at low field. In their early work characterising the phase diagram, Effantin \textit{et~al.} proposed that the magnetic properties of CeB$_{6}$ were dominated by the interplay between single-site Kondo fluctuations and the magnetic and quadrupolar interactions. Here, the dense Kondo state suppresses the indirect interactions which establish AFQ order, leading to the reduction of $T_{\mathrm{Q}}$ in zero field. With increasing field the Kondo state is suppressed and the AFQ ordering is enhanced, leading to the positive slope of $T_{\mathrm{Q}}$ at low field. Contrastingly, Hanzawa and Kasuya \cite{Han84} constructed a mean-field model assuming a ground-state $\mathrm{\Gamma}_{7}$ doublet and an excited $\mathrm{\Gamma}_{8}$ quartet, where the interaction was dominated by quadrupole-quadruple coupling between the neighbouring cerium ions in a simple two-sublattice model. This model was able to reproduce the phase diagram and explain a variety of experimental results, although the authors noted that there were aspects of the known behaviour of CeB$_{6}$ that they were unable to reproduce, such as the $^{11}$B line splitting from NMR measurements, and attributed these discrepancies to over-simplification of their level scheme. As discussed in section~\ref{section:crystal_and_gs}, it was discovered that this level scheme was not correct \cite{Sat84, Shi97}, and so later studies tended to restrict themselves to the $\mathrm{\Gamma}_{8}$ ground state only.

A contrasting explanation, proposed by Ohkawa~\textit{et~al.}, was based on the RKKY interaction between neighbouring $\mathrm{\Gamma}_{8}$ orbitals rather than the quadrupole interaction, derived from a simplified periodic Anderson model \cite{Ohk83, Ohk85}. From this, higher-order interactions originating from spin and orbital degeneracies were responsible for the enhancement of $T_{\mathrm{Q}}$ at low fields. More recently, Uimin~\textit{et~al.} demonstrated that a large fluctuation of the quadrupole moment at low fields would lead to a reduction in $T_{\mathrm{Q}}$ \cite{Uim96}, and it is the suppression of these fluctuations with increasing field that allows the AFQ phase to stabilise. Whilst these proposed mechanisms were able to explain individual aspects of the phase diagram of CeB$_{6}$, none gave a complete description of the known behaviour. Critically, in an applied field in the AFQ phase the cerium moments order antiferromagnetically with an ordering vector of $\textbf{q} = (\frac{1}{2} \frac{1}{2} \frac{1}{2})$. This indicates that the induced AFM is modulated according to the AFQ order, which can be seen by neutron diffraction \cite{Shi97, Eff82, Eff85a, Erk87}. However, NMR results appeared to contradict this showing a splitting of the $^{11}$B signal that could not be simply explained by this ordering vector, but instead suggested an AFQ order with a triple-\textbf{q} structure \cite{Tak83}. Various theoretical studies within the mean-field approach were conducted to resolve such problems, developing a coherent picture of the low-temperature phase diagram \cite{Shi97, Shi98, Sak97, Tha98, Tha03, Tha04}.

\begin{figure}[t]
\begin{center}
	\includegraphics[width=\linewidth]{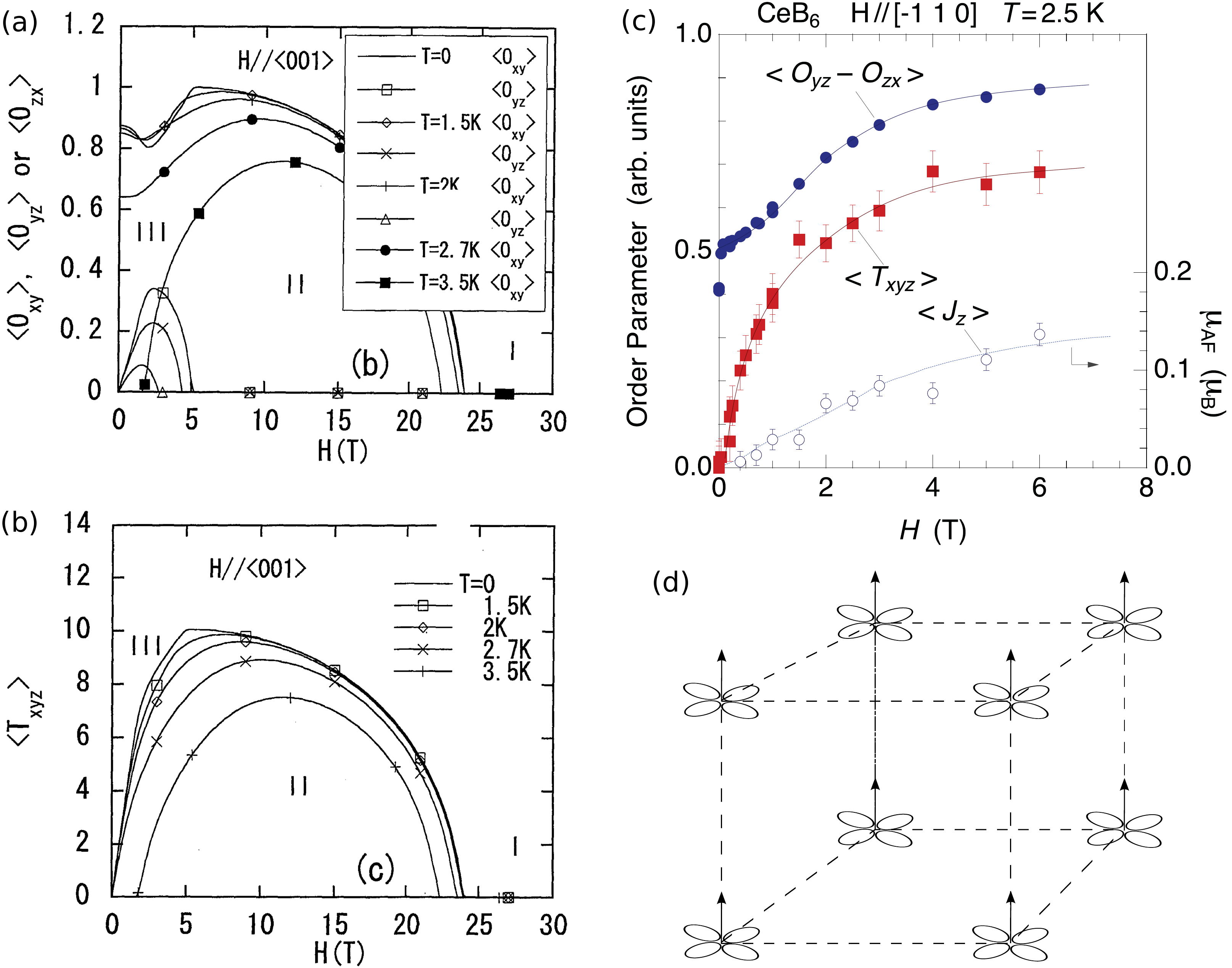}
\end{center}
	\caption{(a)~Field dependence of (a) $O_{xy}$ and $ O_{yz}$ or $O_{zx}$ for magnetic field along the (001) direction. (b)~Field dependence of $T_{xyz}$ for magnetic field along the (001) direction. Both figures reproduced from Ref.~\cite{Ser99}. (c)~Field dependence of the induced dipolar moment, $J_{z}$, the octupolar moment $T_{xyz}$ and the quadrupole moment $O_{yz} - O_{zx}$, determined from RXS. Reproduced from Ref.~\cite{Mat12}, copyright by the American Physical Society. (d)~Orientation of the $O_{xy}$ quadrupoles in an applied magnetic field along the \textit{z} axis. The cube represents a single unit cell of CeB$_{6}$ where Ce ions are at each of the corners. Figure reproduce from Ref.~\cite{Sak97}.
	}
		\label{four_sublattice}
\end{figure}

Within the model of Shiina~\textit{et~al.} which treats the system as a well isolated $\mathrm{\Gamma}_{8}$ ground state there are 15 supported multipolar moments -- 3 dipolar, 5 quadrupolar and 7 octupolar  \cite{Shi97, Shi98}. Of the 5 quadrupolar moments within the $\Gamma_{8}$ quartet, two are of the $\Gamma_{3}$-type and three are of the $\Gamma_{5}$-type, and in zero field it is expected that only the $\Gamma_{5}$-type ($O_{xy}, O_{yz}, O_{zx}$) ordering will be realized in zero field. A schematic diagram of the $O_{xy}$ order is presented in Fig.~\ref{four_sublattice}\,(d), which represents the orientation of the quadrupolar moments in the \textit{xy} plane with the applied magnetic field along the \textit{z} axis. This model also predicts that the application of magnetic field will ``select'' quadrupolar moments, due to the lowering of the symmetry of the system by dipolar ordering, such that a field applied along the (001) direction would result in the realisation of only $O_{xy}$ order, as depicted in Fig.~\ref{four_sublattice}\,(d), whilst $O_{yz} + O_{zx}$ is realised for a field along the (110) direction, and the system remains in an $O_{xy} + O_{yz} + O_{zx}$ state for fields applied along $\langle111\rangle$. To resolve the disagreement between the conclusions from neutron scattering and NMR, Sakai \textit{et~al.} performed a mean-field calculation within this model, considering ordering of the $O_{xy}$ quadrupoles in an applied field along (001), where the coexisting $T_{xyz}$ AF octupolar interaction was present \cite{Sak97}. Here they proposed that the NMR results can be explained by this model if the octupolar moments are included in the calculation, concluding that in this case that the $O_{xy}$ quadrupolar ordering is sufficient to describe both the neutron diffraction and NMR results and the complex triple-\textbf{q} structure initially suggested from the NMR study \cite{Tak83} is not required.

To provide a more complete description for the behaviour of phases II and III, Sera \textit{et~al.} performed a similar mean-field calculation in the `4 sublattice' model, where to consider the consecutive ordering of the quadrupoles and spins, cerium sites were divided into two $O_{xy}$ quadrupolar sublattices, split by the $T_{xyz}$ octupolar interaction, which are then further subdivided into two spin sublattices \cite{Ser99}. This Hamiltonian was also considered in the mean field approach, and from this the authors were able to successfully describe several properties of CeB$_{6}$. The unusual increase of $T_{\mathrm{Q}}(\textbf{B})$ at low field is understood to be a consequence of field-induced octupolar moments, seen in Fig.~\ref{four_sublattice}\,(a), stabilising the AFQ phase. This was also inferred experimentally from the field dependence of the RXS signal at $\mathbf{q}_{\mathrm{AFQ}}$, seen in Fig.~\ref{four_sublattice}\,(c), whereby the $O_{yz} - O_{zx}$ form factor increases with applied field. We can see from a comparison of the theory data in Fig.~\ref{four_sublattice}\,(a,b) and the experimental RXS data in panel (c) of the same figure that this model is able to predict the behaviour of the multipolar moments in CeB$_{6}$. Direct comparison here is not possible as the magnetic field is along different crystal directions, although the general form of the model closely follows the available data. They discovered that the antiferro-ordering of the $O_{xy}$ quadrupoles and the $T_{xyz}$ octupoles favours a ferromagnetic spin alignment. This directly competes with the antiferromagnetic exchange interaction that may explain unusual features of the phase diagram, such as the convex shape of the AFQ Bragg intensity as a function of applied field \cite{Mat12}. In addition, ferromagnetic spin correlations have been observed above $T_{\mathrm{Q}}$ by magnetization measurements and it has been proposed that these are induced by short-range ordering of the quadrupoles and octupoles \cite{Kob99}.

Whilst the mean-field models discussed in this section have provided a reasonable description of much of the behaviour of CeB$_{6}$, they do have difficulties when describing static quantities such as the low-field magnetisation and transport properties. The mean-field calculations of Sera~\textit{et~al.} suggest a zero-field crossing of the field-induced AFM moment as a function of field and temperature, which is not observed \cite{Ser99, Ser01, Mig87}. A metamagnetic-like double step in the low-temperature magnetisation, most pronounced for the field applied along $\langle110\rangle$, is also not explained in these theories \cite{Kun11}. Perhaps most importantly, the model neglects the formation of heavy-fermion quasiparticles in the ground state of CeB$_{6}$, which are observed by transport measurements. Recent neutron experiments by Friemel~\textit{et~al.} and Jang~\textit{et~al.} have shown that, contrary to the expectations of many established theoretical descriptions of CeB$_{6}$, the itinerant electrons also play an important role in the spin dynamics of the system  \cite{Fri12, Jan14}, and we have reserved discussion on this for section \ref{section:spin_dynamics}. This somewhat suggests that the localised \textit{f}-electron picture adopted by many mean-field descriptions may not be appropriate, which will be discussed in section~\ref{Itinerant}.

\subsection{Electronic properties of CeB$_{6}$}
\label{section:electronic_properties}

\begin{figure}[t]
\begin{center}
	\includegraphics[width=0.9\linewidth]{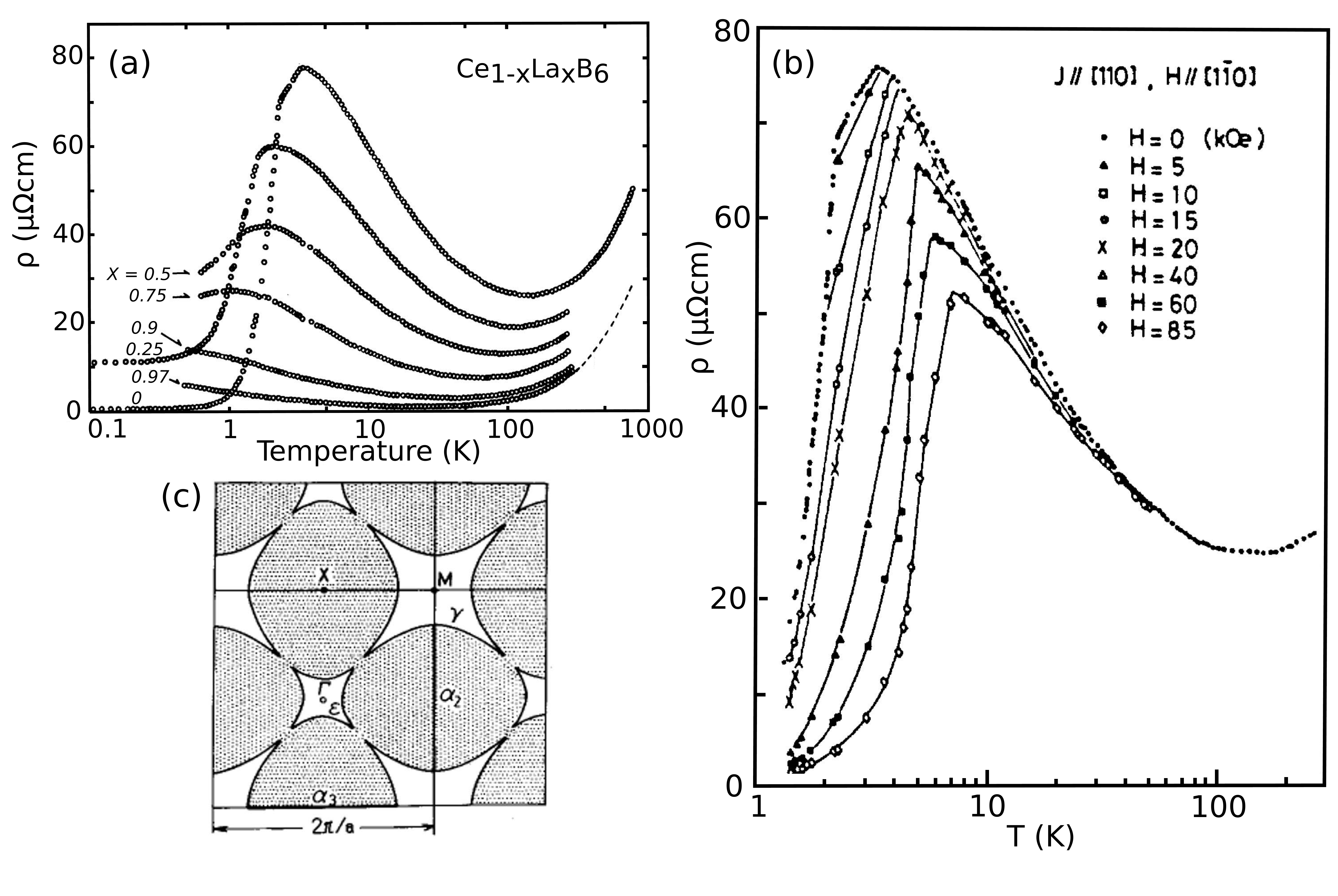}
\end{center}
	\caption{(a) Resistivity as a function of temperature for varying $x$ in Ce$_{1-x}$La$_{x}$B$_{6}$, reproduced from Ref.~\cite{Sat85}. In this figure the doping convention has been changed from Ce$_{x}$La$_{1-x}$B$_{6}$ in the original publication to Ce$_{1-x}$La$_{x}$B$_{6}$, and the concentrations in the figure have been altered accordingly. (b) Resistivity along (110) as a function of temperature for different magnetic fields applied parallel to [$1\bar{1}0$] in CeB$_{6}$, reproduced from Ref.~\cite{Tak80}. (c) Fermi surface of CeB$_{6}$, reproduced from Ref.~\cite{Onu89a}.}
		\label{Resistivity}
\end{figure}

Transport properties provide an important probe for the investigation into the role of heavy-fermion physics in CeB$_{6}$. Resistivity as a function of temperature for six different samples of Ce$_{1-x}$La$_{x}$B$_{6}$ is shown in Fig.~\ref{Resistivity}~(a), with lanthanum concentrations ranging from $x = 0$ to $x = 0.97$ \cite{Sat85}. The behaviour of the resistivity shows a continual change as a function of lanthanum doping, and at intermediate temperatures all concentrations exhibit the logarithmic scaling $\rho(T) = \log(T / T_{\mathrm{K}})$, where $T_{\mathrm{K}}$ is the Kondo temperature, which was found for the dilute samples to be around 1 K. This scaling is expected for the single-ion Kondo effect, and for low concentrations the resistivity drops at low temperatures, signalling the onset of the coherent Kondo state. The decrease in resistivity at intermediate temperatures is coupled to $T_{\mathrm{Q}}$, illustrated for CeB$_{6}$ in Fig.~\ref{Resistivity}\,(b) by measurements in applied magnetic fields, where the temperature of maximum resistivity $T_{\mathrm{max}}$ increases with $T_{\mathrm{Q}}$. At the lowest temperatures the resistivity follows a Fermi-liquid (FL) behaviour, $\rho(T) \propto T^{2}$ \cite{Sat85}. The magnetoresistance shows a decrease for fields $\textbf{B} > \textbf{B}_{\mathrm{N}}$, which would be considered typical for the incoherent Kondo effect since it suppresses scattering of conduction electrons by the \textit{f}-electrons \cite{Slu07, Tak80}. In the AFM phase, the magnetoresistance is weakly positive, which is common for the ground state of heavy-fermion systems. However, the clear feature in the magnetoresistance at the AFM/AFQ boundary and the rapid decrease in resistivity within the AFQ state indicates that the removal of scattering is related to the ordering of the quadrupoles, indicating that a purely heavy-fermion approach to describing the transport properties is insufficient, and that the localised correlated behaviour in CeB$_{6}$ plays and important role.

From the zero-temperature extrapolation of specific heat, the Sommerfield coefficient for CeB$_{6}$ is found to be strongly enhanced when compared to the nonmagnetic LaB$_{6}$ \cite{Mul88}. This indicates correlated quasiparticles, however since the decrease in resistivity follows $T_{\mathrm{Q}}$, it is difficult to define an onset temperature, $T^{\ast}$, for these correlations. Under the application of pressure it is found that the maximum temperature of resistivity, $T_{\mathrm{max}}$, increases beyond $T_{\mathrm{Q}}$ \cite{Kob00}. This is  expected, because pressure increases the hybridisation between \textit{f}-electrons and the conduction electrons and therefore promotes coherence between the \textit{f}-states. However, it also illustrates that the coherent heavy-fermion ground states and the AFQ phase are not necessarily connected, since the AFQ transition temperature $T_{\mathrm{Q}}$ remains nearly constant under pressure \cite{Sat85}.

The Fermi surface of CeB$_{6}$, shown in Fig.~\ref{Resistivity}~(c), is rather similar to other hexaborides such as LaB$_{6}$ and PrB$_{6}$ \cite{Jos87, Deu85}. It consists of ellipsoidal electron pockets centred at the \textit{X} points. Further, it has been systematically studied as a function of doping in Ce$_{1-x}$La$_{x}$B$_{6}$ by quantum-oscillation measurements \cite{End06, Onu89a, MatsuiGoto93, Goo99, Ish77, Deu85, Jos87, Har93, Har98, Tek00}. The carrier density was found to increase with the cerium concentration, indicating at least a partial contribution to the Fermi surface from the \textit{f}-electrons \cite{End06}. The high magnetic fields required for these studies put the system in the AFQ state, and the quasiparticle mass $m^{\ast}$ was found to decrease with increasing field. For varying cerium concentrations, the effective mass diverges as the applied magnetic field is lowered towards the phase II-III$^{\prime}$ boundary, $B_{\mathrm{Q}}$ \cite{End06}, with the effective mass exhibiting a continuous increase upon increasing cerium concentration \cite{Goo99}. Furthermore, the field dependence of the linear specific heat coefficient $\gamma_{0}$, which is proportional to $m^{\ast}$ for a Fermi liquid, also exhibits a similar field dependence, reaching a maximum at $B_{\mathrm{Q}}$ \cite{Mul88}. Owing to the high magnetic fields required for dHvA experiments, they were unable to access either phase III or phase III$^{\prime}$ of CeB$_{6}$, which have low critical fields of $B_{\mathrm{c}} = 1.1$~T and $B_{\mathrm{Q}} = 1.7$~T, respectively. However, the related compounds PrB$_{6}$ and NdB$_{6}$ have much higher critical fields for phases III or phase III$^{\prime}$, and have been studied by this technique \cite{Car80, Goo06, Onu89}. Across the phase III transition the Fermi surface was seen to change drastically, resembling that of LaB$_{6}$ in the paramagnetic state, and their effective masses in the paramagnetic state are comparable to CeB$_{6}$ \cite{Goo06, Onu89}. However, for both compounds no significant increase of the quasiparticle mass towards the critical field $B_{\mathrm{c}}$ was observed, and instead the low effective mass across the whole available field range has been attributed to a much weaker hybridisation between the conduction electrons and \textit{f}-electrons \cite{Goo06, Onu89}. It has been found in CeB$_{6}$ that magnetic ordering in the AFM state reconstructs the Fermi surface. Point-contact spectroscopy and scanning tunnelling spectroscopy have found indications for a small charge gap of $\Delta_{\mathrm{AFM}} \approx 0.6$~meV opening below $T_{\mathrm{N}}$ \cite{Pau85}. The structure of this reconstructed Fermi surface, however, remains elusive. Whilst the main sheet of the Fermi surface in PrB$_{6}$ and NdB$_{6}$ is spherical \cite{Pau85}, the contrasting behaviour of the \textit{f}-electron states and effective masses render us unable to draw a strong comparison between this and the corresponding Fermi surface of CeB$_{6}$.

We see that CeB$_{6}$ shows FL-like behaviour in its transport properties at low temperatures, both in the AFM and AFQ phases. The large linear specific-heat coefficient at low fields and the large effective mass of the charge carriers, determined at high field, indicate that heavy fermions are contributing to the transport. However, the localised \textit{f}-spins magnetically order in the low-temperature AFM phase in a manner analogous to PrB$_{6}$ and NdB$_{6}$, which both have a different behaviour of the \textit{f}-states and the quasiparticle effective mass.

\subsection{Alternative approach to phase II}
\label{section:Alternative_II}

\begin{figure}[b]
\begin{center}
	\includegraphics[width=\linewidth]{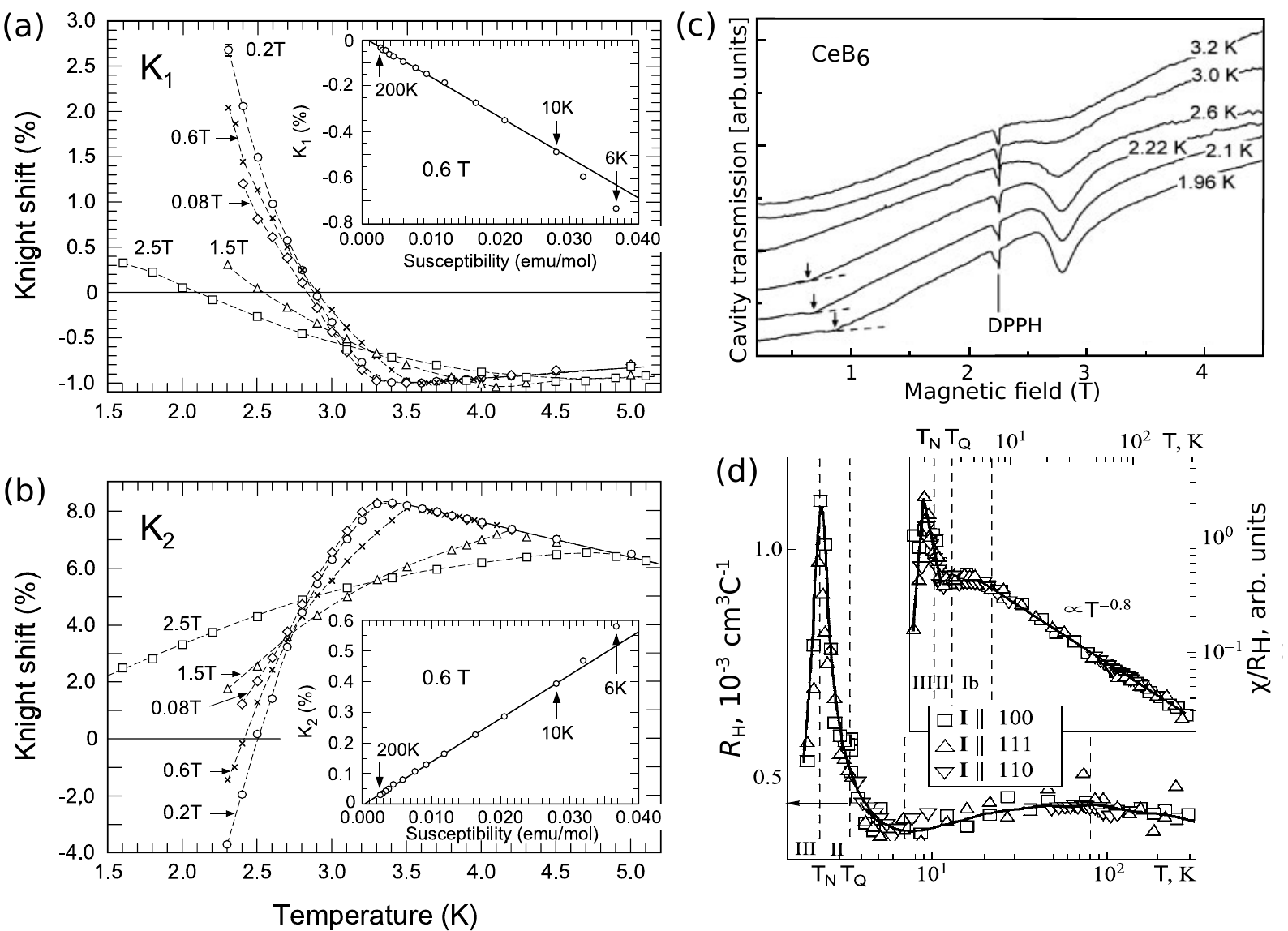}
\end{center}
	\caption{(a,b) Temperature dependence of the $\mu^{+}$ Knight shift. The $\mu$SR signal consisted of two components, $K_{1}$ and $K_{2}$, with the $K_{1}$ signal originating from the $(0 \frac{1}{2} 0)$ and $(\frac{1}{2} 0 0)$ sites and the $K_{2}$ signal from the $(0 0 \frac{1}{2})$ site. Insets show $K_{1,2}$ vs the magnetic susceptibility for an applied field of 0.6~T. Figures reproduced from Ref.~\cite{Sch04}, copyright by the American Physical Society. (c) Magnetic field dependence of the cavity transmission at 60~GHz. One can see a single magnetic resonance peak developing between 2.5~T and 3~T upon entering phase II. The small feature labelled DPPH is from a reference material,  2,2-diphenyl-1-picrylhydrazyl, placed in the cavity alongside CeB$_{6}$, and the arrows mark the deviation in signal which occurs when crossing into the AFM phase. Figure reproduced from Ref.~\cite{Dem05}. (d) Temperature dependence of the Hall coefficient $R_{\mathrm{H}}$, with the inset showing the ratio of magnetic susceptibility $\chi$ and Hall coefficient $R_{\mathrm{H}}$. Figure adapted from Ref.~\cite{Slu07}.}
		\label{Alt_description}
\end{figure}

The disagreement between the commonly accepted model of interacting multipoles, which treats the \textit{f}-electrons as localised states, and results from transport measurements that indicate a coherent heavy-fermion ground state at low temperatures, can be resolved by the two-fluid model of the Kondo lattice \cite{Nak04, Yan12}. Previously, the description of heavy electron materials has treated the system as a set of localised magnetic moments weakly coupled to the itinerant electron band by the AFM interaction. Whilst the single-site impurity problem has been successfully solved to describe the Kondo effect which emerges at low temperatures, complex inter-site interactions hindered the applicability of a model of the Kondo lattice. Nakatsuji~\textit{et~al.} set out to solve this by developing an appropriate two-fluid model \cite{Nak04}. The Kondo system has three important energy scales: the single-ion Kondo temperature $T_{\mathrm{K}}$, the inter-site coupling temperature scale $T^{\ast}$ and the crystal-electric-field splitting. They find that a coherent state emerges when crossing the temperature $T^{\ast}$: a heavy-electron fluid. This it similar to the two-fluid model for superfluid helium, where the heavy electron fluid is analogous to the superfluid and the non-interacting Kondo centres, or Kondo impurity fluid, is analogous to the normal-state fluid. The relative fraction, $f(T)$, of the coherent state undertakes the role of an order parameter, increasing linearly with decreasing temperature until saturating at a value of 0.9 \cite{Nak04}. In this description, both localised and heavy-fermion quasiparticles contribute to the magnetic properties below the inter-site coherence temperature $T^{\ast}$. Various consequences of this can be seen in experiment, such as the deviation from linear scaling between the spin part of the Knight-shift, $K_{\mathrm{spin}}$, and the susceptibility $\chi$ [Fig.~\ref{Alt_description}\,(a,b)]. Here, the local response, $\chi_{\mathrm{KL}}$, and the response of the hybridized quasiparticles $\chi_{\mathrm{HF}}$ contribute differently to the susceptibility $\chi$ and to $K_{\mathrm{spin}}$ below $T^{\ast}$, which has been observed in CeB$_{6}$ below $T^{\ast} \approx 10$~K \cite{Cur04, Sch04}. It has been claimed that this temperature corresponds to the onset of magnetic scattering at $(\frac{1}{2} \frac{1}{2} \frac{1}{2})$ \cite{Pla05}, which has been later shown to originate from spin fluctuations without any sharp transition \cite{Ger12}. At low temperatures, Raman scattering studies have observed the splitting of the $\mathrm{\Gamma}_{8}$ ground state into two doublets, separated by $E_{8,1} - E_{8,2} \approx 30$~K \cite{Zir84}. This splitting was not observed in inelastic neutron scattering measurements for $T > T_{\mathrm{Q}}$, although this is probably due to a vanishing matrix element for the transition \cite{Loe85, Bou93}. Should this be related to short-range quadrupole order, an additional splitting into two transitions upon the application of a magnetic field would be expected from the different Land\'{e} g-factors for the two Ce$^{3+}$ sites. However, only one orbital ordering resonance, in an applied field of $B_{\mathrm{res}} = 2.8$~T, was seen in electron spin resonance (ESR) measurements. This resonance is coincident with the AFQ phase, appearing below $T_{\mathrm{Q}}(B_{\mathrm{res}}) = 4.7$~K, as seen in Fig.~\ref{Alt_description}\,(c) \cite{Dem09, Dem05}. The presence of only a single resonance was attributed to the hybridization of conduction and \textit{f}-electrons, which averages out the two g-factors into a single uniform g-factor, determined by experiment to be $g = 1.6$ \cite{Sch12, Dem09, Dem05}. This ESR study also indicated the presence of ferromagnetic correlations. Attributing the single resonance to hybridised quasiparticles is in accordance with the observation of similar resonances in several other heavy-fermion compounds, such as YbRh$_{2}$Si$_{2}$ and CeRuPO, which also exhibit ferromagnetic correlations \cite{Kre08}, although this interpretation contradicts the multipolar mean-field model that interprets the \textit{f}-electrons as localised states. It is important to note that a second magnetic resonance has been seen at much higher fields \cite{Dem09b}, but is not considered within the discussion of Ref.~\cite{Sch12}. Its origin has yet to be determined, although it is postulated that the appearance of this mode within the high-frequency range may indicate a change in the magnetic structure of CeB$_{6}$ at high field \cite{Dem09b}.

However, the two-fluid Kondo description is not without its problems. Various charge transport \cite{Slu08a, Slu08b, Ign06, Win75, Sat85} and magnetisation \cite{Slu08a} measurements have shown deviations in behaviour from that expected by the Kondo description for both CeB$_{6}$ and lanthanum-doped samples. They find that, in contrast to the majority of the Ce-based Kondo lattices, the Hall coefficient $R_{\mathrm{H}}$ is both negative and nearly independent of temperature and applied magnetic field between $4.2 - 300$~K \cite{Sat84}, which goes against the predictions of the skew-scattering models for dense Kondo systems \cite{Col85, Had86}. An alternative explanation was proposed in the transport and magnetisation study by Sluchanko~\textit{et~al.} \cite{Slu07}. In the temperature region $T_{\mathrm{Q}} < T < 7$~K the susceptibility $\chi$ and the Hall coefficient $R_{\mathrm{H}}$ appear correlated, with the ratio of $ \chi / R_{\mathrm{H}}$ seen in the inset of Fig.~\ref{Alt_description}\,(d) remaining flat at these temperatures. Here, they are successfully described by the relation
\begin{equation}
\chi(T) \propto R_{\mathrm{H}}(T) \propto e^{E_{\mathrm{SP}} / k_{\mathrm{B}}T}.
\end{equation}
The corresponding energy $E_{\mathrm{SP}} / k_{\mathrm{B}} \approx 3.3$~K$^{-1}$ was associated with the formation of spin polarons. They propose, therefore, that below $T_{\mathrm{Q}}$ a transition to a spin-density-wave (SDW) state of the conduction electrons frees the local moments, which leads to the observed enhancement of the magnetic response. In low fields, the AFM state would then be competing with the SDW state due to the antiferromagnetic coupling of the localised spins. This polarises the conduction electrons via the RKKY interaction. The same group has also conducted a more recent study of the transport properties of dilute Ce$_{1-x}$La$_{x}$B$_{6}$ solid solutions with $x \approx 0.99$, where they find the Kondo-impurity model unable to explain their results, and instead argue that heavy-fermion states of the spin-polaron type are formed near the Ce impurities \cite{Slu15}. Inelastic neutron scattering may be able to resolve these questions, since it is able to see charge gaps associated with SDW order, and as discussed in the next section, a possible signature of a SDW has already been seen in neutron diffraction.

\begin{figure}[t]
\begin{center}
	\includegraphics[width=0.9\linewidth]{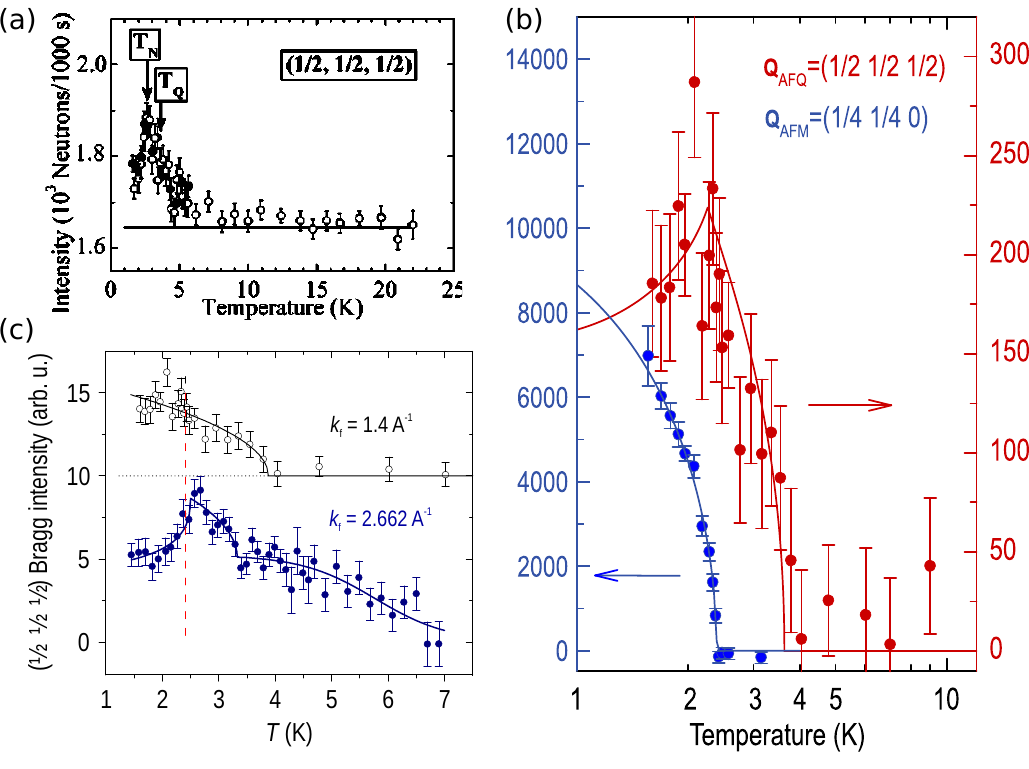}
\end{center}
	\caption{(a) Temperature dependence of the signal at the ($\frac{1}{2} \frac{1}{2} \frac{1}{2}$) peak from spin-flip neutron scattering. The open and filled data points represent two different sets of measurements and the background scattering is indicated by the solid line. Reproduced from Ref. \cite{Pla05}, copyright by the American Physical Society. (b) Temperature dependence of the AFQ and AFM Bragg peaks measured on a cold-neutron triple-axis neutron spectrometer~\cite{Fri12}. (c) The same temperature dependence of the ($\frac{1}{2} \frac{1}{2} \frac{1}{2}$) Bragg peak as in panel (b), compared with a similar measurement performed at a thermal-neutron spectrometer. Data points were taken from three-point scans at final wave vectors $k_{f} = 1.4$\,\AA$^{-1}$ and 2.66\,\AA$^{-1}$, respectively, with the former shifted up by 10 points for clarity. $T_{\mathrm{N}}$ is indicated with a dashed line, having been determined by the temperature dependence of the AFM Bragg peaks \cite{Fri14}.}
		\label{AFQ_vs_T}
\end{figure}

\subsection{Interplay between the competing order parameters of the low-temperature phases}
\label{section:Order_parameters}

Quadrupolar ordering is not visible under neutron scattering in zero magnetic field as the neutron can only be scattered from time-reversal-symmetry breaking fields \cite{Kus08}. Until recently, neutron scattering at the quadrupolar ordering vector, $\textbf{q}_{\mathrm{AFQ}}$, has only been observed under an applied field, attributed to ordering of the dipole moments modulated to the underlying AFQ order \cite{Ser01, Mig87, Eff85a, Bur82, Ohk85}. Surprisingly, however, neutron spin-flip scattering measurements, reproduced here in Fig.~\ref{AFQ_vs_T}\,(a)~\cite{Pla05}, observed a signal at $\textbf{q}_{\mathrm{AFQ}}$ in zero field. This signal appears below $T = 7$~K, rising to a peak at $T_{\mathrm{N}}$ before being suppressed within the AFM phase. Within the statistical fluctuations of the data it appears unperturbed by the AFQ transition. Following this, more recent elastic neutron scattering measurements have also seen a small signal at the AFQ ordering vector, reproduced here in Fig.~\ref{AFQ_vs_T}\,(b). However, the temperature dependence of this signal is in contrast to the earlier spin-flip data, with intensity only appearing below the critical temperature $T_{\mathrm{AFQ}}$ \cite{Fri12}. The Bragg peak at the AFQ ordering vector is more than an order of magnitude weaker than its counterpart at $\textbf{q}_{\mathrm{AFM}}$, and shows a sharp increase from zero between $T_{\mathrm{Q}}$ and $T_{\mathrm{N}}$ before being lightly suppressed in the AFM phase. The apparently contradicting behaviours of the spin-flip scattering data of panel (a) and the nonpolarized elastic neutron scattering data of panel (b) can be reconciled by considering the difference in energy resolutions, as explained in Fig.~\ref{AFQ_vs_T}\,(c) \cite{Fri14}. This figure juxtaposes two data sets from different experiments, performed on the same sample at two triple-axis spectrometers of the Institute Laue Langevin (ILL, Grenoble), one with a neutron wave vector $\textbf{k}_{\mathrm{f}} = 1.4$~\AA$^{-1}$ (cold-neutron IN14 spectrometer) and the other with $\textbf{k}_{\mathrm{f}} = 2.662$~\AA$^{-1}$ (thermal-neutron IN3 spectrometer). Both the old polarized-neutron data and our nonpolarized INS data measured with $\textbf{k}_{\mathrm{f}} = 2.662$~\AA$^{-1}$ show a signal above $T_{\mathrm{Q}}$, whilst cold-neutron measurements (identical to those shown in panel (b) of Fig.~7) show no signal at the AFQ wavevector until passing below $T_{\mathrm{Q}}$. Whilst the presence of an elastic neutron scattering signal at the AFQ wavevector in no applied field is an unexpected result by itself, the apparent disagreement among the two experiments indicates that there is more than one contribution to neutron scattering at this wavevector. When comparing the $\textbf{k}_{\mathrm{f}} = 2.662$~\AA$^{-1}$ and $\textbf{k}_{\mathrm{f}} = 1.4$~\AA$^{-1}$ data in panel (c), it is important to note that the cold-neutron measurement benefited from a better energy resolution. As it is now known that low-energy quasielastic scattering is peaked at the AFQ wave vector and is present above $T_{\rm Q}$, in both thermal-neutron elastic measurements the broad energy resolution leads to an additional diffuse contribution to the peak that is magnetic in origin and therefore seen both in nonpolarized- and polarized-neutron data. However, the resolution of a cold-neutron spectrometer is sufficient to separate the true Bragg scattering below $T_{\rm Q}$ from the quasielastic contribution. Below $T_{\rm N}$, the quasielastic signal is gapped, which leads to a suppression of the intensity and a reported sharpening of the elastic peak~\cite{Pla05}. The higher resolution data from panels (b) and (c), however, have a comparable \textbf{q}-width to the AFM Bragg peaks, implying that this signal is related predominantly to the long-range static magnetic ordering. Assuming that the minor suppression of intensity below $T_{\rm N}$ could be still related to the spin-gap opening in the quasielastic signal, this observation suggests that phase~II is not suppressed by the onset of AFM order and hence coexists with it without competition. Moreover, the presence of a Bragg peak at ($\frac{1}{2} \frac{1}{2} \frac{1}{2}$), albeit weak in comparison to the AFM Bragg peaks, indicates that something in addition to AFQ ordering is taking place in zero field.

\begin{figure}[b]
	\includegraphics[width=\linewidth]{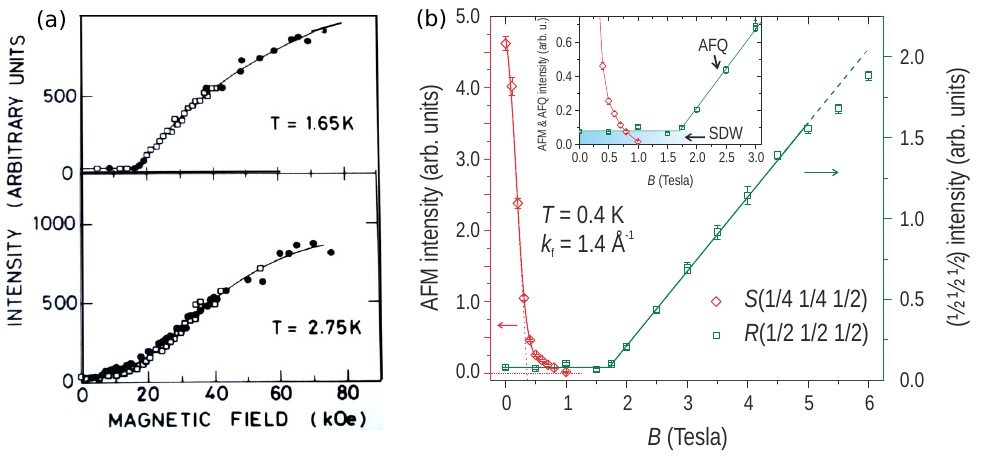}
	\caption{(a) Magnetic field dependence of the AFQ Bragg intensities from neutron scattering at $T = 1.65$~K and $T = 2.75$~K.  At $T = 1.65$~K the sample is initially in the AFM state before undergoing the transition to the AFQ state with increasing field. For $T = 2.75$~K the sample is already in the AFQ state in zero field. Figure reproduced from Ref.~\cite{Mig87}. (b) More recent neutron scattering data on the magnetic field dependence of the AFM and AFQ/SDW Bragg intensities at \textit{T}$= 0.4$~K. The low-field region has been enlarged in the inset, showing that the intensity at the ($\frac{1}{2} \frac{1}{2} \frac{1}{2}$) point does not go completely to zero in the AFM phase. Reproduced from Ref.~\cite{Fri14}.}
		\label{AFQ_vs_B}
\end{figure}

The weak contribution to the magnetic Bragg intensity in the zero-field data of Fig.~\ref{AFQ_vs_T} is also seen in Fig.~\ref{AFQ_vs_B}\,(b). Fig.~\ref{AFQ_vs_B} shows the field dependence of the Bragg intensity at \textbf{q}$_{\mathrm{AFQ}}$ from two separate experiments. The early data in Fig.~\ref{AFQ_vs_B}\,(a)~\cite{Mig87} show the magnetic Bragg intensity at ($\frac{1}{2} \frac{1}{2} \frac{1}{2}$) as a function of field at both 1.65~K and 2.75~K. At 1.65~K the system is in the AFM state at zero field, and initially shows no signal until the AFQ phase is reached, whilst at 2.75~K the system is already in the AFQ phase in zero field and the intensity appears to increase from zero as a function of applied field. This signal arises from the field-induced dipolar moments, modulated to the underlying AFQ order. More recent low-temperature data from the neutron-scattering study of Ref.~\cite{Fri14}, reproduced in Fig.~\ref{AFQ_vs_B}\,(b), show scattering at $T=0.4$\,K from both the AFM and AFQ wavevectors, where the inset shows an expanded low-field region. Here we can see that there is actually very weak magnetic Bragg scattering in the AFM state at ($\frac{1}{2} \frac{1}{2} \frac{1}{2}$), now visible in the newer data because of a better signal-to-noise ratio. It is independent of applied field within the AFM phase, being unperturbed by the phase III\,--\,III$^{\prime}$ phase boundary at $\textbf{B}_{\rm c} = 1.1$~T. Only above 1.5~T, upon entering phase II, the intensity of this peak begins to increase linearly with applied field, as reported previously. Although the localised multipolar ordering model is able to explain the increasing intensity in the AFQ state in applied magnetic field as originating from dipole ordering modulated to the AFQ order, it is unable to explain the flat intensity at the AFQ wavevector seen within phase III \cite{Ser01, Bur82, Fri14}. This suggests that this signal may have a different origin, and has been suggested to come from a SDW formed from itinerant quasiparticles as opposed to the localised \textit{f}-electrons \cite{Fri14}, which would correlate with the conclusions of Sluchanko \textit{et~al.} discussed in section \ref{section:Alternative_II}. Still, the disappearance of the SDW peak at $T_{\rm Q}$ strongly indicates that it is linked to the AFQ order parameter or is even induced by it. It is evident here that there is a multitude of competing order parameters within the low temperature region of CeB$_{6}$, which in some cases appear to co-exist or compete, arising from the complex interplay of the itinerant and more localised \textit{f}-electron physics.

\begin{figure}[t]
\begin{center}
\includegraphics[width=0.9\linewidth]{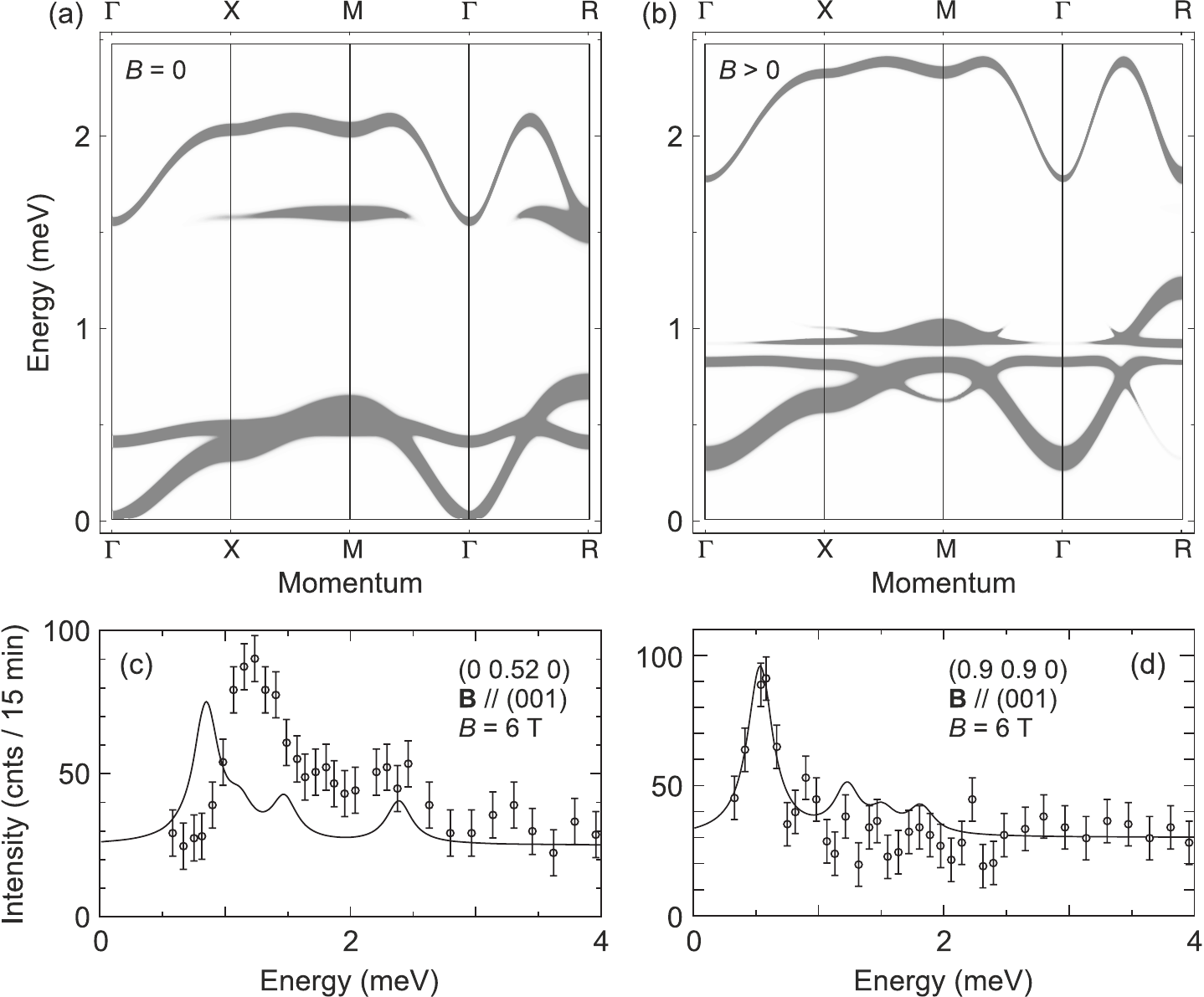}
\end{center}
	\caption{(a,b) The dipolar excitation spectrum, $S(\mathbf{q},\omega)$, calculated in the RPA for the AFQ phase at zero field and at a finite magnetic field applied along the $\langle001\rangle$ direction, respectively. The figure is adapted from Ref.\,\cite{Tha03}. (c,d) Comparison of the measured (data points) and calculated (solid lines) spectra for the (0~0.52~0) and (0.9 0.9 0) wave vectors (near the $X$ and $\Gamma$ points), respectively. Magnetic field of 6~T is applied along the $(001)$ direction in both measurements. The figure is reproduced from Ref.~\cite{Shi03}.}
		\label{INS_Theory}
\end{figure}

\subsection{Spin dynamics in CeB$_{6}$}
\label{section:spin_dynamics}

The mean-field descriptions, discussed in section \ref{Mean_field}, are able to predict the formation of collective multipolar modes associated with the ordering of the quadrupoles. Thalmeier \textit{et~al.} considered this model, assuming a symmetric interaction term and calculated the dipolar scattering function in both the random-phase approximation (RPA) from the local multipolar susceptibility \cite{Tha98, Tha03, Tha04} and in the Holstein-Primakoff approach \cite{Shi03}. These calculations predict, in zero field, a Goldstone mode at the $\Gamma$ point from a dispersion branch that extends across the whole Brillouin zone \cite{Tha98, Tha03, Tha04}, as shown in Fig.\,\ref{INS_Theory}\,(a). All modes were found to increase linearly with field to the first approximation \cite{Tha98}, an example is given in Fig.\,\ref{INS_Theory}\,(b). INS measurements by Bouvet \textit{et~al.} \cite{Bou93} and Regnault \textit{et~al.} \cite{Reg88} find dispersing excitations along lines of high symmetry in applied magnetic fields of 4~T and 6~T parallel to the $(001)$ direction. Two representative comparisons between the measurements of Reganult~\textit{et~al.} and the calculations of Shiina~\textit{et~al.} are shown in Fig.~\ref{INS_Theory}\,(c,d) \cite{Shi03}, demonstrating a reasonable agreement near the $\Gamma$ point that becomes less satisfactory near the $X$ point at the zone boundary. It was concluded that the theory describes the ``leading'' mode in the spectrum in field, however zero-field experiments found a featureless quasielastic response \cite{Bou93} which disagrees with the calculations. Until recently, however, the INS data were limited in their coverage of $\mathbf{Q}$-space and, furthermore, these studies focussed on the AFQ phase, with few measurements taken in the AFM state.

\begin{figure}[t]
\begin{center}
\includegraphics[width=0.7\linewidth]{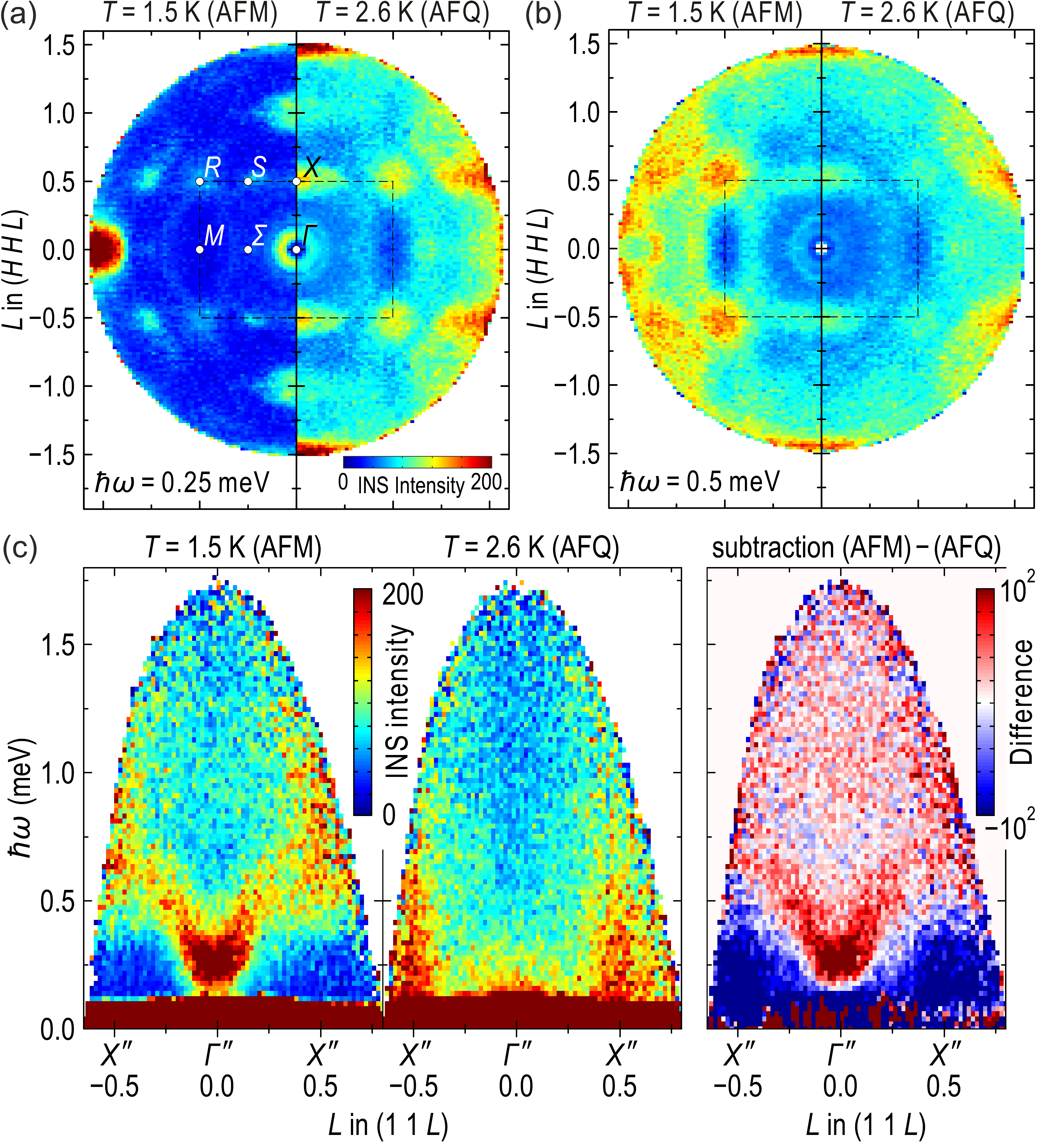}
\end{center}
	\caption{(a,b) Constant energy maps in the (\textit{H H L}) plane of reciprocal space, obtained from TOF neutron scattering data, corresponding to energies of $\hbar \omega =$ 0.25 and 0.5~meV, respectively. The integration range in energy was $\pm0.1$~meV about these values and data were symmetrised about the natural mirror planes of reciprocal space, with the left and right sides of each panel measured in the AFM (1.5~K) and AFQ (2.6~K) phases respectively. Locations of high symmetry in the $(H\,H\,L)$ plane are shown in panel (a). (c) Energy-momentum cuts in the $(1\,1\,L)$ plane. From left to right: Unprocessed data at 1.5~K (AFM state), unprocessed data at 2.6~K (AFQ state) and the subtraction the two data sets. Figures reproduced from Ref.~\cite{Jan14}.}
		\label{INS}
\end{figure}

\begin{figure}[t]
\includegraphics[width=\textwidth]{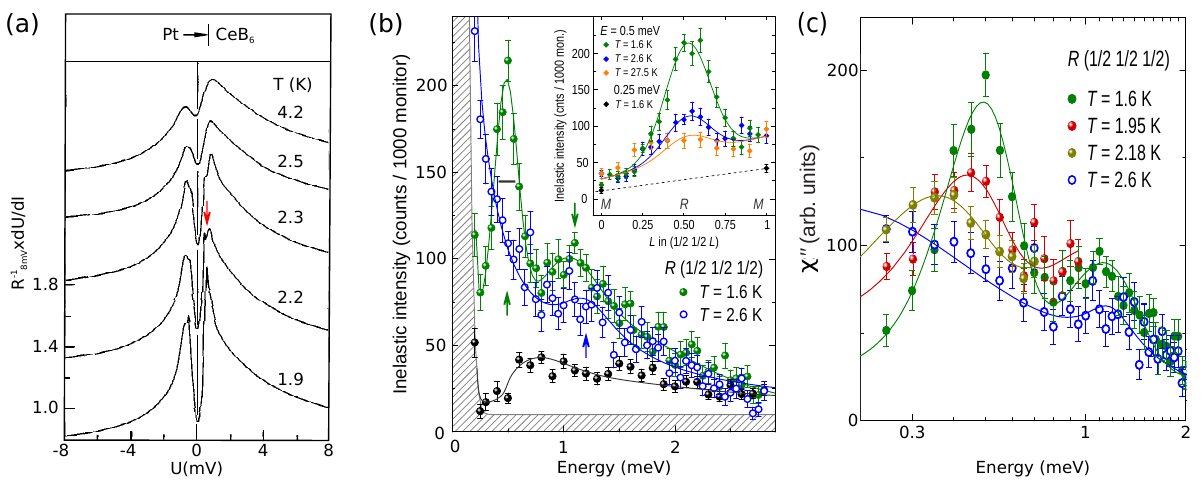}
\caption{(a) Temperature dependence of $\mathrm{d}U/\mathrm{d}I$ of a Pt/CeB$_{6}$ point contact, for temperatures between 1.9~K and 4.2~K. Figure reproduced from Ref.~\cite{Pau85}. (b) Inelastic scattering spectrum at $R(\frac{1}{2} \frac{1}{2} \frac{1}{2})$ in both the AFM state ($T = 1.5$~K) and AFQ state ($T = 2.5$~K). Figure reproduced from Ref.~\cite{Fri14}. (c) Temperature evolution of the spectra at $R(\frac{1}{2} \frac{1}{2} \frac{1}{2})$ from within the AFM state to the AFQ state, focusing on the charge gap. Figure reproduced from Ref.~\cite{Fri14}.}
\label{Charge_gap}
\end{figure}

More recent neutron scattering measurements provided a more complete description of the spin excitation spectrum of CeB$_{6}$. Recent developments in neutron-spectrometer instrumentations have allowed Friemel~\textit{et~al.} and Jang~\textit{et~al.} to re-investigate the spectra under improved energy resolution throughout a greater proportion of reciprocal space, both in the AFM and AFQ states \cite{Fri12, Jan14}. A broad summary of their data, consisting of constant-energy maps in reciprocal space along the $(HHL)$ plane is shown in Fig.~\ref{INS}\,(a,b) at energies of 0.25 and 0.5 meV, respectively. This spectrum consists of several key features. Firstly, within the AFM state there is a strong signal at the \textit{R} points, most intense in the 0.5~meV map, which comes from the resonant exciton mode initially reported by Friemel~\textit{et~al.} \cite{Fri12}. The appearance of this resonant exciton is contradictory to the naive expectation that the spin dynamics in phase III would be dominated by collective modes emanating from the AFM ordering vectors $\textbf{q}_{1} = (\frac{1}{4} \frac{1}{4} 0)$ and $\textbf{q}_{1}^{\prime} = (\frac{1}{4} \frac{1}{4} \frac{1}{2})$. This exciton peak, centred at an energy of $\hbar \omega_{\mathrm{R}} = 0.48$~meV, develops with an order-parameter-like behaviour below $T_{\mathrm{N}}$ with a simultaneous broadening of the resonance. The development of this strong mode at the \textit{R}($\frac{1}{2}\frac{1}{2}\frac{1}{2}$) point, is accompanied by the opening of a spin gap below 0.35~meV. This \textit{R}-point mode bears a strong resemblance to the resonant modes observed in various heavy-fermion and high-$T_{\mathrm{c}}$ superconductors \cite{Fon95, Ino10, Sto08, Sto11}. These modes are understood to be excitons that form below the onset of a particle-hole continuum (within the particle-hole energy gap) due to a divergence in the dynamical spin susceptibility derived from itinerant models within the RPA. For CeB$_{6}$, a charge gap has been observed in the AFM state with a size $2\Delta_{\mathrm{AFM}}$ $\approx 1.2$~meV from point-contact spectroscopy, reproduced in Fig.~\ref{Charge_gap}\,(a) \cite{Pau85}. This suggests that the mode at $\hbar \omega_{R}$ lies inside this charge gap and is therefore consistent with the exciton picture. However, the resonant mode in unconventional superconductors is confined to the wavevector of the magnetic Bragg peak, whereas within the AFM phase of CeB$_{6}$ an exciton-like feature can be seen at a number of wavevectors, reaching a maximum of intensity at the ($\frac{1}{2}\frac{1}{2}\frac{1}{2})$ point.

Further support for the opening of the charge gap comes from the neutron-spectroscopy data themselves, showing that in the AFM state a second, broader peak at the \textit{R} point, indicated by the arrow in Fig.~\ref{Charge_gap}\,(b), may signal the onset of the particle-hole continuum \cite{Fri15}. This peak, found between 0.95~meV and 1.05~meV, is close in energy to the gap measured by point-contact spectroscopy \cite{Pau85}. An alternative explanation for this peak such as a crystal-field excitation can be excluded as the peak is seen to disappear as the temperature passes above $T_{\mathrm{N}}$ [Fig.~\ref{Charge_gap}\,(c)], with the spectra in the AFQ state taking on a quasielastic lineshape with no high-energy mode \cite{Ger12}.

The development of a resonant exciton mode in the AFM phase is not predicted by the standard interacting-multipole model of CeB$_{6}$, which would have anticipated that the magnetic response in the AFM phase be dominated by the magnetic dipole moments of the localised Ce electrons \cite{Shi97, Tha03, Akb12}. However, the presence of this resonant mode can be understood in terms of a two-component model that includes strong interactions between the localised magnetic moments and the itinerant electrons \cite{Sat01}. Here, the formation of the AFM state results in the opening of a gap in the electronic density of states, leading to a nearly isotropic spin gap in the magnetic spectrum originating from strongly hybridised crystal-field excitations of the Ce 4\textit{f}-electrons and the itinerant conduction electrons. The resonant mode observed in the neutron scattering measurements can then be considered as a collective mode below the onset of a particle-hole continuum, in a similar manner to the current models describing the resonant modes in unconventional superconductors \cite{Fon95, Sto08, Ino10, Sto11}. Further, CeB$_{6}$ is not the only hexaboride to show a resonant exciton in its magnetic excitation spectrum, with recent neutron scattering measurements on SmB$_{6}$ showing similar 14~meV exciton modes at the \textit{X} and \textit{R} high symmetry points \cite{Fuh15}.

A further surprising development in the exploration of the spin dynamics of the AFM phase was the observation of ferromagnetic fluctuations by time-of-flight neutron scattering. Jang \textit{et~al.} discovered an intense and dispersive low-energy mode originating from the $(110)$ and $(001)$ points \cite{Jan14}, which can be seen clearly in the reciprocal space map of Fig.~\ref{INS}\,(a), with its dispersion along the $X-\Gamma-X$ line reproduced in panel (c) of the same figure. Its temperature dependence indicates that it cannot originate from the acoustic phonon modes, but rather is magnetic in origin, and it has a parabolic dispersion which is expected for ferromagnetic spin waves. The dispersion of this mode, spanning the whole Brillouin zone, is continuously connected to the exciton-like peaks that were observed previously for a number of wavevectors using triple-axis spectrometers \cite{Ger12}. Within the AFQ state the scattering becomes quasielastic throughout the Brillouin zone, with intensity peaked at the $\Gamma$, $X$ and \textit{R} points rather than the AFM ordering vectors $\textbf{q}_{1} = (\frac{1}{4} \frac{1}{4} 0)$ and $\textbf{q}_{1}^{\prime} = (\frac{1}{4} \frac{1}{4} \frac{1}{2})$. The intense dispersing modes seen in the AFM phase should therefore be associated with the itinerant quasiparticles, whose scattering function in {\textbf{q}-space is defined by the nesting vectors of the paramagnetic Fermi surface. This is similar to the spin dynamics in the hidden-order compound URu$_{2}$Si$_{2}$, where spin excitations pre-exist as paramagnons in the paramagnetic phase before becoming visible as dispersing modes upon transition into the hidden-order phase \cite{Wie07}.

In zero field, the proposed AFQ ordering of the $O_{xy}$ quadrupole moments with the antiferro-coupling of the $T_{xyz}$ octupolar moments favours a ferromagnetic alignment of the dipoles \cite{Tha03, Shi03}. However, this theory proposes that the dispersing magnon modes would already be visible in the AFQ state, at odds with the neutron-scattering observations. Similarly, the exciton at ($\frac{1}{2}\frac{1}{2}\frac{1}{2}$) would be expected to be a damped crystal-field-like multipolar excitation within these descriptions, instead of the observed change to quasielastic scattering upon entering the AFQ state. Furthermore, this weakly \textbf{q}-dependent quasielastic scattering has its linewidth minima at the $R$, $\Gamma$ and $X$ points. This suggests a correlation between the quasielastic linewidth in the AFQ state and magnon energy in the AFM state, indicating that the strongly damped signal in the AFQ phase already carries important information about the collective modes that form below $T_{\rm N}$. These observations call for a re-examination of the theoretical description, which is discussed in section \ref{Itinerant}.

The neutron scattering studies were also able to investigate the spin waves associated with the AFM ordering \cite{Jan14}. They found a cone-shaped dispersion emanating from the ($\frac{1}{4}\frac{1}{4}\frac{1}{2}$) and equivalent vectors, resembling the spin waves expected for a Heisenberg antiferromagnet. The spin-wave bands are restricted to low energies ($<0.8$~meV) and a narrow \textbf{\textit{Q}}-space range around the ordering wave vectors.

In the most recent measurements, Friemel~\textit{et~al.} have extended their investigation of the resonant mode at the \textit{R} point to include the effects of both lanthanum doping and applied field \cite{Fri15}. Whilst their initial findings of regarding this resonant mode in pure CeB$_{6}$ under no applied field led to the suggestion that the localised description of the spin dynamics was insufficient and it was necessary to account for the effect of itinerant heavy-quasiparticles in the system (see section \ref{Itinerant}), under applied field they find a crossover from an itinerant regime in the AFM phase to a more localised one above $B_{\mathrm{Q}}$. Their data show that the resonant mode at the \textit{R} point disappears above the critical field of the AFM phase, and that this is followed by the formation of two other modes whose energies grow linearly with magnetic field when entering the AFQ phase. This field behaviour occurs for all the lanthanum dopings, $x = 0, 0.18, 0.23, 0.28$ in Ce$_{1-x}$La$_{x}$B$_{6}$, and resembles a transition between two Zeeman-split energy levels, consistent with a purely localised description of the spin dynamics within a mean-field approach of ordered multipoles \cite{Tha98}.

Concluding, the zero-field neutron scattering studies of Friemel~\textit{et~al.} and Jang~\textit{et~al.} uncovered some surprising behaviour. In total, one can distinguish three distinct types of low-energy magnetic excitations, possessing their own local maxima of intensity in momentum space: the ferromagnetic mode at the $\Gamma$ point, the resonant exciton mode at the \textit{R} point and spin-wave modes at the AFM wavevectors. Both the resonant mode at the \textit{R} point and the low-energy ferromagnetic fluctuations at the $\Gamma$ point were unexpected, as they were not predicted by prevailing theories of the interacting multipolar moments of CeB$_{6}$ and call for a re-examination of the interplay between the competing interactions within this material.

\subsection{Itinerant description of the spin dynamics.}
\label{Itinerant}

The above-mentioned INS results \cite{Fri12, Jan14} are at odds with the expected behaviour from established theories. In the AFQ state only quasielastic scattering is observed, instead of the dispersive collective modes expected from the multipolar model \cite{Tha98} (see Fig.\,\ref{INS_Theory}). In zero field, spin excitations only appear within the AFM phase, revealing both spin waves emanating from the AFM Bragg peaks and a strong dispersing ferromagnetic mode at the $\Gamma$ point alongside exciton-like modes at the \textit{R} and \textit{X} points. Similar modes are recurring phenomena, found in the superconducting state of heavy-fermion superconductors \cite{Sto08, Sto11}, Kondo insulators such as SmB$_{6}$ \cite{Ale95} and YbB$_{12}$ \cite{Nem07, Mig05} as well as the hidden order phase of URu$_{2}$Si$_{2}$ \cite{Wie07, Bou10} and now CeB$_{6}$. It also appears that the appearance of these modes is somewhat independent of the exact ground state of the system in question: From the ``1-2-10" family both CeRu$_{2}$Al$_{10}$ and CeFe$_{2}$Al$_{10}$ display such modes, with the former being an antiferromagnet and the latter a Kondo insulator. Most importantly, following the work of Ohkawa \cite{Ohk85}, the ordering phenomena in CeB$_{6}$ have been considered in the localised picture of \textit{f}-electron multipoles. Whilst this approach has been able to describe a large body of the experimental data (see section~\ref{Mean_field}), the recent INS measurements by Friemel \textit{et~al.}~\cite{Ger12} and Jang \textit{et~al.}~\cite{Jan14} indicate that the spectral weight of spin fluctuations predominantly originates from itinerant magnetic moments. This challenge to the localised picture is in accordance with the heavy-fermion character of CeB$_{6}$: Its large effective mass $m^{\ast} > 20{\kern.5pt}m_0$ \cite{MatsuiGoto93, Jos87, End06}, where $m_0$ is the bare electron mass, and its Kondo temperature of $\sim\!1$\,K that is of a similar order to $T_{\mathrm{N}}$ and $T_{\mathrm{Q}}$ may cast doubt on the validity of treating CeB$_{6}$ with a localised 4\textit{f} approach.

To overcome these shortcomings in the existing theories, Akbari \textit{et~al.}~\cite{Akb12} have considered CeB$_{6}$ in terms of the microscopic fourfold degenerate $\mathrm{\Gamma}_{8}$-type Anderson lattice model. Here, the AFQ and AFM order parameters are treated as particle-hole condensates in the itinerant heavy-quasiparticle picture, and the resonant mode from the INS measurements \cite{Ger12} is interpreted as a feedback spin exciton, where the feedback effect results in a change in magnetic spectral properties across a respective transition due to the appearance of an order parameter. To ensure single occupancy of the \textit{f}-orbitals, they assume an infinite on-site Coulomb repulsion, and find the main effect of the AFQ and AFM ordering on the conduction electrons is to create charge gaps at the respective ordering vectors. Within the RPA formalism, they evaluate an expression for the dynamic susceptibility for the interacting quasiparticles $\chi_{\mathrm{RPA}}(\mathbf{q}, \omega)$:
\begin{equation}
\chi_{\mathrm{RPA}}(\mathbf{q}, \omega) = \frac{\chi_{0}(\mathbf{q}, \omega)}{1 - J_{\!\mathbf{q}\,}\chi_{0}(\mathbf{q}, \omega)},
\end{equation}
where they take the heavy quasiparticle interaction term $J_{\mathbf{q}}$ to have a Lorentzian form centred at the exciton wavevector. The charge gaps that appear in the AFQ and AFM phases have the effect of pushing the quasiparticle response to higher energies, and if the condition $J_{\!\mathbf{\mathbf{q}_0}\,}\chi_{0}(\mathbf{\mathbf{q}_0}, \omega) = 1$ is met, then the RPA susceptibility manifests a pole at $\textbf{q} = \mathbf{q}_0$. It is suggested that this condition is satisfied in the AFM state of CeB$_{6}$, which explains the sharp exciton feature seen at the \textit{R}-point in inelastic neutron scattering.

Whilst this model is successful in explaining the observation of the \textit{R}-point resonance in the INS measurements, there are still several open questions regarding these results. Firstly, the origin of the interaction $J_{\mathbf{q}}$ is left open, being either direct exchange or an RKKY-mediated dipolar or quadrupolar interaction. Secondly, an alternative approach is conceivable, which was used to describe the resonant mode in CeCoIn$_{5}$ \cite{Chu08}. It describes excitons such as the one observed by the INS study as localised magnon-like excitations. This description of the resonant modes in unconventional superconductors finds the magnon-like excitations appearing at the antiferromagnetic wavevector only after the opening of the energy gap in the charge channel, as this removes damping by particle-hole excitations and thus allows them to form as sharp modes from an overdamped continuum. For CeB$_{6}$, the gap is a charge gap generated by the AFM order, and the localised soft mode could be of the multipolar type predicted by Thalmeier \textit{et~al.}~\cite{Tha98} or a crystal-field excitation. As the model by Akbari~\textit{et~al.} was constructed to explain the early INS results by Friemel \textit{et~al.} \cite{Fri12}, the Q-dependence of $J_{\mathbf{q}}$ used in this model generates an exciton only at the \textit{R}-point. However, the more complete mapping in the study by Jang \textit{et~al.} has shown that the spin dynamics across the Brillouin zone is more complicated than this, and the more recent in-field measurements on Ce$_{1-x}$La$_{x}$B$_{6}$ have illustrated the possibility of a crossover between itinerant and localised regimes. Whether this model can be adapted to explain these new results remains to be seen.

\section{Doped compound -- Ce$_{1-x}$La$_{x}$B$_{6}$}
\label{Sec:La_Doped}

A significant body of research has been dedicated to investigating the effect of substituting cerium with non-magnetic lanthanum in CeB$_{6}$, as the absence of an \textit{f} electron makes it an ideal tuning parameter to investigate the heavy-fermion physics in this system. The outer orbital structure of lanthanum is [Xe]$5d^{1}6s^{2}$, similar to cerium but lacking any occupation of the $4f^{1}$ states. In the same manner as with cerium, the $6s^{2}$ electrons are donated to the boron octahedra, and with the $5d^{1}$ electrons acting as itinerant conduction electrons we find that the substitution of cerium with lanthanum dilutes the spins, resulting in the suppression of $T_{\mathrm{N}}$ and $T_{\mathrm{Q}}$. We can see in Fig.~\ref{Phase_diagram}\,(c) that these do not fall at the same rate, and $T_{\mathrm{Q}}$ falls faster than $T_{\mathrm{N}}$ until they coincide at $x \approx 0.2$  \cite{Fur85}. Here, a new phase termed ``phase IV'' forms above $T_{\rm N}$, which then persists as the ground state after the AFM phase is suppressed above the critical La concentration of $x \approx 0.3$. The magnetic phase diagram, illustrated in Fig.~\ref{Phase_diagram}\,(b), shows that above the critical concentration phase II is now only accessible in finite field \cite{Kob03, Suz98}. Further doping towards pure LaB$_{6}$ eventually suppresses the correlated electron physics, replacing phase IV with the paramagnetic phase I as the ground state of the system at around $x = 0.7$.

Whilst the introduction of lanthanum adds chemical disorder into the system, it was noted by Goodrich~\textit{et~al.} that its scattering potential appears negligible, as evidenced by the observation of dHvA oscillations within a range of lanthanum doped samples \cite{Goo99}. They found that both the Fermi surface topology and the quasiparticle effective masses transform gradually as a function of the doping parameter \textit{x}, in contrast to the rapid dephasing of quasiparticles at low concentrations of the dopant atoms in previously studied dilute Kondo alloys \cite{Low73, Sho84}. Furthermore, recent neutron scattering measurements on a variety of lanthanum doped samples find a common behaviour for the excitation spectrum of all measured samples, with a smooth evolution of characteristic parameters as a function of doping \cite{Fri15}. These observations indicate that experimental results from Ce$_{1-x}$La$_{x}$B$_{6}$ should reflect intrinsic effects, rather than being dominated by chemical disorder, making this a model system for the study of diluting \textit{f}-electron systems.

\subsection{An update to the known phase diagrams}

\begin{figure}[b]
\begin{center}
	\includegraphics[width=1\linewidth]{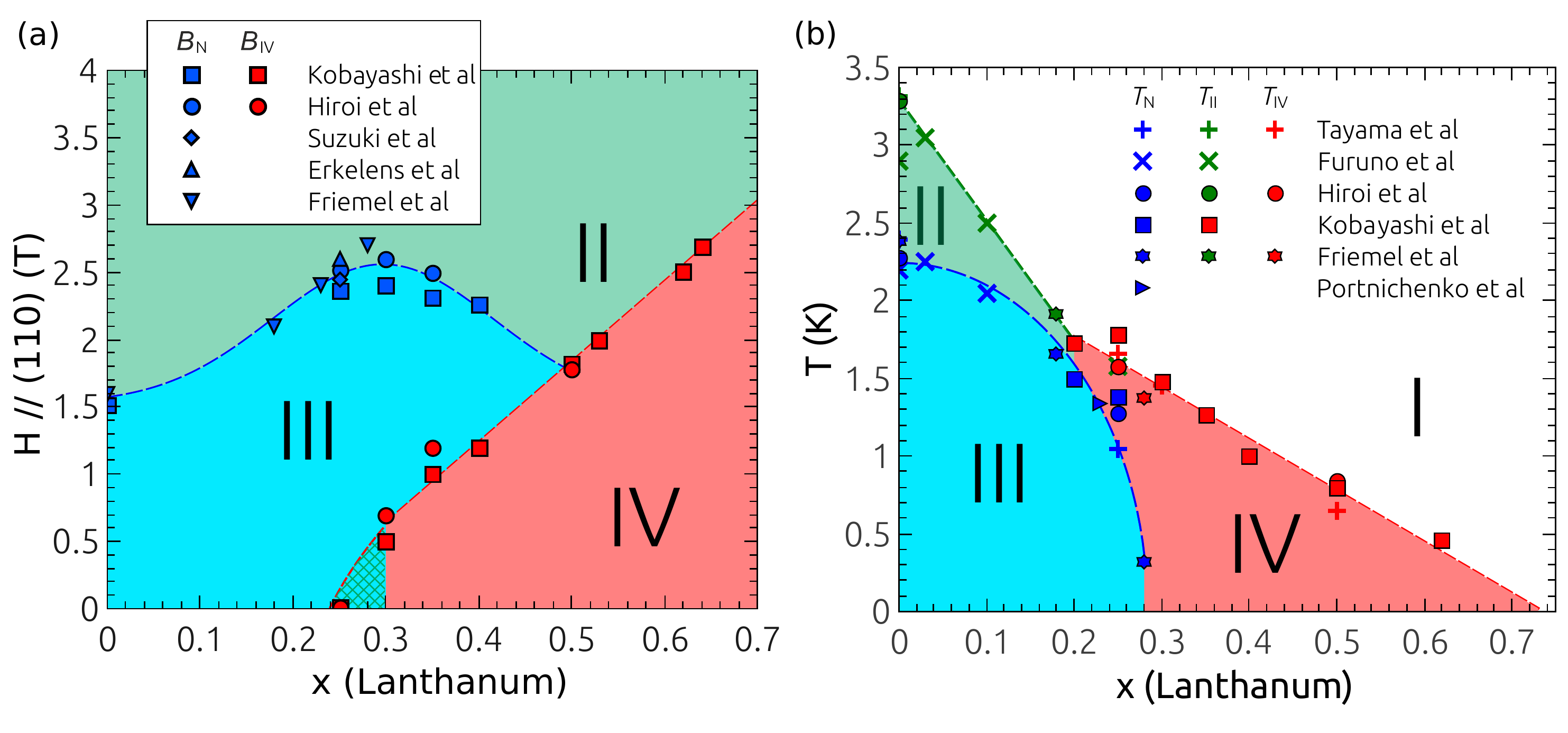}
\end{center}
	\caption{(a)~Zero-temperature phase diagram of Ce$_{1-x}$La$_{x}$B$_{6}$ constructed from data from references \cite{Hir97, Hir98, Kob00a, Erk87, Suz05, Fri15} for magnetic field applied parallel to the (110) axis. The lines indicating the phase boundaries are guides for the eyes. (b)~Zero field phase diagram of Ce$_{1-x}$La$_{x}$B$_{6}$ constructed from data published in references \cite{Hir98, Tay97, Fur85,Kob03,Hir97,Kob00a, Fri14, Por15a}.}
		\label{ls_phase_diagram}
\end{figure}

Figure \ref{Phase_diagram}\,(b) shows the magnetic field vs. lanthanum concentration \textit{x} phase diagram for Ce$_{1-x}$La$_{x}$B$_{6}$, for magnetic fields applied along the $\langle110\rangle$ and $\langle100\rangle$ directions, extrapolated to zero temperature. A qualitatively identical phase diagram is found for magnetic field applied along the $\langle111\rangle$ direction although the exact locations of the phase transitions will vary. The lines indicating the phase boundaries are a guide for the eyes, and neither the QCP of the AFM phase (phase III) nor the location of the quantum tri-critical point between phases II, III and IV is well known. It is evident that this tri-critical point will change its location depending on the direction of applied field, and that at higher doping levels, above $x \approx 0.7$, phase IV gives way to the paramagnetic state, although this region of the phase diagram is not well studied. In Fig.~\ref{ls_phase_diagram}\,(a) we have significantly updated this phase diagram for fields applied along the (110) direction, collating data from references \cite{Hir97, Hir98, Kob00a, Erk87, Suz05, Fri15}. In this figure, we can see that the exact location of the AFM QCP is somewhere between $x = 0.25$ and $x = 0.3$, and that the tri-critical point between phases II, III and IV appears to be very close to $x = 0.5$. There is a slight difference between the behaviour of different samples from different studies, although this is perhaps expected given that the location of $T_{\mathrm{N}}$ and $T_{\rm Q}$ was seen to vary between samples in early studies on pure CeB$_{6}$.

Figure \ref{ls_phase_diagram}\,(b) is an updated version of the temperature vs. lanthanum concentration phase diagram for Ce$_{1-x}$La$_{x}$B$_{6}$ at zero field originally seen in Fig.~\ref{Phase_diagram}\,(c). It is compiled predominantly from specific heat, magnetisation and transport measurements \cite{Hir98,Tay97,Fur85,Kob03,Hir97,Kob00a} as well as some recent neutron scattering data \cite{Fri14, Por15a}. There are some significant changes to our previous understanding of the phase diagram. It is now clear that the continuity in the phase II-I and phase III-IV transition lines and the phase II-III and I-IV transition lines seen in Fig.~\ref{Phase_diagram}\,(c) is incorrect, and that this was purely coincidental in the earlier study. The updated phase diagram also strongly suggests that the QCP of the AFM phase is at $x = 0.28$, and whilst there are no measurements near the termination of phase IV at high lanthanum concentrations, we can see from the linear extrapolation of the phase IV transition temperature, $T_{\rm IV}$, that it probably ends around $x = 0.75$, where quantum critical behaviour in the resistivity has been reported \cite{Nak06}. It also appears that phases II and IV do not terminate in connection to the phase III line at $x = 0.2$ but in fact meet just above the AFM phase boundary, which itself forms a smooth dome. This would mean that the transition between the AFM and paramagnetic phases is avoided, although this is inferred by following the transition lines and there potentially could still be a doping level where all of the phases meet.

\subsection{The order parameter of phase IV in Ce$_{1-x}$La$_{x}$B$_{6}$}

\begin{figure}[t]
\begin{center}
	\includegraphics[width=1\linewidth]{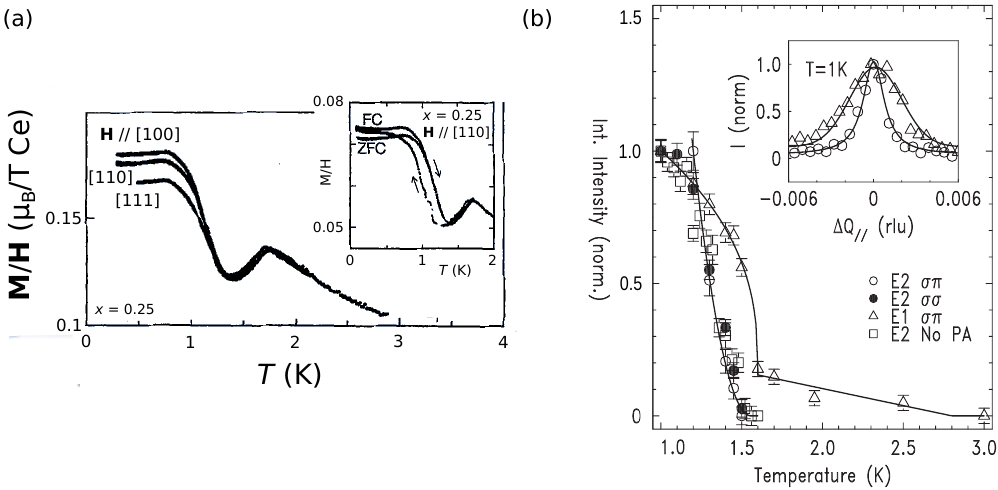}
\end{center}
	\caption{(a) Magnetic susceptibility in a field of $\textbf{B} = 0.1$~T of a sample with lanthanum doping $x = 0.25$. Reproduced from Ref.~\cite{Tay97}. (b) Temperature dependence of the RXS signal at the \textit{R} point, for the E1 and E2 transitions at the $L_{2}$ edge in a sample with doping $x = 0.3$. Reproduced from Ref.~\cite{Man05}, copyright by the American Physical Society.}
		\label{phase_IV_evidence}
\end{figure}

In a similar manner to the order parameter of phase II, the mechanism behind phase IV has been difficult to determine. Experimental consensus remained elusive, with a variety of propositions as to the nature of phase IV stemming from seemingly incompatible results. On one hand, studies of the magnetic and transport properties \cite{Kob03, Suz98} have observed  a continuous transition line between $T_{\mathrm{N}}$ and  $T_{\mathrm{IV}}$, suggesting that phase IV, like phase III, is a magnetically ordered state. The sharp peak in the heat capacity suggests a long-range ordered phase, whilst the cusp in magnetic susceptibility across the transition in Fig.~\ref{phase_IV_evidence}\,(a) suggests a quench in the magnetic degrees of freedom, which would imply that this is magnetic in nature \cite{Tay97, Suz98}. However, no such magnetic signal has been found from neutron diffraction, with both powder and single crystal studies ruling out long range magnetic order and providing support to multipolar ordering as a candidate \cite{Fis05, Iwa03}. Furthermore, $\mu$SR results show only a random distribution of internal fields of the order of 0.1~T, displaying no anomaly across the phase III -- phase IV boundary, perhaps suggesting that phase IV is a short-range-order region of the AFM phase \cite{Sch07}. An anomalously large softening of the $C_{44}$ elastic constant has been observed when entering phase IV~\cite{Kob03}, which is connected to a minute contraction ($\Delta{L} / L \approx 10^{-6}$) of the lattice along the $\langle111\rangle$ direction. This is not seen in phases II or III, and it has been argued that it is incompatible with long range AFQ order \cite{Suz98, Suz05}. Evidence for multipolar ordering is provided by RXS experiments \cite{Kus05, Man05, Kuw07, Mat14}, and data from Ref.~\cite{Man05}, on a sample with doping parameter $x = 0.3$, are reproduced in Fig.~\ref{phase_IV_evidence}~(b). These experiments, which monitored the $L_{2}$ edge of cerium, have observed dipolar and octupolar symmetry for the E1 and E2 transitions respectively. Both show a sudden onset when entering phase IV, with some dipolar signal remaining above $T_{\mathrm{IV}}$ \cite{Man05}. Recent elastic neutron scattering has revealed a weak Bragg peak forming below $T_{\mathrm{IV}}$ at $(\frac{1}{2} \frac{1}{2} \frac{1}{2} )$ \cite{Kuw07}, which is suppressed with the application of magnetic field. These data also revealed an increase in the Bragg intensity $I(\textbf{Q})$ with $|\mathbf{Q}|$, and this was interpreted as lending further evidence for an octupolar order parameter of phase IV.

Multipolar order was suggested from a mean-field theory study \cite{Kub04}, which explained the (111) lattice distortion as a result a ferroquadrupolar ordering induced by an antiferrooctupolar (AFO) ordering, the latter playing the role of the order parameter of phase IV. This model was also able to account for the cusp in magnetisation at $T_{\mathrm{IV}}$ \cite{Suz98} and the softening of the elastic constant which were observed in experiment.  From this, a ground state of $\Gamma_{5u}$ octupolar ordering was proposed. This is somewhat consistent with the RXS data reproduced in Fig.~\ref{phase_IV_evidence}~(b), which show a strong onset of the E2 edge below $T_{\mathrm{IV}}$. However, these results also suggest the onset of dipolar order, with the \textbf{q}-width of the octupolar peak suggesting long-range order whilst the comparatively broader dipolar peak indicates a short-range order. The temperature dependencies of these peaks are markedly different, and as a result the authors of the RXS study suggest a novel segregated order parameter phase of long range AFO order alongside mesoscopic AFM order \cite{Man05}, which also accounts for some of the discrepancies in the earlier experimental data.

Despite the disparity in early experimental results, the emerging consensus is that $\Gamma_{5u}$ AFO ordering is the ground state of phase IV. However, it is evident that this is not the only correlated behaviour to be found in phase IV. More recent RXS experiments, coupled with mean-field theory calculations, have shown the presence of field-induced quadrupolar moments within the AFO phase IV \cite{Mat14}. Both $\mathrm{\Gamma}_{3\mathrm{g}}$ and $\mathrm{\Gamma}_{5\mathrm{g}}$ quadrupolar moments are observed to be induced by field within the AFO phase, however in contrast to the mean-field calculation the $\mathrm{\Gamma}_{5\mathrm{g}}$ quadrupoles are induced far more strongly than the $\mathrm{\Gamma}_{3\mathrm{g}}$ quadrupoles. The authors conclude that a large fluctuation of the $\mathrm{\Gamma}_{5\mathrm{g}}$ order is hidden behind the $\mathrm{\Gamma}_{5\mathrm{u}}$ octupolar order, and speculate that quantum fluctuations from the close degeneracy of the AFM, AFQ and AFO orders play an important role in a region where mean-field models are unable to explain their experimental observations. This makes Ce$_{1-x}$La$_{x}$B$_{6}$ one of only a handful of compounds where octupolar ordering has been reported, alongside NpO$_{2}$ \cite{Pai02, Cac03}, SmRu$_{4}$P$_{12}$ \cite{Yos05, Aok08} and URu$_{2}$Si$_{2}$ \cite{Myd11, Shi12}.

\begin{figure}[b]
\begin{center}
    \includegraphics[width=\textwidth]{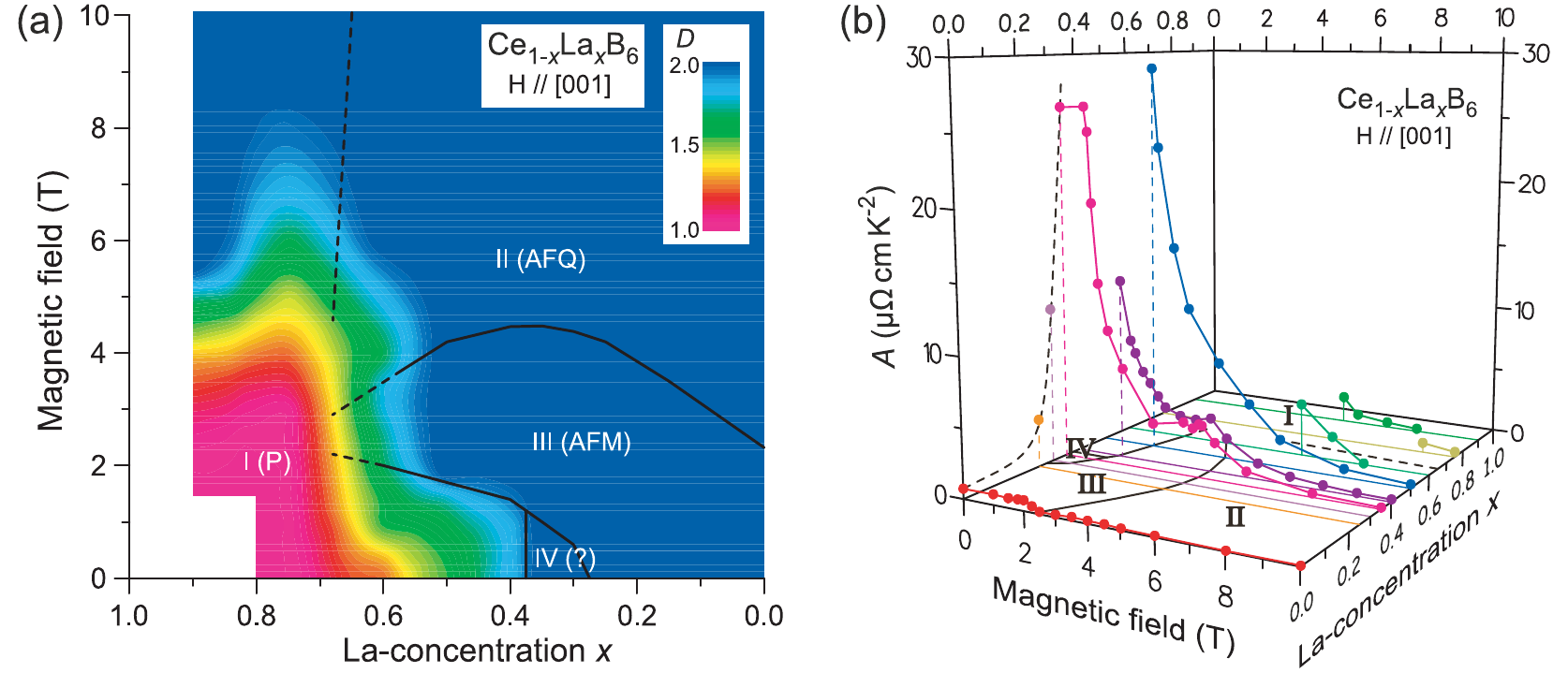}
\end{center}
	\caption{(a)~Magnetic field vs. lanthanum concentration phase diagram of Ce$_{1-x}$La$_{x}$B$_{6}$ for $\mathbf{B} \parallel \langle001\rangle$ at low temperatures. Colour denotes the value of the exponent \textit{D} in the temperature dependence of resistivity $\rho (T) = \rho (0) + AT^{D}$. (b)~Field dependence of the $A$ coefficient for various samples. Adapted from Ref.~\cite{Nak06}, copyright by the American Physical Society.}
		\label{CeLaB6_NFL}
\end{figure}

\subsection{Non-Fermi-liquid regime and quantum criticality in Ce$_{1-x}$La$_{x}$B$_{6}$}

Following the evolution of $T_{\mathrm{N}}$ with doping in Fig.~\ref{ls_phase_diagram}, at $x_{\rm c} \approx 0.3$ there is a QCP for the AFM phase, and it is expected that critical fluctuations in this region may result in deviations from FL behaviour. However, a transport study found the FL behaviour at low temperatures to persist up to a doping of $x = 0.35$, beyond the critical doping \cite{Nak06}, as illustrated in Fig.\,\ref{CeLaB6_NFL}\,(a). In the FL regime the resistivity follows the law $\rho = \rho_{0} + AT^{2}$, which was observed in these measurements. As can be further seen in Fig.\,\ref{CeLaB6_NFL}\,(b), the parameter \textit{A}, which is proportional to the square of the effective mass $m^{\ast}$, increases divergently on the approach to $x_{\rm c}$ in zero field, and at $x_{\rm c}$ it assumes a large but finite value which remains unchanged up to $x = 0.5$. This indicates that whilst FL behaviour is being adhered to in the region of the AFM QCP, the system is experiencing a dramatic mass enhancement with increasing lanthanum doping. This could be explained in a 3D SDW quantum critical picture, where only a fraction of the quasiparticles are scattered from the critical fluctuations, which are confined to ``hot spot'' wavevectors \cite{Sil06, Nak06, Col01}. Quasiparticles from the contrasting ``cold'' regions of the Fermi surface remain unaffected, and the averaged effective mass remains finite, although these results may also suggest that the mass enhancement is a property of phase IV rather than a phenomenon emerging from criticality. Further doping beyond $x_{\rm c}$ brings the system into a situation where the AFM phase is no longer the ground state and can only be realised in an applied field, while the ground state is now phase IV.

For the region $0.35 < x < 0.8$, non-Fermi liquid (NFL) behaviour has been observed in the low-temperature resistivity of Ce$_{1-x}$La$_{x}$B$_{6}$}, which follows the relation $\rho = \rho_{0} + AT^{D}$ with $D < 2$ \cite{Nak06}, as seen in Fig.~\ref{CeLaB6_NFL}. There is a distinct region at low fields and high lanthanum concentrations where \textit{D} has decreased from the value of 2 predicted for Fermi liquids, reaching $D = 1$ at $x \approx 0.7$. However, within the NFL region of Ce$_{1-x}$La$_{x}$B$_{6}$ the specific heat in phase I does not show diverging behaviour with decreasing temperature \cite{Nak00, Nak03} as expected for some NFL models \cite{Ste01, Mir05}. Here the heat capacity can be described as $C_{p} \propto T^{\gamma}$, with $\gamma= 1.6$ for $x = 0.4$ \cite{Nak00}. Whilst this \textit{is} a deviation from FL behaviour (which assumes $\gamma = 1$), it could not be explained yet in any known quantum-critical scenario, where an increase of specific heat with decreasing temperature is typically expected \cite{Col01}. Furthermore, the magnetic susceptibility shows non-typical behaviour within the NFL regime with signs of short-range AFM correlations surviving to low concentrations of cerium, despite the NFL transport properties, and high lanthanum concentrations show local-FL behaviour \cite{Nak06}.

Whilst the coefficient \textit{D} has decreased from 2 at concentrations below $x = 0.35$ to 1 at $x = 0.7$ across the NFL region as the lanthanum concentration increases, further increasing the lanthanum concentration beyond $x \approx 0.8$ reverses this trend, and \textit{D} increases to 1.8 at Ce$_{0.1}$La$_{0.9}$B$_{6}$ in zero field (not shown in Fig.~\ref{CeLaB6_NFL}), with the value for $A$ becoming negative at this point \cite{Nak06}. It is also important to note that within the doping-field phase diagram, the effective mass undergoes an enhancement from the paramagnetic phase I on the approach to any of the ordered phases. This effect, when measured as a function of applied magnetic field, appears to scale with the concentration of cerium ions, and it is speculated that it therefore originates from a single-site effect, such as the Kondo effect \cite{Nak06}. The transition from FL to NFL behaviour may normally be associated with the QCP, however in this system the same transition takes place away from the QCP, and the NFL region extends over a wide range of doping, suggesting that the two are not necessarily associated.

Potential signs of quantum criticality near the critical doping, $x_{\rm c}$, have also been observed in neutron-scattering measurements \cite{Fri15}. Here, the authors investigate the behaviour of magnetic excitations in Ce$_{1-x}$La$_{x}$B$_{6}$ as a function of both magnetic field and lanthanum concentration. Whilst the spin excitations within phase IV are quasielastic in no applied field, it is found that the ``AFQ$_{1}$'' mode at the \textit{R} point, induced by magnetic field within the AFQ phase, is significantly broader in energy near the critical doping than in the other samples with lower doping levels. This increase in linewidth could indicate the onset of critical fluctuations, which arise from the suppression of the exciton energy close to zero as one approaches the critical doping. In contrast to other heavy-fermion systems with quantum-critical behaviour such as YbRh$_{2}$Si$_{2}$ \cite{Cus03} and Ce$_{3}$Pd$_{20}$Si$_{6}$ \cite{Cus12}, no critical fluctuations are observed near $B_{\rm c}$ or $B_{\rm Q}$. Friemel~\textit{et~al.}~\cite{Fri15} argue that the absence of a field-induced QCP can be explained by considering the itinerant magnetic moments to be ferromagnetically coupled \cite{Jan14}, leading to a stabilisation of the associated spin dynamics by the application of magnetic field. The observation that the energy scale of the AFQ$_1$ mode vanishes as $T_{\mathrm{N}}$ is driven to zero leads the authors to conclude that the AFM QCP near $x_{\rm c} = 0.3$ is coincident in doping with the zero-field extrapolated QCP of the AFQ phase.

The evolution from the FL state of LaB$_{6}$ to the FL state of CeB$_{6}$, passing through the local-FL and NFL states, is in contrast to the path taken for typical non-magnetic HF systems, such as Ce$_{1-x}$La$_{x}$Cu$_{6}$. Here, the system follows the $\rho = \rho_{0} + AT^{2}$ throughout the entire doping range, going from the local-FL at the lanthanum end of the spectrum to the FL region at the cerium end without passing through the NFL regime \cite{Sum86}. The behaviour of the \textit{f}-electrons in these systems is considered to be quite different, being notably itinerant in the CeCu$_{6}$ in contrast to more localised in the CeB$_{6}$ system. However, as has been discussed previously in this review, the localised picture is not able to fully describe the behaviour of CeB$_{6}$ either, with the two-fluid heavy electron picture providing a more satisfactory description. Whilst one might expect the NFL behaviour to originate from the quantum criticality associated with the disappearing AFM phase, its persistence only at the lanthanum-rich side of the phase diagram would perhaps point to a separate origin, possibly requiring proximity to phase~IV.

\subsection{Superconductivity in LaB$_{6}$}

For the metallic hexaboride LaB$_{6}$ superconductivity has been reported in several studies. Initially found to have a transition temperature of $T_{\mathrm{c}} = 5.7$~K \cite{Mat68}, this result has proven difficult to reproduce. Another study reported the transition to take place at 0.1~K \cite{Sob79}, whilst resistivity measurements on several samples found no superconducting transition down to a temperature of 5~mK \cite{Bat95}. The most commonly accepted value appears to be 0.45~K \cite{Sch82}. Several explanations have been suggested for this disparity of results and the low critical temperature. The resistivity studies of Ref.~\cite{Bat95} surprisingly found a slight increase of resistivity at low temperatures in this otherwise metallic compound, and suggested that Kondo scattering from magnetic impurity atoms in their samples could be responsible, and thus also explained the lack of superconductivity in their sample. It has also been suggested that interactions between the boron octahedra and the conduction electrons result in a deformation of the phonon modes and thus a suppression of the superconducting critical temperature \cite{Sch82, Sam88}. To our knowledge, determination of the true intrinsic value of $T_{\rm c}$ as well as its behaviour upon Ce doping still remains an open problem.

\subsection{Magnetic ion doping}

Recently, Kondo~\textit{et~al.} and Matsumura~\textit{et~al.} have experimented with co-doping Ce$_{1-x}$La$_{x}$B$_{6}$ with either Pr or Nd \cite{Kon07, Kon09, Mat14a}. In their initial study, it appeared that in Ce$_{0.7}$Nd$_{y}$La$_{0.3-y}$B$_{6}$ the phase IV transition temperature is rapidly suppressed with Nd doping, whilst phase III is on the contrary stabilised. However, in their following study they found contrasting behaviour. These experiments involved doping Ce$_{1-x}$La$_{x}$B$_{6}$ with either Pd or Nd, and they found that what appeared to be the transition temperature $T_{\mathrm{IV}}$ showed a notable \textit{increase} when the sample is doped with 10\% Pr or Nd when the cerium concentration is less than 0.5. For cerium concentrations of 0.6 or higher, however, they found no effect. These two experiments, therefore, appeared to be contradictory. From their specific heat, magnetisation and resistivity measurements in this study, they rule out the pressure effect of doping with these ions, as their relative size and the change in transition temperature would lead to an opposite effect. Further, they are able to rule out the contribution of either octupolar or quadrupolar moments from the dopant ions. This leads them to conclude that it is likely the dipole moments in these dopant ions which cause the stabilisation of the ground state. However, they also note that dipolar moments cannot couple to the $\mathrm{\Gamma}_{5\mathrm{u}}$ octupolar moment favoured by the prevalent mean-field theories describing phase IV. In order to shed further light on this problem, they performed neutron-scattering measurements on Ce$_{0.5}$Pr$_{0.1}$La$_{0.4}$B$_{6}$ and discovered that rather than the AFO order, which is expected for phase IV, it instead exhibits AFM dipole order. The AFO order associated with phase IV has an ordering vector $\mathbf{q} = (\frac{1}{2} \frac{1}{2} \frac{1}{2})$. However, the neutron diffraction measurements clearly showed a magnetic peak at $\mathbf{q} = (\frac{1}{4} \frac{1}{4} \frac{1}{2})$, the \textbf{q}-vector for AFM order, with no intensity at the expected position for the AFO order. This perhaps explains the disparity between their earlier studies, and shows that the AFM dipole order is one of the competing order parameters within lanthanum doped CeB$_{6}$, and that it is close enough in energy to be realised under a weak perturbation.

\begin{figure}[b]
    \includegraphics[width=\textwidth]{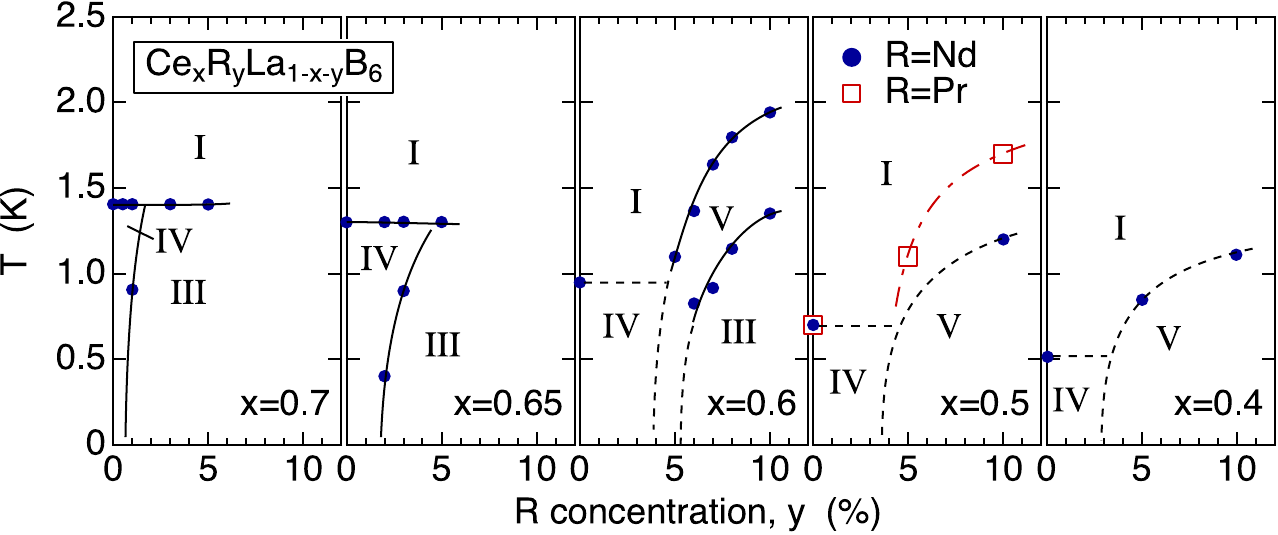}
	\caption{Proposed $T-x-y$ phase diagrams by Matsumura~\textit{et~al.} for Ce$_{x}$$R_{y}$La$_{1-x-y}$B$_{6}$ for \textit{R} = Nd, Pr. Data points are shown, with known phase boundaries marked in solid lines and speculative phase boundaries in dashed lines. Reproduced from Ref.~\cite{Mat14a}.\vspace{-2pt}}
		\label{Pr_dope_phase}
\end{figure}

Phase diagrams for these systems are reproduced here in Fig.~\ref{Pr_dope_phase}. The AFM order seen here is labelled as phase V by the authors, to distinguish it from phase III, and is seen only for cerium concentrations of 0.6 or less. For the cerium concentration of 0.6, the authors find that phase V is not the ground state of the system, but is developing above phase III (in temperature) as the \textit{R} concentration is increased. Their results show that for cerium concentrations of less than 0.6, phase III, the double-\textbf{q} structure seen in the ground state of CeB$_{6}$, is only present at intermediate fields, with the ground state becoming phase IV, that is realised below 1.8~T. At higher fields the development of phases III and II in a similar manner to CeB$_{6}$ is found. Above $\sim$\,4~T the AFQ order is realised, indicated by the presence of a magnetic Bragg peak with $\textbf{q} = (\frac{1}{2} \frac{1}{2} \frac{1}{2})$, induced by field and modulated by the underlying AFQ order.

It is worth noting here that the neutron diffraction measurements showed an unusual temperature dependence of the scattered intensity, not characteristic of a typical developing order parameter. The authors point out that this is concurrent with the characteristic shape of the specific heat anomaly reported in their earlier measurements \cite{Kon09} and suggests that the development of long-range order is hindered by by Pr-doping. The orbital structure of Pr possesses a different ground state from cerium, and the differing magnetic interactions originating from these two ions may be hindering the development of magnetic order, in a manner that may be similar to that observed in Ce$_{1-x}$Pr$_{x}$B$_{6}$ \cite{Mig08}.

\section{Related Materials}

\subsection{Ce$_{x}$Pr$_{1-x}$B$_{6}$}

Recently, the compound Ce$_{x}$Pr$_{1-x}$B$_{6}$ has attracted attention, in addition to the mentioned experiments where Pr is co-doped into Ce$_{1-x}$La$_{x}$B$_{6}$. A variety of publications have investigated the low-temperature magnetically ordered phases of this compound across the complete range of doping levels \cite{Mig08,Kis05,Nag08,End06}. The crystal-field ground state of the Pr$^{3+}$ ions is a $\mathrm{\Gamma}_{3}$ triplet state, as opposed to the $\mathrm{\Gamma}_{8}$ ground state of cerium in CeB$_{6}$, and this triplet state can support  quadrupolar moments, but not the octupolar moments which play an active role in the physics of the cerium based hexaborides. PrB$_{6}$ exhibits no AFQ order in the manner of CeB$_{6}$, but instead supports two low-temperature AFM phases. Firstly it enters an incommensurate state passing below $T_{\mathrm{N}} = 7$~K, and then undergoes a lock-in transition to commensurate antiferromagnetic order at $T_{\mathrm{C}} = 4.2$~K. The commensurate phase (C-phase) is of the same double-k structure as CeB$_{6}$, and it has been argued that the similarity with CeB$_{6}$ suggests an association with the $O_{xy}$ type AFQ order that plays a role in CeB$_{6}$. However, as a counterpoint to this, GdB$_{6}$ possesses an AFM phase with the same ordering vectors as CeB$_{6}$ and PrB$_{6}$, but has no 4\textit{f} orbital degeneracy, and so it has been suggested instead that this particular AFM structure is a result of features of the Fermi surface that are common to all light-earth hexaborides \cite{Kur02}.

\begin{figure}
\begin{center}
    \includegraphics[width=0.5\linewidth]{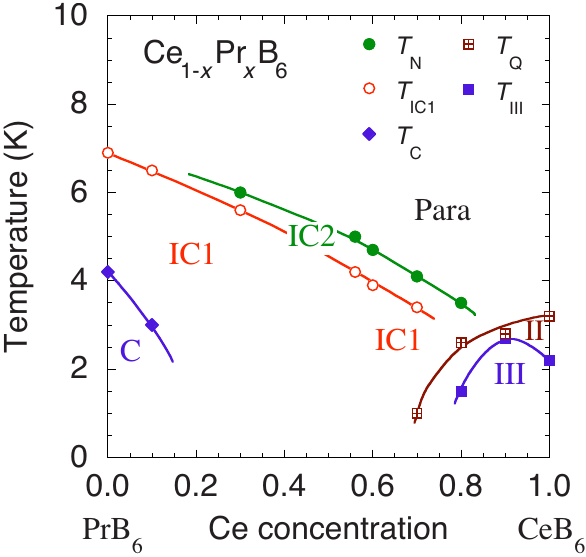}
\end{center}
	\caption{Phase diagram for Ce$_{x}$Pr$_{1-x}$B$_{6}$. Reproduced from \cite{Mig08}, copyright by the American Physical Society.}
		\label{CePrB6}
\end{figure}

The phase diagram of Ce$_{x}$Pr$_{1-x}$B$_{6}$ from the neutron scattering study of Mignot~\textit{et~al.} is reproduced in Fig.~\ref{CePrB6}. They find that the system displays a variety of magnetic structures as a function of doping and temperature, with the PrB$_{6}$-like region of the phase diagram extending to remarkably high cerium concentrations of around $x = 0.8$. The commensurate magnetic order characteristic of the ground state of PrB$_{6}$ is directly observed in their measurements only until $x = 0.1$, whilst two incommensurate magnetic states also characteristic of PrB$_{6}$-like behaviour are seen all the way up to $x=0.8$. This is a crossover situation, as both PrB$_{6}$ and CeB$_{6}$ behaviour exists at this doping. Indeed, data from Ref.~\cite{Kis05} at $x=0.7$, whilst showing a Pr-like ground state, also exhibits the Ce-like phase II under applied field. At the highest cerium concentrations, phases II and III are observed, and it appears that light doping by Pr ions actually stabilises phase III at the expense of phase II.

Whilst the commensurate phase is only indicated for the lowest Ce concentrations, it is important to note that these results are based on both single-crystal and powder diffraction, which show slightly different behaviour. For the $x = 0.2$ and 0.4 concentrations, reflections associated with the C-phase were present alongside the incommensurate order all the way up to $T_{\mathrm{N}}$ in the powder data, whilst the single-crystal data showed no C-phase at these concentrations. This behaviour is also seen in the pure compound, with commensurate reflections are seen immediately below $T_{\mathrm{N}}$ alongside the incommensurate peaks, and no distinct lock-in to a commensurate phase is observed. The origin of this behaviour is not yet known. The IC2 phase seen in these data is not observed in pure PrB$_{6}$ in zero field, but is related to a distinct type of incommensurate magnetic order which is seen in applied magnetic field, and has also been observed in La-doped PrB$_{6}$. This series is an illustration of the importance of octupolar interactions in CeB$_{6}$ in stabilising the AFQ phase, as the $O_{xy}$ quadrupolar interactions are known to play a role across this entire series, however it is only in the cerium-dominant compounds where the octupolar moments are prevalent that quadrupolar order is expressed.

\subsection{Ce$_{3}$Pd$_{20}$Si$_{6}$}

\begin{figure}[t]
\begin{center}
    \includegraphics[width=\textwidth]{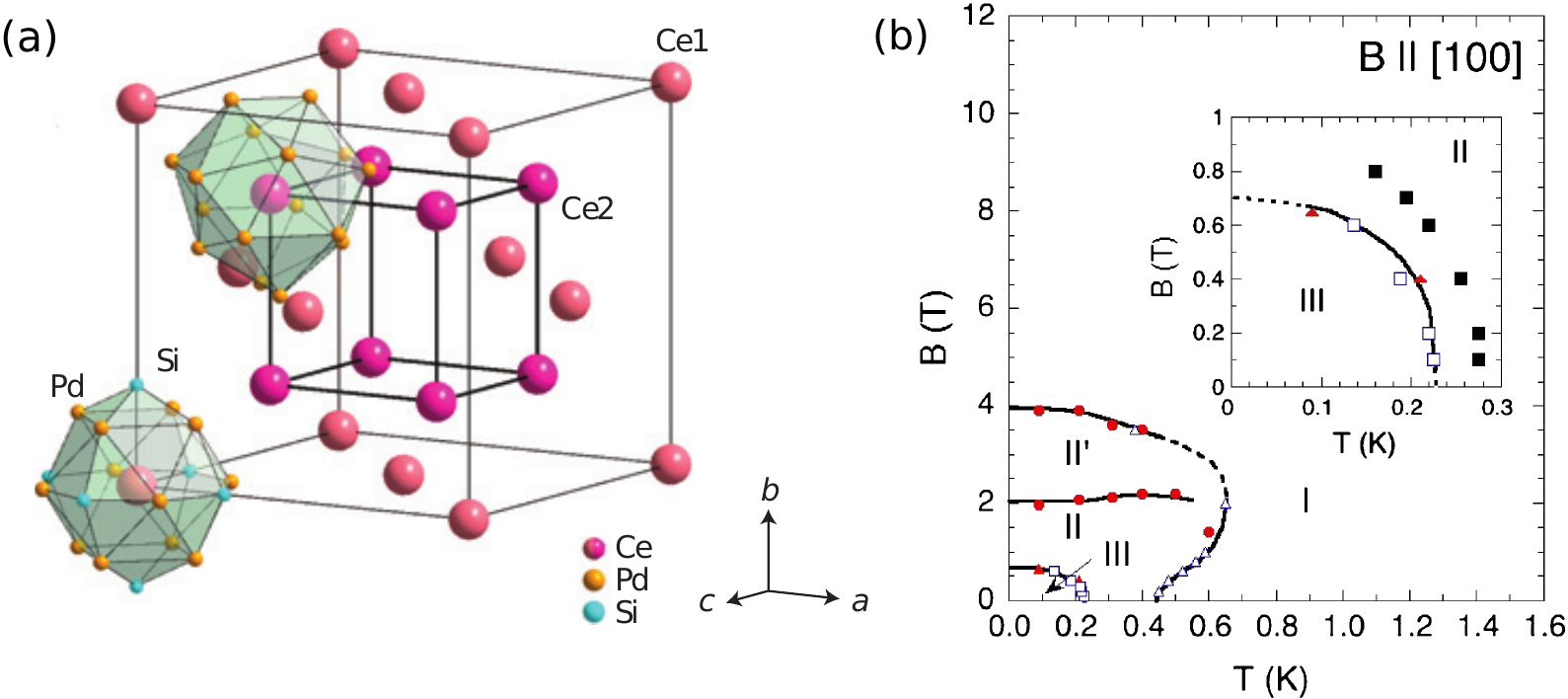}
\end{center}
	\caption{(a) Crystal structure of Ce$_{3}$Pd$_{20}$Si$_{6}$, showing two inter-penetrating sub-lattices of cerium ions. Reproduced with permission from Ref.~\cite{Cus12}. (b) Magnetic field and temperature phase diagram, for fields applied parallel to the $\langle100\rangle$ crystal direction, showing three distinct low temperature phases. Reproduced from Ref.~\cite{Mit10}.}
		\label{CePdSi}
\end{figure}

Ce$_{3}$Pd$_{20}$Si$_{6}$ is a relatively new clathrate compound ($R_{3}$Pd$_{20}X_{6}$, \textit{R} = rare earth, \textit{X} = Si, Ge) which has attracted recent attention due to indications of multipolar ordering in its low-temperature phases. It consists of two inter-penetrating sub-lattices of cerium, as shown in Fig.~\ref{CePdSi}\,(a). The structure possesses one simple-cubic and one face-centred-cubic (fcc) lattice, with each cerium ion surrounded by a cage of either palladium or silicon-palladium depending on which sub-lattice the Ce ion belongs to. Whilst one might expect that the significantly different crystal structures of CeB$_{6}$ and Ce$_{3}$Pd$_{20}$Si$_{6}$ would lead to different low temperature behaviour for the two compounds, inspection of the temperature-field phase diagram shown in Fig.~\ref{CePdSi}\,(b) indicates that Ce$_{3}$Pd$_{20}$Si$_{6}$ may be remarkably similar in behaviour to CeB$_{6}$, potentially displaying the same low-temperature phases albeit at greatly reduced temperatures and fields. It is clear that Ce$_{3}$Pd$_{20}$Si$_{6}$, alongside Ce$_{3}$Pd$_{20}$Gd$_{6}$ \cite{Doe00}, differs from many of its relatives in the isostructural $R_{3}$Pd$_{20}X_{6}$ group in that it doesn't exhibit a pair of antiferromagnetic transitions. These have been seen, one for each sub-lattice, in its relatives Tb$_3$Pd$_{20}$Si$_6$ \cite{Her99}, Dy$_3$Pd$_{20}$Si$_6$ \cite{Her00}, Nd$_3$Pd$_{20}$Ge$_6$ \cite{Don98} and Nd$_3$Pd$_{20}$Si$_6$ \cite{Don00a}. It does however exhibit two low-temperature ordered phases, II and III, with the characteristic enhancement with field of the AFQ phase in CeB$_{6}$ also seen in the phase II of this material, which in turn surrounds a similarly placed ground-state phase III. Magnetisation measurements have indicated that phase III in this material may be either an AFM or ordered octupolar phase \cite{Mit10}, with neutron scattering showing the presence of static AFM order \cite{Lor12, Lor15}. Magnetic susceptibility \cite{Pas07} and ultrasound \cite {Got09, Wat07} measurements have strongly suggested that phase II in this material is an AFQ phase. This situation looks remarkably similar to CeB$_{6}$, and should phase II in this material be the same as it is in CeB$_{6}$, residing on the simple-cubic cerium sublattice, then we would expect an AFQ phase with a propagation vector of $(111)$.

Perhaps one of the most significant complications when investigating Ce$_{3}$Pd$_{20}$Si$_{6}$ with diffraction experiments arises from its crystal structure. Whilst recent neutron measurements have indicated that the fcc sublattice of cerium is ``magnetically silent'' \cite{Por15}, it still provides complications for investigating the nature of the low-temperature phases. Although recent macroscopic measurements have suggested that phase II of Ce$_{3}$Pd$_{20}$Si$_{6}$ is an AFQ phase as in CeB$_{6}$, should the ordering vector of this phase be also be at $(\frac{1}{2} \frac{1}{2} \frac{1}{2})$ for the simple-cubic lattice, as in CeB$_{6}$, then under neutron scattering it would fall on the $(111)$ nuclear Bragg peak arising from the fcc sublattice. The tactic that was successful in CeB$_{6}$, of observing an AFM order induced by applied magnetic field but modulated to the AFQ order, would in this case be more complicated, since the disparate strength of the nuclear and magnetic signals would render the latter difficult to distinguish. Recent measurements are consistent with the cerium ions in the 8c simple-cubic sublattice [Ce2 in Fig.~\ref{CePdSi}\,(a)] having a $\mathrm{\Gamma}_{8}$ ground state with a $\mathrm{\Gamma}_{7}$ excited state some 3.9~meV above this \cite{Pas08, Por15}, whilst the 4a fcc lattice cerium ions seem to possess a $\mathrm{\Gamma}_{7}$ ground state \cite{Got09, Ben11, Yam15}. Interestingly enough, when taking into account the lack of contribution of the fcc lattice, the magnetic environment of Ce$_{3}$Pd$_{20}$Si$_{6}$ echoes that of CeB$_{6}$ \cite{Por15, Yam15}, and Goto~\textit{et~al.} have noted that with the inter-site distance of the simple cubic lattice being larger in Ce$_{3}$Pd$_{20}$Si$_{6}$ than for CeB$_{6}$, one might expect the inter-site coupling to be correspondingly weaker, which may account for the reduced temperature scales in this material.

\section{Summary}

Despite  more than 50 years of research on CeB$_{6}$ and its doped analogues, such as Ce$_{1-x}$La$_{x}$B$_{6}$, it continues to display interesting and unexplained physics, with more modern techniques unearthing behaviour not predicted by theories which were able to describe earlier results. The variety of low-temperature ordering phenomena it displays, resulting from a complex interplay of both competing and cooperating interactions, has proven difficult to model and, with a complete understanding remaining an elusive goal, demands further investigation from both theory and experiment. It is evident that the interesting low-temperature physics of this system predominantly arises from the $\Gamma_{8}$ crystal-field ground state of the Ce$^{3+}$ ion. This supports a variety of multipolar moments that are able to interact and order, with some acting as competing order parameters whilst others seem to cooperate to undergo ordering. Being metallic, the system also displays a variety of interesting heavy-fermion phenomena, and it is clear that the \textit{f}-electrons are not always a purely localised entity, so that in order to explain the behaviour of this system it is necessary to account for their two-faced nature. Furthermore, recent results from a variety of experiments have unearthed a host of novel phenomena, ranging from the presence of a SDW to the system's proximity to a ferromagnetic instability. These unexplained results, the continued characterisation of phase IV and the new and interesting physics arising from the doping of CeB$_{6}$ with other magnetic ions indicate that this model system will remain an interesting and active research topic for many years to come.

\section*{Acknowledgments}

We are very grateful to N.~E. Sluchanko, S.\,V. Demishev, N.~Yu.~Shitsevalova, B. Keimer, J.-M. Mignot and P.~Thalmeier for stimulating our initial interest to this project and for many fruitful and encouraging discussions in its course. This work was funded by the German Research Foundation (DFG) under Grant No. IN 209/3-1.

\section*{References}
\bibliographystyle{iopart-num}
\bibliography{CeB6}

\providecommand{\newblock}{}
\begin{thebibliography}{100}
\expandafter\ifx\csname url\endcsname\relax
  \def\url#1{{\tt #1}}\fi
\expandafter\ifx\csname urlprefix\endcsname\relax\def\urlprefix{URL }\fi
\providecommand{\eprint}[2][]{\url{#2}}

\bibitem{Sta32}
Stackelberg M~V and Neumann F 1932 {\em Z. Physik. Chem.\/} {\bf B19} 314

\bibitem{Laf51}
Lafferty J~M 1951 {\em J. Appl. Phys.\/} {\bf 22} 299

\bibitem{Pad61}
Paderno Y and Samsonow G 1961 {\em Doklady Akad. Nauk USSR\/} {\bf 137} 646

\bibitem{Geb68}
Geballe T~H, Matthias B~T, Andres K, Maita J~P, Cooper A~S and Corenzwit E 1968
  {\em Science\/} {\bf 160} 1443--1444

\bibitem{Nic69}
Nickerson J~C and White R~M 1969 {\em J. Appl. Phys.\/} {\bf 40} 1011--1012

\bibitem{Win75}
Winzer K 1975 {\em Solid State Commun.\/} {\bf 16} 521--524

\bibitem{Tak80}
Takase A, Kojima K, Komatsubara T and Kasuya T 1980 {\em Solid State Commun.\/}
  {\bf 36} 461--464

\bibitem{Zir84}
Zirngiebl E, Hillebrands B, Blumenr\"{o}der S, G\"{u}ntherodt G, Loewenhaupt M,
  Carpenter J~M, Winzer K and Fisk Z 1984 {\em Phys. Rev. B\/} {\bf 30}
  4052--4054

\bibitem{Lee72}
Lee K and Bell B 1972 {\em Phys. Rev. B\/} {\bf 6} 1032--1040

\bibitem{Fuj80}
Fujita T, Suzuki M, Komatsubara T, Kunii S, Kasuya T and Ohtsuka T 1980 {\em
  Solid State Commun.\/} {\bf 35} 569--572

\bibitem{Bur82}
Burlet P, Rossat-Mignod J, Effantin J~M, Kasuya T, Kunii S and Komatsubara T
  1982 {\em J. Appl. Phys.\/} {\bf 53} 2149--2151

\bibitem{Eff85a}
Effantin J, Rossat-Mignod J, Burlet P, Bartholin H, Kunii S and Kasuya T 1985
  {\em J. Magn. Magn. Mater.\/} {\bf 47--48} 145--148

\bibitem{Kob00}
Kobayashi T, Hashimoto K, Eda S, Shimizu K, Amaya K and Onuki Y 2000 {\em
  Physica B: Condens. Matter\/} {\bf 281--282} 553--554

\bibitem{Tay97}
Tayama T, Sakakibara T, Tenya K, Amitsuka H and Kunii S 1997 {\em J. Phys. Soc.
  Jpn.\/} {\bf 66} 2268

\bibitem{Fri15}
Friemel G, Jang H, Schneidewind A, Ivanov A, Dukhnenko A~V, Shitsevalova N~Y,
  Filipov V~B, Keimer B and Inosov D~S 2015 {\em Phys. Rev. B\/} {\bf 92}(1)
  014410

\bibitem{Kob00a}
Kobayashi S, Sera M, Hiroi M, Kobayashi N and Kunii S 2000 {\em J. Phys. Soc.
  Jpn.\/} {\bf 69} 926--936

\bibitem{Mat68}
Matthias B~T, Geballe T~H, Andres K, Corenzwit E, Hull G~W and Maita J~P 1968
  {\em Science\/} {\bf 159} 530

\bibitem{Kun10}
Kunimori K, Tanida H, Matsumura T, Sera M and Iga F 2010 {\em J. Phys. Soc.
  Jpn.\/} {\bf 79} 073703

\bibitem{Erk87}
Erkelens W, Regnault L, Burlet P, Rossat-Mignod J, Kunii S and Kasuya T 1987
  {\em J. Magn. Magn. Mater.\/} {\bf 63--64} 61--63

\bibitem{Suz05}
Suzuki O, Nakamura S, Akatsu M, Nemoto Y, Goto T and Kunii S 2005 {\em J. Phys.
  Soc. Jpn.\/} {\bf 74} 735--741

\bibitem{Hir98}
Hiroi M, Kobayashi S~I, Sera M, Kobayashi N and Kunii S 1998 {\em J. Phys. Soc.
  Jpn.\/} {\bf 67} 53--56

\bibitem{Eff85}
Effantin J~M 1985 {\em \'{E}tude par {D}iffusion des {N}eutrons des composes\/}
  Ph.D. thesis L'Universit\'e Scientifique et Medical de Grenoble

\bibitem{Bur82a}
{Burlet, P}, {Boucherle, J X}, {Rossat-Mignod, J}, {Cable, J W}, {Koehler, W
  C}, {Kunii, S} and {Kasuya, T} 1982 {\em J. Physics Colloques\/} {\bf 43}
  273--278

\bibitem{Goo04}
Goodrich R~G, Young D~P, Hall D, Balicas L, Fisk Z, Harrison N, Betts J,
  Migliori A, Woodward F~M and Lynn J~W 2004 {\em Phys. Rev. B\/} {\bf 69}
  054415

\bibitem{Hor81}
Horn S, Steglich F, Loewenhaupt M, Scheuer H, Felsch W and Winzer K 1981 {\em
  Zeitschrift f\"{u}r Physik B Condens. Matter\/} {\bf 42} 125--134

\bibitem{Bog06}
Bogach A, Glushkov V, Demishev S, Samarin N, Paderno Y, Dukhnenko A,
  Shitsevalova N and Sluchanko N 2006 {\em J. Sol. State Chem.\/} {\bf 179}
  2819--2822 special Issue: Boron, Borides and Related Compounds: Proceedings
  of the 15th International Symposium on Boron, Borides and Related Compounds

\bibitem{Kaw80}
Kawakami M, Kunii S, Komatsubara T and Kasuya T 1980 {\em Solid State
  Commun.\/} {\bf 36} 435--439

\bibitem{Slu07}
Sluchanko N, Bogach A, Glushkov V, Demishev S, Ivanov V, Ignatov M, Kuznetsov
  A, Samarin N, Semeno A and Shitsevalova N 2007 {\em J. Exp. Theor. Phys.\/}
  {\bf 104} 120--138

\bibitem{Bra85}
Brandt N, Moshchalkov V, Pashkevich S, Vybornov M, Semenov M, Kolobyanina T,
  Konovalova E and Paderno Y 1985 {\em Solid State Commun.\/} {\bf 56} 937--941

\bibitem{Sul94}
Sullow S, Trappe V, Eichler A and Winzer K 1994 {\em J. Phys.: Condens.
  Matter\/} {\bf 6} 10121

\bibitem{Kom83}
Komatsubara T, Sato N, Kunii S, Oguro I, Furukawa Y, Onuki Y and Kasuya T 1983
  {\em J. Magn. Magn. Mater.\/} {\bf 31-34, Part 1} 368--372

\bibitem{Han01}
Hane S, Goto T, Mitamura H and Kunii S 2001 {\em J. Magn. Magn. Mater.\/} {\bf
  226--230, Part 1} 98--100 proceedings of the International Conference on
  Magnetism (ICM 2000)

\bibitem{Mat14}
Matsumura T, Michimura S, Inami T, Otsubo T, Tanida H, Iga F and Sera M 2014
  {\em Phys. Rev. B\/} {\bf 89} 014422

\bibitem{Suz98}
Suzuki O, Goto T, Nakamura S, Matsumura T and Kunii S 1998 {\em J. Phys. Soc.
  Jpn.\/} {\bf 67} 4243--4250

\bibitem{Kob03}
Kobayashi S, Yoshino Y, Tsuji S, Tou H, Sera M and Iga F 2003 {\em J. Phys.
  Soc. Jpn.\/} {\bf 72} 2947--2954

\bibitem{Jan14}
Jang H, Friemel G, Ollivier J, Dukhnenko A~V, Shitsevalova N~Y, Filipov V~B,
  Keimer B and Inosov D~S 2014 {\em Nature Mater.\/} {\bf 13} 682--687

\bibitem{Gru85}
Grushko Y~S, Paderno Y~B, Ya~Mishin K, Molkanov L~I, Shadrina G~A, Konovalova
  E~S and Dudnik E~M 1985 {\em phys. stat. sol. (b)\/} {\bf 128} 591--597

\bibitem{Lan54}
Languet-Higgins H~C and de~V~Roberts M 1954 {\em Proc. Roy. Soc.\/} {\bf A224}
  336

\bibitem{Loe85}
Loewenhaupt M, Carpenter J and Loong C~K 1985 {\em J. Magn. Magn. Mater.\/}
  {\bf 52} 245--249

\bibitem{Sat84}
Sato N, Kunii S, Oguro I, Komatsubara T and Kasuya T 1984 {\em J. Phys. Soc.
  Jpn.\/} {\bf 53} 3967--3979

\bibitem{Kas81}
Kasuya T, Takegahara K, Aoki Y, Hanzawa K~, MKasaya, Kunii S, Fujita T, Sato N,
  Kimura H, Komatsubara T, Furuno T and Rossat-Mignod J 1981 {\em Valance
  Fluctuations in Solds\/} (North-Holland, Amsterdam)

\bibitem{Got83}
Goto T, Tamaki A, Kunii S, Nakajima T, Fujimura T, Kasuya T, Komatsubarra T and
  Woods S 1983 {\em J. Magn. Magn. Mater.\/} {\bf 31--34, Part 1} 419--420

\bibitem{Eff82}
Effantin J, Burlet P, Rossat-Mignod J, Kunii S and Kasuya T 1982 {\em Proc.
  Int. Conf. Val. Inst., Zurich\/} Edited by P. Wachter and H. Boppart

\bibitem{Zah03}
Zaharko O, Fischer P, Schenck A, Kunii S, Brown P~J, Tasset F and Hansen T 2003
  {\em Phys. Rev. B\/} {\bf 68} 214401

\bibitem{Fey94}
Feyerherm R, Amato A, Gygax F, Schenck A, \={O}nuki Y and Sato N 1994 {\em
  Physica B: Condens. Matter\/} {\bf 194--196, Part 1} 357--358

\bibitem{Fey95}
Feyerherm R, Amato A, Gygax F, Schenck A, \={O}nuki Y and Sato N 1995 {\em J.
  Magn. Magn. Mater.\/} {\bf 140--144, Part 2} 1175--1176 international
  Conference on Magnetism

\bibitem{Kus01}
Kusunose H and Kuramoto Y 2001 {\em J. Phys. Soc. Jpn.\/} {\bf 70} 1751--1761

\bibitem{Kun11}
Kunimori K, Kotani M, Funaki H, Tanida H, Sera M, Matsumura T and Iga F 2011
  {\em J. Phys. Soc. Jpn.\/} {\bf 80} SA056

\bibitem{Mig87}
Rossat-Mignod J 1987 {\em Methods of Experimental Physics: Neutron Scattering
  in Condens. Matter Research, volume 23C\/} (Academic Press Inc.)

\bibitem{Han84}
Hanzawa K and Kasuya T 1984 {\em J. Phys. Soc. Jpn.\/} {\bf 53} 1809--1818

\bibitem{Ohk83}
Ohkawa J~F 1983 {\em J. Phys. Soc. Jpn.\/} {\bf 52} 3897--3906

\bibitem{Ohk85}
Ohkawa J~F 1985 {\em J. Phys. Soc. Jpn.\/} {\bf 54} 3909--3914

\bibitem{Shi97}
Shiina R, Shiba H and Thalmeier P 1997 {\em J. Phys. Soc. Jpn.\/} {\bf 66}
  1741--1755

\bibitem{Kaw81}
Kawakami M, Kunii S, Mizuno K, Sugita M, Kasuya T and Kume K 1981 {\em J. Phys.
  Soc. Jpn.\/} {\bf 50} 432--437

\bibitem{Tak83}
Takigawa M, Yasuoka H, Tanaka T and Ishizawa Y 1983 {\em J. Phys. Soc. Jpn.\/}
  {\bf 52} 728--731

\bibitem{Ser01}
Sera M, Ichikawa H, Yokoo T, Akimitsu J, Nishi M, Kakurai K and Kunii S 2001
  {\em Phys. Rev. Lett.\/} {\bf 86} 1578--1581

\bibitem{Nak01}
Nakao H, Magishi K~i, Wakabayashi Y, Murakami Y, Koyama K, Hirota K, Endoh Y
  and Kunii S 2001 {\em J. Phys. Soc. Jpn.\/} {\bf 70} 1857--1860

\bibitem{Yak01}
Yakhou F, Plakhty V, Suzuki H, Gavrilov S, Burlet P, Paolasini L, Vettier C and
  Kunii S 2001 {\em Phys. Lett. A\/} {\bf 285} 191--196

\bibitem{Mur98}
Murakami Y, Kawada H, Kawata H, Tanaka M, Arima T, Moritomo Y and Tokura Y 1998
  {\em Phys. Rev. Lett.\/} {\bf 80} 1932--1935

\bibitem{Shi98}
Shiina R, Sakai O, Shiba H and Thalmeier P 1998 {\em J. Phys. Soc. Jpn.\/} {\bf
  67} 941--949

\bibitem{Kas97}
Kasuya T 1997 {\em J. Magn. Magn. Mater.\/} {\bf 174} L28--L32

\bibitem{Lov02}
Lovesey S~W 2002 {\em J. Phys.: Condens. Matter\/} {\bf 14} 4415

\bibitem{Ser88}
Sera M, Sato N and Kasuya T 1988 {\em J. Phys. Soc. Jpn.\/} {\bf 57} 1412--1423

\bibitem{San09}
Santini P, Carretta S, Amoretti G, Caciuffo R, Magnani N and Lander G~H 2009
  {\em Rev. Mod. Phys.\/} {\bf 81} 807--863

\bibitem{Kur09}
Kuramoto Y, Kusunose H and Kiss A 2009 {\em J. Phys. Soc. Jpn.\/} {\bf 78}
  072001

\bibitem{Uim96}
Uimin G, Kuramoto Y and Fukushima N 1996 {\em Solid State Commun.\/} {\bf 97}
  595--598

\bibitem{Sak97}
Sakai O, Shiina R, Shiba H and Thalmeier P 1997 {\em J. Phys. Soc. Jpn.\/} {\bf
  66} 3005--3007

\bibitem{Tha98}
Thalmeier P, Shiina R, Shiba H and Sakai O 1998 {\em J. Phys. Soc. Jpn.\/} {\bf
  67} 2363--2371

\bibitem{Tha03}
Thalmeier P, Shiina R, Shiba H, Takahashi A and Sakai O 2003 {\em J. Phys. Soc.
  Jpn.\/} {\bf 72} 3219--3225

\bibitem{Tha04}
Thalmeier P, Shiina R, Shiba H, Takahashi A and Sakai O 2004 {\em Physica B:
  Condens. Matter\/} {\bf 350} E35--E38 proceedings of the Third European
  Conference on Neutron Scattering

\bibitem{Ser99}
Sera M and Kobayashi S 1999 {\em J. Phys. Soc. Jpn.\/} {\bf 68} 1664--1678

\bibitem{Mat12}
Matsumura T, Yonemura T, Kunimori K, Sera M, Iga F, Nagao T and Igarashi J~I
  2012 {\em Phys. Rev. B\/} {\bf 85} 174417

\bibitem{Kob99}
Kobayashi S~i, Sera M, Hiroi M, Kobayashi N and Kunii S 1999 {\em J. Phys. Soc.
  Jpn.\/} {\bf 68} 3407--3412

\bibitem{Fri12}
Friemel G, Li Y, Dukhnenko A, Shitsevalova N, Sluchanko N, Ivanov A, Filipov V,
  Keimer B and Inosov D 2012 {\em Nature Commun.\/} {\bf 3} 830

\bibitem{Sat85}
Sato N, Sumiyama A, Kunii S, Nagano H and Kasuya T 1985 {\em J. Phys. Soc.
  Jpn.\/} {\bf 54} 1923--1932

\bibitem{Onu89a}
\={O}nuki Y, Komatsubara T, H~P~Reinders P and Springford M 1989 {\em J. Phys.
  Soc. Jpn.\/} {\bf 58} 3698--3704

\bibitem{Mul88}
M\"{u}ller T, Joss W, van Ruitenbeek J, Welp U, Wyder P and Fisk Z 1988 {\em J.
  Magn. Magn. Mater.\/} {\bf 76-77} 35--36

\bibitem{Jos87}
Joss W, van Ruitenbeek J~M, Crabtree G~W, Tholence J~L, van Deursen A~P~J and
  Fisk Z 1987 {\em Phys. Rev. Lett.\/} {\bf 59} 1609--1612

\bibitem{Deu85}
van Deursen A~P~J, Pols R~E, de~Vroomen A~R and Fisk Z 1985 {\em J. Less Common
  Metals\/} {\bf 111} 331--334

\bibitem{End06}
Endo M, Nakamura S, Isshiki T, Kimura N, Nojima T, Aoki H, Harima H and Kunii S
  2006 {\em J. Phys. Soc. Jpn.\/} {\bf 75} 114704

\bibitem{MatsuiGoto93}
Matsui H, Goto T, Kunii S and Sakatsume S 1993 {\em Physica B: Condens.
  Matter\/} {\bf 186--188} 126--128

\bibitem{Goo99}
Goodrich R~G, Harrison N, Teklu A, Young D and Fisk Z 1999 {\em Phys. Rev.
  Lett.\/} {\bf 82} 3669--3672

\bibitem{Ish77}
Ishizawa Y, Tanaka T, Bannai E and Kawai S 1977 {\em J. Phys. Soc. Jpn.\/} {\bf
  42} 112--118

\bibitem{Har93}
Harrison N, Meeson P, Probst P~A and Springford M 1993 {\em J. Phys.: Condens.
  Matter\/} {\bf 5} 7435

\bibitem{Har98}
Harrison N, Hall D~W, Goodrich R~G, Vuillemin J~J and Fisk Z 1998 {\em Phys.
  Rev. Lett.\/} {\bf 81} 870--873

\bibitem{Tek00}
Teklu A~A, Goodrich R~G, Harrison N, Hall D, Fisk Z and Young D 2000 {\em Phys.
  Rev. B\/} {\bf 62} 12875--12881

\bibitem{Car80}
McCarthy C and Tompson C 1980 {\em J. Phys. Chem. Solids\/} {\bf 41} 1319--1321

\bibitem{Goo06}
Goodrich R~G, Harrison N and Fisk Z 2006 {\em Phys. Rev. Lett.\/} {\bf 97}
  146404

\bibitem{Onu89}
\={O}nuki Y, Umezawa A, Kwok W~K, Crabtree G~W, Nishihara M, Yamazaki T, Omi T
  and Komatsubara T 1989 {\em Phys. Rev. B\/} {\bf 40} 11195--11207

\bibitem{Pau85}
Paulus E and Voss G 1985 {\em J. Magn. Magn. Mater.\/} {\bf 47-48} 539--541

\bibitem{Sch04}
Schenck A, Gygax F~N, Solt G, Zaharko O and Kunii S 2004 {\em Phys. Rev.
  Lett.\/} {\bf 93} 257601

\bibitem{Dem05}
Demishev S~V, Semeno A~V, Paderno Y~B, Shitsevalova N~Y and Sluchanko N~E 2005
  {\em phys. stat. sol. (b)\/} {\bf 242} R27--R29

\bibitem{Nak04}
Nakatsuji S, Pines D and Fisk Z 2004 {\em Phys. Rev. Lett.\/} {\bf 92} 016401

\bibitem{Yan12}
Yang Y and Pines D 2012 {\em Proc. Nat. Acad. Sci.\/} {\bf 109} E3060--E3066

\bibitem{Cur04}
Curro N~J, Young B~L, Schmalian J and Pines D 2004 {\em Phys. Rev. B\/} {\bf
  70} 235117

\bibitem{Pla05}
Plakhty V~P, Regnault L~P, Goltsev A~V, Gavrilov S~V, Yakhou F, Flouquet J,
  Vettier C and Kunii S 2005 {\em Phys. Rev. B\/} {\bf 71} 100407

\bibitem{Ger12}
Friemel G, Li Y, Dukhnenko A, Shitsevalova N, Sluchanko N, Ivanov A, Filipov V,
  Keimer B and Inosov D 2012 {\em Nature Commun.\/} {\bf 3} 830

\bibitem{Bou93}
Bouvet A 1993 {\em \'{E}tude par diffusion in\'elastique des neutrons des
  propri\'et\'es magn\'etiques des borures de terre rare: {C}e{B}$_{6}$,
  {P}r{B}$_{6}$ et {Y}b{B}$_{12}$\/} Ph.D. thesis L'Universit\'e Joseph
  Fourrier

\bibitem{Dem09}
Demishev S~V, Semeno A~V, Bogach A~V, Samarin N~A, Ishchenko T~V, Filipov V~B,
  Shitsevalova N~Y and Sluchanko N~E 2009 {\em Phys. Rev. B\/} {\bf 80} 245106

\bibitem{Sch12}
Schlottmann P 2012 {\em Phys. Rev. B\/} {\bf 86} 075135

\bibitem{Kre08}
Krellner C, F\"orster T, Jeevan H, Geibel C and Sichelschmidt J 2008 {\em Phys.
  Rev. Lett.\/} {\bf 100} 066401

\bibitem{Dem09b}
Demishev S, Semeno A, Ohta H, Okubo S, Paderno Y, Shitsevalova N and Sluchanko
  N 2009 {\em Appl. Magn. Res.\/} {\bf 35} 319--326

\bibitem{Slu08a}
Sluchanko N, Bogach A, Glushkov V, Demishev S, Ignatov M, Kuznetsov A and
  Shitsevalova N 2008 {\em Physica B: Condens. Matter\/} {\bf 403} 742--743
  proceedings of the International Conference on Strongly Correlated Electron
  Systems

\bibitem{Slu08b}
Sluchanko N, Glushkov V, Demishev S, Samarin N, Bogach A, Gon'kov K, Khayrullin
  E, Filipov V and Shitsevalova N 2008 {\em Physica B: Condens. Matter\/} {\bf
  403} 1393--1394 proceedings of the International Conference on Strongly
  Correlated Electron Systems

\bibitem{Ign06}
Ignatov M, Bogach A, Glushkov V, Demishev S, Paderno Y, Shitsevalova N and
  Sluchanko N 2006 {\em Physica B: Condens. Matter\/} {\bf 378--380} 780--781
  proceedings of the International Conference on Strongly Correlated Electron
  Systems

\bibitem{Col85}
Coleman P, Anderson P~W and Ramakrishnan T~V 1985 {\em Phys. Rev. Lett.\/} {\bf
  55} 414--417

\bibitem{Had86}
Had\v{z}i\'{c}-Leroux M, Hamzi\'{c} A, Fert A, Haen P, Lapierre F and Laborde O
  1986 {\em Europhys. Lett.\/} {\bf 1} 579

\bibitem{Slu15}
Sluchanko N, Anisimov M, Bogach A, Voronov V, Gavrilkin S, Glushkov V, Demishev
  S, Krasnorusskii V, Filippov V and Shitsevalova N 2015 {\em JETP Lett.\/}
  {\bf 101} 36

\bibitem{Fri14}
Friemel G 2014 {\em Itinerant Spin Dynamics in Iron-based Superconductors and
  Cerium-based Heavy-Fermion Antiferromagnets\/} Ph.D. thesis Fakult\"at
  Mathematik und Physik, Universit\"at Stuttgart

\bibitem{Kus08}
Kusunose H 2008 {\em J. Phys. Soc. Jpn.\/} {\bf 77} 064710

\bibitem{Shi03}
Shiina R, Shiba H, Thalmeier P, Takahashi A and Sakai O 2003 {\em J. Phys. Soc.
  Jpn.\/} {\bf 72} 1216--1225

\bibitem{Reg88}
Regnault L, Erkelens W, Rossat-Mignod J, Vettier C, Kunii S and Kasuya T 1988
  {\em J. Magn. Magn. Mater.\/} {\bf 76-77} 413--414

\bibitem{Fon95}
Fong H~F, Keimer B, Anderson P~W, Reznik D, Do\ifmmode~\breve{g}\else
  \u{g}\fi{}an F and Aksay I~A 1995 {\em Phys. Rev. Lett.\/} {\bf 75} 316--319

\bibitem{Ino10}
Inosov D~S, Park J~T, Bourges P, Sun D~L, Sidis Y, Schneidewind A, Hradil K,
  Haug D, Lin C~T, Keimer B and Hinkov V 2010 {\em Nature Phys.\/} {\bf 6}
  178--181

\bibitem{Sto08}
Stock C, Broholm C, Hudis J, Kang H~J and Petrovic C 2008 {\em Phys. Rev.
  Lett.\/} {\bf 100} 087001

\bibitem{Sto11}
Stockert O, Arndt J, Faulhaber E, Geibel C, Jeevan H~S, Kirchner S, Loewenhaupt
  M, Schmalzl K, Schmidt W, Si Q and Steglich F 2011 {\em Nature Phys.\/} {\bf
  7} 119--124

\bibitem{Akb12}
Akbari A and Thalmeier P 2012 {\em Phys. Rev. Lett.\/} {\bf 108} 146403

\bibitem{Sat01}
Sato N~K, Aso N, Miyake K, Shiina R, Thalmeier P, Varelogiannis G, Geibel C,
  Steglich F, Fulde P and Komatsubara T 2001 {\em Nature\/} {\bf 410} 340--343

\bibitem{Fuh15}
Fuhrman W\ T, Leiner J, Nikoli\ifmmode~\acute{c}\else \'{c}\fi{} P, Granroth G\
  E, Stone M\ B, Lumsden M\ D, DeBeer-Schmitt L, Alekseev P\ A, Mignot J~M,
  Koohpayeh S\ M, Cottingham P, Phelan W~A, Schoop L, McQueen T\ M and Broholm
  C 2015 {\em Phys. Rev. Lett.\/} {\bf 114} 036401

\bibitem{Wie07}
Wiebe C~R, Janik J~A, MacDougall G~J, Luke G~M, Garrett J~D, Zhou H~D, Jo Y~J,
  Balicas L, Qiu Y, Copley J~R~D, Yamani Z and Buyers W~J~L 2007 {\em Nature
  Phys.\/} {\bf 3} 96

\bibitem{Ale95}
Alekseev P~A, Mignot J~M, Rossat-Mignod J, Lazukov V~N, Sadikov I~P, Konovalova
  E~S and Paderno Y~B 1995 {\em J. Phys.: Condens. Matter\/} {\bf 7} 289

\bibitem{Nem07}
Nemkovski K~S, Mignot J~M, Alekseev P~A, Ivanov A~S, Nefeodova E~V, Rybina A~V,
  Regnault L~P, Iga F and Takabatake T 2007 {\em Phys. Rev. Lett.\/} {\bf 99}
  137204

\bibitem{Mig05}
Mignot J~M, Alekseev P~A, Nemkovski K~S, Regnault L~P, Iga F and Takabatake T
  2005 {\em Phys. Rev. Lett.\/} {\bf 94} 247204

\bibitem{Bou10}
Bourdarot F, Hassinger E, Raymond S, Aoki D, Taufour V, Regnault L~P and
  Flouquet J 2010 {\em J. Phys. Soc. Jpn.\/} {\bf 79} 064719

\bibitem{Chu08}
Chubukov A~V and Gor'kov L~P 2008 {\em Phys. Rev. Lett.\/} {\bf 101} 147004

\bibitem{Fur85}
Furuno T, Sato N, Kunii S, Kasuya T and Sasaki W 1985 {\em J. Phys. Soc.
  Jpn.\/} {\bf 54} 1899--1905

\bibitem{Low73}
Lowndes D~H, Miller K~M, Poulsen R~G and Springford M 1973 {\em Proc. Roy. Soc.
  A\/} {\bf 331} 497--523

\bibitem{Sho84}
Shoenberg D 1984 {\em Magnetic Oscillations in Metals\/} (Cambridge University
  Press)

\bibitem{Hir97}
Hiroi M, Sera M, Kobayashi N and Kunii S 1997 {\em Phys. Rev. B\/} {\bf 55}
  8339--8346

\bibitem{Por15a}
Portnichenko P 2015 {\em Unpublished\/}

\bibitem{Nak06}
Nakamura S, Endo M, Yamamoto H, Isshiki T, Kimura N, Aoki H, Nojima T, Otani S
  and Kunii S 2006 {\em Phys. Rev. Lett.\/} {\bf 97} 237204

\bibitem{Man05}
Mannix D, Tanaka Y, Carbone D, Bernhoeft N and Kunii S 2005 {\em Phys. Rev.
  Lett.\/} {\bf 95} 117206

\bibitem{Fis05}
Fischer P, Iwasa K, Kuwahara K, Kohgi M, Hansen T and Kunii S 2005 {\em Phys.
  Rev. B\/} {\bf 72} 014414

\bibitem{Iwa03}
Iwasa K, Kuwahara K, Kohgi M, Fischer P, D\"{o}nni A, Keller L, Hansen T, Kunii
  S, Metoki N, Koike Y and Ohoyama K 2003 {\em Physica B: Condens. Matter\/}
  {\bf 329--333, Part 2} 582--583 proceedings of the 23rd International
  Conference on Low Temperature Physics

\bibitem{Sch07}
Schenck A, Gygax F~N and Solt G 2007 {\em Phys. Rev. B\/} {\bf 75} 024428

\bibitem{Kus05}
Kusunose H and Kuramoto Y 2005 {\em J. Phys. Soc. Jpn.\/} {\bf 74} 3139--3142

\bibitem{Kuw07}
Kuwahara K, Iwasa K, Kohgi M, Aso N, Sera M and Iga F 2007 {\em J. Phys. Soc.
  Jpn.\/} {\bf 76} 093702

\bibitem{Kub04}
Kubo K and Kuramoto Y 2004 {\em J. Phys. Soc. Jpn.\/} {\bf 73} 216--224

\bibitem{Pai02}
Paix\~ao J~A, Detlefs C, Longfield M~J, Caciuffo R, Santini P, Bernhoeft N,
  Rebizant J and Lander G~H 2002 {\em Phys. Rev. Lett.\/} {\bf 89} 187202

\bibitem{Cac03}
Caciuffo R, {Paix\~ao} J~A, Detlefs C, Longfield M~J, Santini P, Bernhoeft N,
  Rebizant J and Lander G~H 2003 {\em J. Phys.: Condens. Matter\/} {\bf 15}
  S2287

\bibitem{Yos05}
Yoshizawa M, Nakanishi Y, Oikawa M, Sekine C, Shirotani I, Saha S~R, Sugawara H
  and Sato H 2005 {\em J. Phys. Soc. Jpn.\/} {\bf 74} 2141--2144

\bibitem{Aok08}
Aoki Y, Sanada S, Kikuchi D, Sugawara H and Sato H 2008 {\em Physica B:
  Condens. Matter\/} {\bf 403} 1574--1576 {P}roceedings of the International
  Conference on Strongly Correlated Electron Systems

\bibitem{Myd11}
Mydosh J~A and Oppeneer P~M 2011 {\em Rev. Mod. Phys.\/} {\bf 83}(4) 1301--1322

\bibitem{Shi12}
Takagi S, Ishihara S, Yokoyama M and Amitsuka H 2012 {\em J. Phys. Soc. Jpn.\/}
  {\bf 81} 114710

\bibitem{Sil06}
Silhanek A~V, Ebihara T, Harrison N, Jaime M, Tezuka K, Fanelli V and Batista
  C~D 2006 {\em Phys. Rev. Lett.\/} {\bf 96} 206401

\bibitem{Col01}
Coleman P, P\'epin C, Si Q and Ramazashvili R 2001 {\em J. Phys.: Condens.
  Matter\/} {\bf 13} R723

\bibitem{Nak00}
Nakamura S, Goto T, Suzuki O, Kunii S and Sakatsume S 2000 {\em Phys. Rev. B\/}
  {\bf 61}(22) 15203

\bibitem{Nak03}
Nakamura S, Endo M, Aoki H, Kimura N, Nojima T and Kunii S 2003 {\em Phys. Rev.
  B\/} {\bf 68}(10) 100402

\bibitem{Ste01}
Stewart G~R 2001 {\em Rev. Mod. Phys.\/} {\bf 73} 797--855

\bibitem{Mir05}
Miranda E and Dobrosavljevi\'c V 2005 {\em Rep. Prog. Phys.\/} {\bf 68} 2337

\bibitem{Cus03}
Custers J, Gegenwart P, Wilhelm H, Neumaier K, Tokiwa Y, Trovarelli O, Geibel
  C, Steglich F, Pepin C and Coleman P 2003 {\em Nature\/} {\bf 424} 524--527

\bibitem{Cus12}
Custers J, Lorenzer K~A, M\"uller M, Prokofiev A, Sidorenko A, Winkler H,
  Strydom A~M, Shimura Y, Sakakibara T, Yu R, Si Q and Paschen S 2012 {\em
  Nature Mater.\/} {\bf 11} 189--194

\bibitem{Sum86}
Sumiyama A, Oda Y, Nagano H, \={O}nuki Y, Shibutani K and Komatsubara T 1986
  {\em J. Phys. Soc. Jpn.\/} {\bf 55} 1294--1304

\bibitem{Sob79}
Sobczak R~J and Sienko M 1979 {\em J. Less Common Metals\/} {\bf 67} 167--171

\bibitem{Bat95}
Bat'ko I, Bat'kov\'a M, Flachbart K, Filippov V, Paderno Y, Shicevalova N and
  Wagner T 1995 {\em J. Alloys Compd.\/} {\bf 217} L1--L3

\bibitem{Sch82}
Schell G, Winter H, Rietschel H and Gompf F 1982 {\em Phys. Rev. B\/} {\bf 25}
  1589

\bibitem{Sam88}
Samuely P, Reiffers M, Flachbart K, Akimenko A, Yanson I, Ponomarenko N and
  Paderno Y 1988 {\em J. Low Temp. Phys.\/} {\bf 71} 49--61

\bibitem{Kon07}
Kondo A, Tou H, Sera M, Iga F and Sakakibara T 2007 {\em J. Phys. Soc. Jpn.\/}
  {\bf 76} 103708

\bibitem{Kon09}
Kondo A, Taniguchi T, Tanida H, Matsumura T, Sera M, Iga F, Tou H, Sakakibara T
  and Kunii S 2009 {\em J. Phys. Soc. Jpn.\/} {\bf 78} 093708

\bibitem{Mat14a}
Matsumura T, Kunimori K, Kondo A, Soejima K, Tanida H, Mignot J~M, Iga F and
  Sera M 2014 {\em J. Phys. Soc. Jpn.\/} {\bf 83} 094724

\bibitem{Mig08}
Mignot J~M, Andr\'e G, Robert J, Sera M and Iga F 2008 {\em Phys. Rev. B\/}
  {\bf 78} 014415

\bibitem{Kis05}
Kishimoto S, Kondo A, Kim M~S, Tou H, Sera M and Iga F 2005 {\em J. Phys. Soc.
  Jpn.\/} {\bf 74} 2913--2916

\bibitem{Nag08}
Nagai S, Ikeda S, Iwakubo H, Tou H, Sera M and Iga F 2008 {\em J. Phys. Soc.
  Jpn.\/} {\bf 77} 288--290

\bibitem{Kur02}
Kuramoto Y and Kubo K 2002 {\em J. Phys. Soc. Jpn.\/} {\bf 71} 2633--2636

\bibitem{Doe00}
D\"onni A, Herrmannsd\"orfer T, Fischer P, Keller L, Fauth F, McEwen K~A, Goto
  T and Komatsubara T 2000 {\em J. Phys.: Condens. Matter\/} {\bf 12} 9441

\bibitem{Her99}
Herrmannsd\"orfer T, D\"onni A, Fischer P, Keller L, B\"ottger G, Gutmann M,
  Kitazawa H and Tang J 1999 {\em J. Phys.: Condens. Matter\/} {\bf 11} 2929

\bibitem{Her00}
Herrmannsd\"orfer T, D\"onni A, Fischer P, Keller L and Kitazawa H 2000 {\em
  Physica B: Condens. Matter\/} {\bf 281--282} 167--168

\bibitem{Don98}
D\"onni A, Keller L, Fischer P, Aoki Y, Sato H, Fauth F, Zolliker M,
  Komatsubara T and Endoh Y 1998 {\em J. Phys.: Condens. Matter\/} {\bf 10}
  7219

\bibitem{Don00a}
D\"onni A, Fauth F, Fischer P, Herrmannsd\"orfer T, Keller L and Komatsubara T
  2000 {\em J. Alloys Compd.\/} {\bf 306} 40--46

\bibitem{Mit10}
Mitamura H, Tayama T, Sakakibara T, Tsuduku S, Ano G, Ishii I, Akatsu M, Nemoto
  Y, Goto T, Kikkawa A and Kitazawa H 2010 {\em J. Phys. Soc. Jpn.\/} {\bf 79}
  074712

\bibitem{Lor12}
Lorenzer K~A 2012 Ph.D. thesis Vienna University of Technology

\bibitem{Lor15}
Lorenzer K~A, Strydom A, Deen P, Mignot J~M and Buehler-Paschen S {\em ILL Exp.
  Report\/}  \#5--41--601 (unpublished).

\bibitem{Pas07}
Paschen S, M\"uller M, Custers J, Kriegisch M, Prokofiev A, Hilscher G, Steiner
  W, Pikul A, Steglich F and Strydom A 2007 {\em J. Magn. Magn. Mater.\/} {\bf
  316} 90--92

\bibitem{Got09}
Goto T, Watanabe T, Tsuduku S, Kobayashi H, Nemoto Y, Yanagisawa T, Akatsu M,
  Ano G, Suzuki O, Takeda N, D\"onni A and Kitazawa H 2009 {\em J. Phys. Soc.
  Jpn.\/} {\bf 78} 024716

\bibitem{Wat07}
Watanabe T, Yamaguchi T, Nemoto Y, Goto T, Takeda N, Suzuki O and Kitazawa H
  2007 {\em J. Magn. Magn. Mater.\/} {\bf 310} 280--282 proceedings of the 17th
  International Conference on Magnetism

\bibitem{Por15}
Portnichenko P~Y, Cameron A~S, Surmach M~A, Deen P~P, Paschen S, Prokofiev A,
  Mignot J~M, Strydom A~M, Telling M~T~F, Podlesnyak A and Inosov D~S 2015 {\em
  Phys. Rev. B\/} {\bf 91} 094412

\bibitem{Pas08}
Paschen S, Laumann S, Prokofiev A, Strydom A, Deen P, Stewart J, Neumaier K,
  Goukassov A and Mignot J~M 2008 {\em Physica B: Condens. Matter\/} {\bf 403}
  1306--1308 proceedings of the International Conference on Strongly Correlated
  Electron Systems

\bibitem{Ben11}
Benlagra A, Fritz L and Vojta M 2011 {\em Phys. Rev. B\/} {\bf 84}(7) 075126

\bibitem{Yam15}
Yamaoka H, Schwier E~F, Arita M, Shimada K, Tsujii N, Jarrige I, Jiang J,
  Hayashi H, Iwasawa H, Namatame H, Taniguchi M and Kitazawa H 2015 {\em Phys.
  Rev. B\/} {\bf 91}(11) 115139

\end{thebibliography}

\end{document}